\documentclass[10pt,letterpaper]{article}
\usepackage[top=0.85in,left=2.75in,footskip=0.75in]{geometry}

\usepackage{amsmath,amssymb}

\usepackage{changepage}

\usepackage{textcomp,marvosym}

\usepackage{cite}

\usepackage{nameref,hyperref}

\usepackage[right]{lineno}

\usepackage[nopatch=eqnum]{microtype}
\DisableLigatures[f]{encoding = *, family = * }

\usepackage[table]{xcolor}

\usepackage{array}

\newcolumntype{+}{!{\vrule width 2pt}}

\newlength\savedwidth

\usepackage{setspace} 

\raggedright
\usepackage[aboveskip=1pt,labelfont=bf,labelsep=period,justification=raggedright,singlelinecheck=off]{caption}

\makeatletter
\renewcommand{\@biblabel}[1]{\quad#1.}
\makeatother

\usepackage{lastpage,fancyhdr,graphicx}
\usepackage{epstopdf}
\fancyheadoffset[L]{2.25in}
\fancyfootoffset[L]{2.25in}
\usepackage{subcaption}
\usepackage{mathtools}
\usepackage{gensymb}
\usepackage{color}
\usepackage{wrapfig}
\usepackage{cancel}
\usepackage{MnSymbol,wasysym}
\usepackage{comment}
\usepackage{algorithm}
\usepackage{algpseudocode}
\usepackage{longtable}
\usepackage{multirow}
\usepackage{etoolbox}
\usepackage{booktabs}
\usepackage[noabbrev]{cleveref}
\usepackage{xfrac}
\usepackage{nicefrac}
\usepackage{placeins}

\usepackage[T1]{fontenc}    
\usepackage{lmodern}        

\usepackage{svg}

\begin{document}
\vspace*{0.2in}

\begin{flushleft}
{\Large
\textbf\newline{Uncertainty Quantification of Bacterial Microcompartment Permeability} 
}
\newline
\\
Andre G. Archer\textsuperscript{1},
Brett J. Palmero\textsuperscript{3},
Charlotte H. Abrahamson\textsuperscript{2},
Carolyn E. Mills\textsuperscript{2},
Nolan W. Kennedy\textsuperscript{3},
Danielle Tullman-Ercek\textsuperscript{2,3,4},
Niall M. Mangan\textsuperscript{1,4,*},
\\
\bigskip
\textbf{1} Department of Engineering Sciences and Applied Mathematics, Northwestern University, Evanston, IL, USA
\\
\textbf{2} Department of Chemical and Biological Engineering, Northwestern University, Evanston, IL, USA
\\
\textbf{3} Interdisciplinary Biological Sciences Program, Northwestern University, Evanston, IL, USA
\\
\textbf{4} Center for Synthetic Biology, Northwestern University, Evanston, IL, USA
\bigskip

%
%





* niall.mangan@northwestern.edu

\end{flushleft}
\section*{Abstract}
{\it Salmonella enterica serovar Typhimurium LT2} expresses bacterial microcompartments (MCPs) upon 1,2-propanediol exposure. MCPs are nanoscale protein-bound shells that encase enzymes that facilitate the cofactor-dependent metabolism of 1,2-propanediol. They are hypothesized to limit exposure to the toxic intermediate, propionaldehyde, decrease cofactor involvement in competing reactions, and enhance reaction flux. However, there has been limited research in quantifying these hypothesized MCP organizational benefits. In this work, we construct a mass-action mathematical model of purified MCPs and calibrate parameters to measured metabolite concentrations. To ensure model fidelity to existing research, we constrain mass-action kinetic parameters to previously estimated Michaelis-Menten parameters. After calibration, we identified two distinct fits with different dynamics in the pathway product, propionate, but similar goodness of fit. Visual inspection of the model fits indicates that the collection of additional data points between 6 and 24 hours would eliminate posterior bimodality and increase inference precision. Across fits, we inferred that the MCP 1,2-propanediol and propionaldehyde permeability should be greater than 10\textsuperscript{-6} and 10\textsuperscript{-8} m/s, respectively. Our results identify parameter ranges consistent with prevailing theories that MCPs impose preferential diffusion to 1,2-propanediol over propionaldehyde, and function to sequester toxic propionaldehyde away from the cell cytosol. The bimodality of the posterior distribution arises from bimodality in the estimated coenzyme-A (CoA) permeability and inhibition rates. The MCP permeability to CoA was inferred to be either less than 10\textsuperscript{-8.8} m/s or greater than 10\textsuperscript{-7.3} m/s, affecting the inferred activity of CoA-dependent enzymes. In a high CoA permeability environment with low rates of CoA inhibition, enzymes produced metabolites by recycling nicotinamide adenine dinucleotide (NAD+)/nicotinamide adenine dinucleotide hydride (NADH). In a low CoA permeability environment with high rates of CoA inhibition, enzymes required external NAD+/H to produce metabolites. Our analysis suggests that dynamics are consistent with prevailing hypotheses about MCP function to sequester toxic propionaldehyde, and that measurements of metabolites at additional time points or characterization of enzyme inhibition rates could further reduce uncertainty and provide better estimates of compartment permeability.

\section*{Author summary}

{\it Salmonella enterica} proliferate in oxygen- and nutrient-limited environments in part due to their ability to express bacterial microcompartments upon exposure to 1,2-propanediol. The functional properties by which MCPs confer a growth advantage to {\it Salmonella} have yet to be quantified. In this study, we construct and calibrate a high-fidelity model of 1,2-propanediol MCP metabolism. We report metabolite permeabilities estimated from experiments and include uncertainty bounds on these predictions. Our fits identify parameter regions consistent with the hypothesized Pdu MCP functions: reduction of cytosolic propionaldehyde burden and interaction with external cofactors. We also develop a framework to constrain mass-action kinetic parameters with commonly performed Michaelis-Menten measurements. Thus, future researchers could incorporate new kinetic measurements, update the model, and potentially expand our {\it in vitro} to an {\it in vivo} model of 1,2-propanediol growth.  

\section*{Introduction}

\begin{figure}[!ht]
\includegraphics[width=0.7\linewidth]{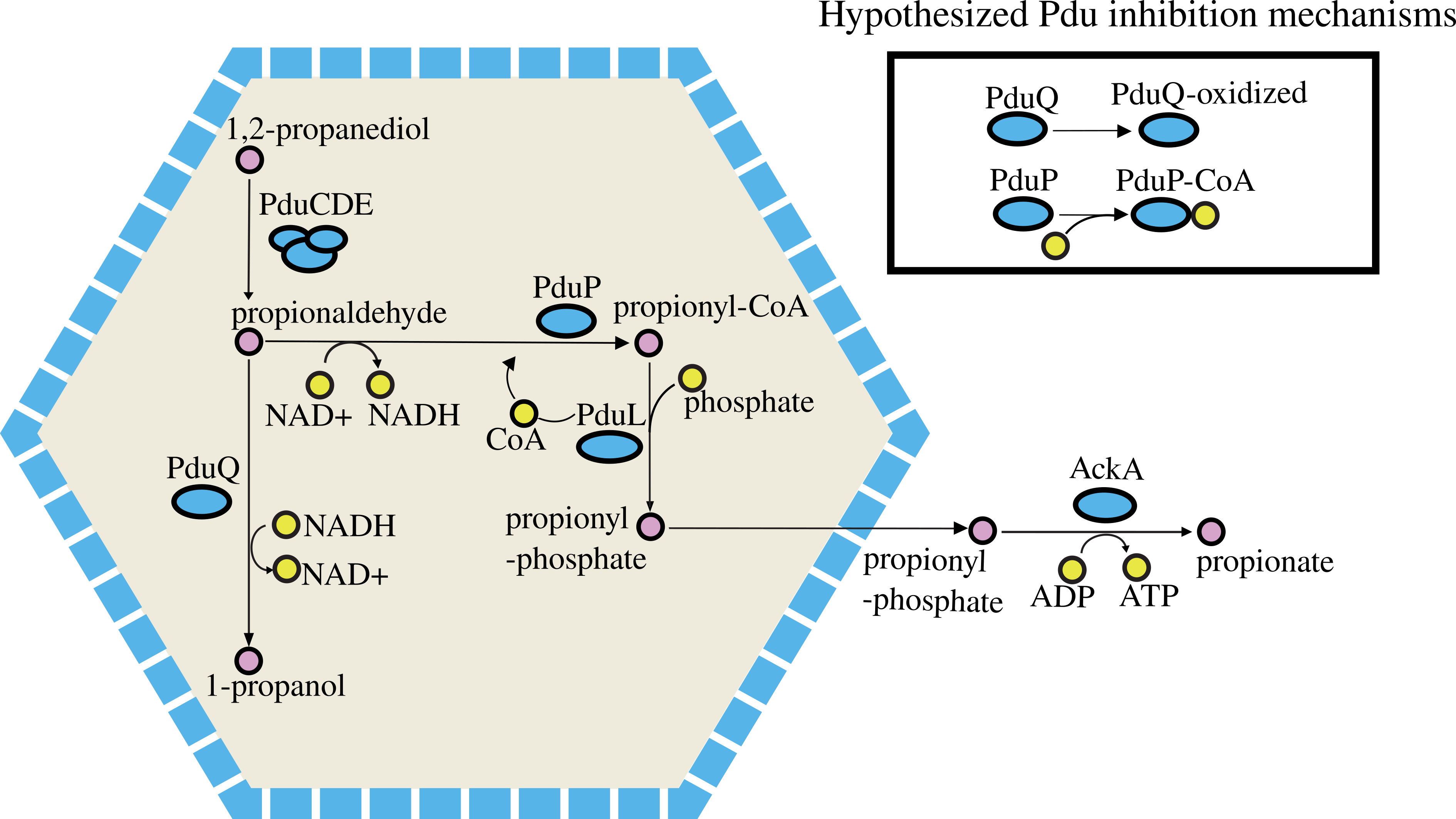}
\caption{Pdu Metabolic Pathway Network. Pdu MCPs house metabolic reactions that convert 1,2-propanediol  to 1-propanol and propionyl-CoA while producing propionaldehyde, a growth-inhibiting intermediate. Downstream PduL-facilitated reactions convert propionyl-CoA to propionate. Enzyme-inhibiting mechanisms, such as PduQ oxidation and PduP inhibition through complex formation, may occur in the compartment.  Cobalamin activation and diol dehydratase reactivation encapsulated pathways also occur in the compartment (not shown or modeled explicitly).}
\label{Fig1}
\end{figure}

{\it Salmonella enterica} serovar Typhimurium LT2 expresses a special class of MCPs upon 1,2-propanediol exposure called Pdu (1,2-\textbf{P}ropane\textbf{d}iol \textbf{u}tilization) MCPs. Pdu MCPs encapsulate a 5-step 1,2-propanediol metabolic reaction \cite{Yang2020} (Figure \ref{Fig1}). PduCDE, a diol dehydratase enzyme, first converts 1,2-propanediol to propionaldehyde, a toxic intermediate, in the presence of adenosylcobalamin (AdoB\textsubscript{12}) \cite{Bobik1997,Toraya2000}. Through NAD+/NADH electron transfer, propionaldehyde is then either oxidized by PduP to propionyl-CoA while converting NAD+ to NADH \cite{Leal2003} or reduced by PduQ to 1-propanol while converting NADH to NAD+ \cite{Cheng2012}. Lastly, PduL phosphorylates propionyl-CoA to propionyl-phosphate \cite{Liu2007} and PduW converts propionyl-phosphate and adenosine diphosphate (ADP) to propionate and adenosine triphosphate (ATP) \cite{Palacios2003}. Propionyl-CoA acts as a carbon source {\it in vivo}, allowing {\it Salmonella} to produce ATP through the methylcitrate and tricarboxylic cycles \cite{Palacios2003}.

Pdu MCPs increase LT2 growth on 1,2-propanediol by providing a selective barrier to metabolites and encasing ancillary reactions to facilitate PduCDE activation. Pdu MCPs limit cytosolic propionaldehyde exposure; LT2 mutants lacking MCP encapsulation undergo growth arrest correlated with an increase in extracellular propionaldehyde \cite{Sampson2008, Kennedy2022, Mills2022}. Compartment shell protein pores have been implicated in reducing propionaldehyde burden while still allowing substrate diffusion \cite{Park2017}. MCPs are also hypothesized to restrict cofactor diffusion, thereby enhancing PduQ and PduP flux and increasing downstream metabolite production \cite{Jakobson2017, Cheng2012}. Lastly, Pdu MCPs encapsulate ancillary reaction pathways for optimal 1,2-propanediol consumption. These pathways include cobalamin activation (PduSO) \cite{Cheng2010, Johnson2004, Yang2020}, which produces AdoB\textsubscript{12} from inactive B\textsubscript{12}, and diol dehydratase reactivation (PduGH), which repairs inactivated PduCDE \cite{Bobik1999, Yang2020}. Removal of PduSO enzymes resulted in impaired growth when exposed to 1,2-propanediol \cite{Cheng2010, Johnson2001}.

Pdu MCPs serve as a model system to study enzyme organization and metabolite sequestration in prokaryotes \cite{Lawrence2014,Nichols2020}. The efficacy of metabolite sequestration in the MCP depends strongly on its permeability. Understanding and quantifying MCP properties are also paramount to its further deployment as a metabolic engineering platform \cite{Jakobson2017, Jakobson2018}. Park et al. 2017 estimated that the PduA pore is 3-10 times more permeable to 1,2-propanediol than propionaldehyde using molecular dynamic simulations \cite{Park2017}. Young et al. 2025 quantified the bacterial microcompartment shell to be between $10^{-9}$- $10^{-6}$ m/s for ATP and ATP-related complexes \cite{Young2025}. Raza et al. 2025 quantified MCP pore permeability to be as low as 5 x 10\textsuperscript{-5} m/s for glyceraldehyde-3-phosphate and dihydroxyacetone phosphate \cite{Raza2025}. The latter two analyses were restricted to select reactants. 

Lastly, Palmero et al. 2025 recently quantified the absolute Pdu MCP permeability to be around $10^{-7}$ m/s by fitting a kinetic model to a metabolite profile of purified Pdu MCPs \cite{Palmero2025}. However, this analysis assumed a single permeability of the MCP to all metabolites and did not include parameter uncertainty when fitting their kinetic model. Differentiating and quantifying the uncertainty in the MCP permeability parameters would allow us to identify those constrained by data, identify mechanistic rules of metabolite interaction with the MCP environment, and infer parameter ranges consistent with existing kinetic measurements. 

In this work, we extend the analysis of Palmero et al. 2025 \cite{Palmero2025} to quantify uncertainty in MCP properties and Pdu pathway kinetics. Prior systems biology modeling efforts have incorporated uncertainty quantification into dynamic kinetic metabolic modeling using ensemble methods. Given a reaction network, ensemble modeling generates a set of mass-action kinetic parameters that satisfy hard fluxomics and thermodynamic constraints \cite{Tran2008}. Forward simulations of these parameterized models are used to produce flux predictions and rank optimal strains with uncertainty. While the predictive range of ensemble modeling is an improvement from standard point-wise fluxomic techniques\cite{Greene2017}, ensemble modeling suffers from inefficient sampling, the inability to fit dynamic data, and a lack of transferability of parameter fits across models. There have been extensions to the ensemble framework to include {\it in vitro} kinetics \cite{Khodayari2016, Martin2023}, particle filters \cite{Martin2023}, sensitivity analysis\cite{Martin2023, Greene2017}, and dynamic data\cite{Martin2023}. Despite offering improvements to the original ensemble framework, these methods still suffer from inefficient sampling and are not amenable to including {\it in vitro} kinetic measurements as a hard constraint. 

Bayesian methods provide comprehensive uncertainty quantification of model parameters and have been used recently in the calibration of kinetic metabolic \cite{saa2017formulation} and, more broadly, biological systems models \cite{linden2024increasing}, making them a promising approach.  For example, groups have applied Bayesian techniques to estimate kinetic parameters in central carbon metabolism \cite{murabito2014monte}, protein degradation \cite{liepe2015quantitative}, the methionine cycle \cite{saa2016probabilistic},  and the pentose phosphate pathway \cite{hurbain2022quantitative}. Most of these studies employ Michaelis-Menten or other quasi-steady-state formulations to simplify the model.  Uncertainty quantification can support Bayesian model selection and discriminate between underlying mechanisms \cite{liepe2015quantitative}. Approximate Bayesian Computation is a common strategy to reduce computational time \cite{saa2016probabilistic, liepe2015quantitative}. Notably, Saa and Nielsen developed a probabilistic framework using Approximate Bayesian Computation to estimate enzyme kinetics and regulatory control, and integrating thermodynamic constraints through normalization of parameters to a reference state and existing knowledge of partial kinetic information into the priors \cite{saa2016probabilistic}. As Bayesian methods improve, higher-complexity systems become accessible for uncertainty quantification.

We present a framework to fit kinetic models to metabolite time series data while maintaining strict consistency with prior {\it in vitro} kinetic measurements. We calibrate our model to a metabolite profile of purified Pdu MCPs with external AckA, similar to that published in Palmero et al. 2025 \cite{Palmero2025}. Exogenous AckA was necessary to ensure propionate kinase activity, as PduW is likely localized to the cytosol and not contained within purified MCPs \cite{Yang2020}. Our kinetic models utilize mass-action kinetics and, thus, remove the restrictive assumption that all enzymes are under quasi-steady state. We have previously used related uncalibrated Michaelis-Menten mechanistic models to describe LT2 Pdu compartmentalization \cite{Kennedy2022,Mills2022}. We calibrate our kinetic model using No U-Turns (NUTS) \cite{Hoffman2014} Hamiltonian Monte Carlo implementation in PyMC, a Python Bayesian calibration package \cite{pymc2023}. We derived quasi-steady-state approximation (QSSA) expressions to relate Michaelis-Menten measurements to mass-action kinetic parameters. We integrated the QSSA expressions in our PyMC sampler, which seamlessly facilitated the transformation between Michaelis-Menten and mass-action parameters when fitting our kinetic model to metabolite time series.

From our PyMC inference, we identified two model fits (mode 1 and mode 2) with quantitatively similar data fits but qualitatively different propionate predictions. The model fits reproduced {\it in vitro} time series and, thus, showed that previously estimated Michaelis-Menten parameters are consistent with {\it in vitro} MCP metabolite profiles. Mode 1 produced propionate with an initial lag and a diminishing slope, while mode 2 produced propionate linearly with an abrupt halt in production. The differences were due to differences in inferred inhibition mechanisms and MCP cofactor permeability. Across modes, we inferred the MCP 1,2-propanediol and propionaldehyde permeability to be approximately greater than 10\textsuperscript{-6}  and 10\textsuperscript{-8} m/s, respectively. Due to the bimodality of our posterior distribution, we could not appreciably reduce uncertainty in the cofactor permeabilities. However, mode 1 is most consistent with the hypothesis that MCPs provide a private NAD(H) pool to facilitate propionaldehyde transformation by PduP and PduQ \cite{Cheng2012}.

\section*{Materials and Methods}

\subsection*{Mathematical Models} 

\subsubsection*{{\it in vitro} Model} 

We modeled the assay of purified MCPs as a well-mixed system of spherical shells in an external volume,
\begin{equation}
\label{Eq:InVitroModel}
\begin{aligned}
\frac{dX_{\text{MCP}}}{dt} &= R^X_{\text{MCP}}(X_{\text{MCP}}) - P_{\text{MCP}, X} \times \frac{\text{Surface Area of MCP}}{\text{external volume}}( X_{\text{external}} - X_{\text{MCP}})\\
\frac{dX_{\text{external}}}{dt} &= R^X_{\text{external}}(X_{\text{external}}) \\
&\quad+ P_{\text{MCP}, X}\times \text{number of MCPs} \times \frac{\text{Surface Area of MCP}}{\text{external volume}}\times(X_{\text{external}}-X_{\text{MCP}})
\end{aligned}
\end{equation}
where P\textsubscript{MCP,X} is the permeability of species X to the MCP, and X=\{1,2-propanediol, propionaldehyde, 1-propanol, propionyl-CoA, and propionate\}. R\textsuperscript{X}\textsubscript{MCP/external}(X\textsubscript{MCP/external}) describes the reaction kinetics of the enzyme acting on the reactant, X, and subscripts indicate the location of the reaction in the MCP or external volume.

The well-mixed assumption implies all compartments are identical and independent. This conveniently allowed us to model all MCPs as a single effective MCP. Each reactant has an MCP and an external volume component. Both components have a diffusion term, $$P_{\text{MCP}, X} \times \text{Surface Area of MCP},$$ linked to mass flux across the MCP shell, and a reaction term, R\textsubscript{X}(X), linked to mass flux within a component. We scaled the diffusion flux in the external component by the number of interacting MCPs, 2 $\times$ 10\textsuperscript{9} MCPs (0.067 mg/mL of MCP in the 30$\mu$L volume and Pdu MCP mass is, approximately, 600 MDa \cite{Cheng2008}).

We only modeled reactions involving PduCDE, PduP, PduQ, PduL, and AckA. The PduCDE, PduP, PduQ, and PduL catalyzed reactions take place in the MCP volume. The AckA-catalyzed reaction took place in the external volume. See \nameref{Text:S1_Text} for the mass-action decomposition of the reactions associated with each enzyme and their respective parameters. A previous study using fluorescence microscopy and mass spectroscopy on purified MCPs found that PduW was primarily cytosolic \cite{Yang2020}. As our experiments used purified microcompartments, we modeled the AckA added to the external volume. Inclusion of a separate PduW enzyme did not impact results.

\subsubsection*{Reaction Function, R\textsubscript{X}(X)}

\begin{figure}[!ht]
\includegraphics[width=0.85\linewidth]{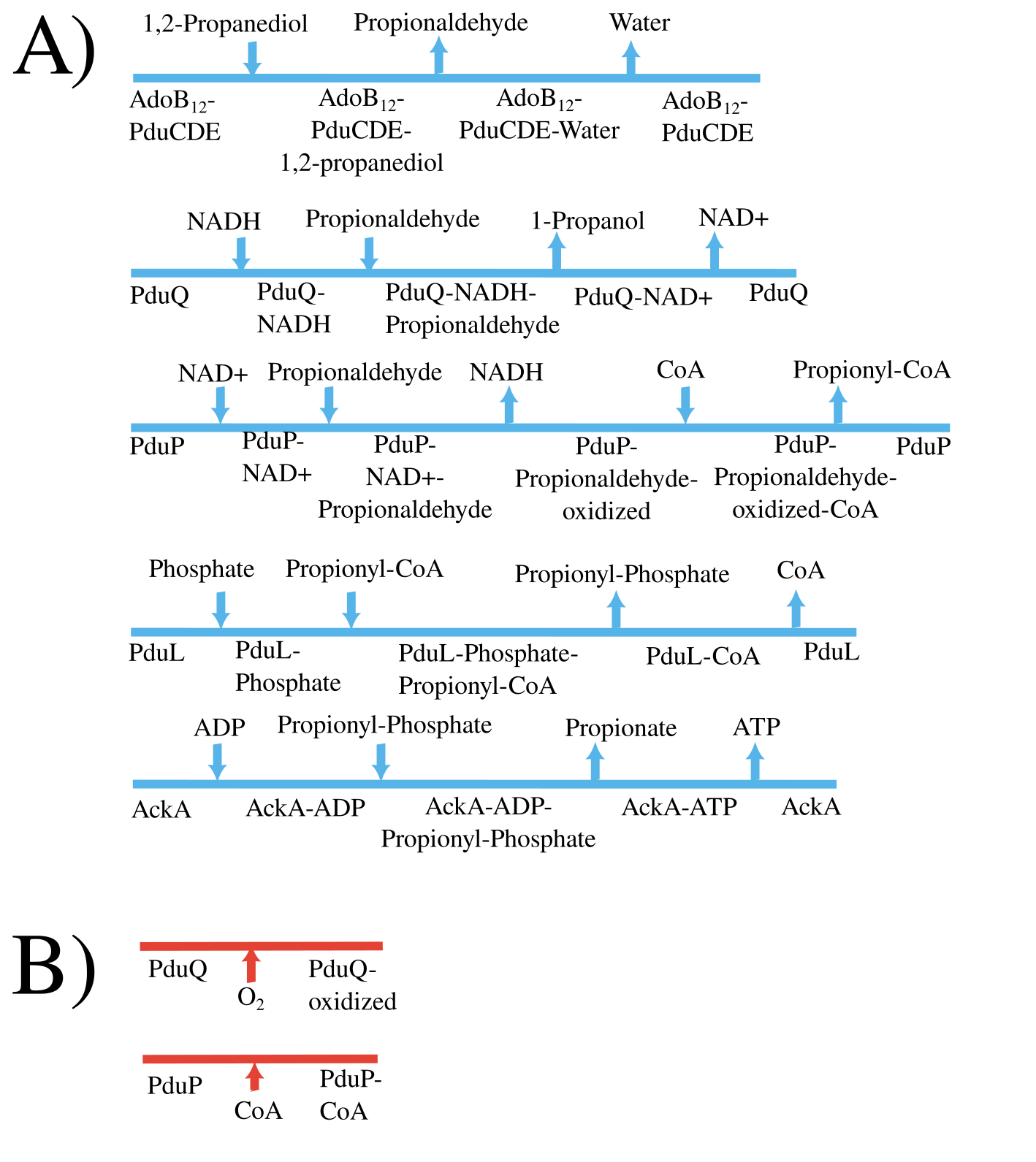}
\caption{Mass-action decomposition reaction of A) Pdu enzymes and B) inhibition of PduQ and PduP. Reactions occur from left to right, with the sequence of substrates, intermediate complexes, and products indicated (below the line).  A vertical downward-facing arrow indicates a binding event, and a vertical upward arrow indicates an unbinding event. From top to bottom, Figure A details PduCDE conversion of 1,2-propanediol to propionaldehyde \cite{Toraya2000}, PduQ reduction of propionaldehyde to 1-propanol, PduP oxidation of propionaldehyde to propionyl-CoA, PduL convert of propionyl-CoA to propionyl-phosphate \cite{Smith1980} and AckA conversion of propionyl-phosphate to propionate. We assume that PduQ, PduL, and AckA sequentially bind to cofactors first and their organic substrate second. From top to bottom, Figure B outlines the mass-action decomposition of PduQ oxidation and PduP-CoA inhibition \cite{Smith1980}. We assume that PduQ oxidation follows first-order elimination.}
\label{Fig2}
\end{figure}

We used first-principle mass-action kinetics to model all enzyme-catalyzed reactions. Mass-action kinetics decomposes an enzymatic reaction into a sequence of first-order reactions in which reactants interact with enzyme complexes in a specified order. The binding order of all mass-action reactions is outlined in Figure \ref{Fig2}A. Mass-action kinetics allowed us to integrate measured Michaelis-Menten constants in our model without assuming that all reactions are under a Michaelis-Menten or quasi-steady-state regime. While quasi-steady-state holds under many conditions \cite{schnell2003century}, it will not hold when the relative concentrations of metabolites, cofactors, and active enzymes vary dramatically.

Based on experimental observations and previous work where reactions do not run to completion  (Figure \ref{Fig4}C-D, \cite{Palmero2025}), we modeled PduQ and PduP inhibition mechanisms. The final 1-propanol and propionate are far from the values expected in thermodynamic equilibrium with propionaldehyde (K\textsubscript{eq,PduP}, K\textsubscript{eq,PduQ} $>$ 0), indicating the enzyme-mediated reactions have slowed due to external factors. From existing literature, we identified two possible mechanisms: oxidation of PduQ and CoA inhibition of PduP. Cheng et al. 2008 reported that PduQ is oxygen-sensitive, with the enzyme losing much of its activity within two hours. We modeled PduQ oxidation as a first-order reaction (Figure \ref{Fig2}B). CoA-acylating aldehyde dehydrogenases (ALDH) have been observed to undergo substrate inhibition with CoA \cite{Lei2008, Sabet-Azad2013, Smith1980, Palosaari1988}, including a homologous PduP in {\it Lactobacillus reuteri} \cite{Sabet-Azad2013}. Smith et al. 1980 \cite{Smith1980} proposed that CoA formed a dead-end substrate-enzyme complex with ALDH when bound as the first substrate. We modeled PduP-CoA formation as a second-order process as described in Smith et al. 1980 (Figure \ref{Fig2}B).

\subsection*{Uncertainty Quantification}
\subsubsection*{Bayesian Inference}\label{ssec:BayesianInference}

We used Bayesian inference to calibrate differential equation models to metabolite time series. Bayesian inference requires a prior distribution, $p(\theta)$, which describes initial uncertainty in model parameters, and a likelihood function, $\ell(\theta, \text{data})$, which relates the model time series to the data time series. Taken together, both define a posterior distribution, $p(\theta, \text{data}) \propto \ell(\theta, \text{data})p(\theta)$, that weighs between data fitting and prior belief integration \cite{Smith2014}. 

We formulated prior distributions based on parameter estimates from the literature (\nameref{S1_Table}). However, given the large order of magnitude ranges and many unknown parameters, the prior distributions were often uninformative and only acted as a regularizer with bounded support. We defined a Chi-square likelihood function,

$$\log \ell(\theta, \hat{\mu}, \hat{\sigma}) = -\sum_{\substack{X \in \{\text{1,2-propanediol, propionaldehyde,}\\\text{ 1-propanol, propionate}\}}}\,\,\sum_{i=1}^{N} \frac{(X_{\text{ext}}(t_i, \theta) + \beta X_{\text{MCP}}(t_i, \theta) - \hat{\mu}_{X,i})^2}{2\hat{\sigma}_{X,i}^2}, $$

where $X_{\text{ext/MCP}}(t_i, \theta)$ is the external/MCP X state variable at time $t_i$ \ref{Eq:InVitroModel}, $X\in \{$1,2-propanediol, propionaldehyde, 1-propanol, propionate$\}$,  $\hat{\mu}_{X,i}$ and $\hat{\sigma}_{X,i}$ were the sample mean and sample standard deviation computed from three biological replicates at time $t_i$ , and  $\beta$ accounts for the mass contributed from MCPs after aliquot processing, 
$$\beta = \substack{\text{Concentration of MCPs}\\\text{in external volume}} \times V_{\text{MCP}} \approx 10^{-4}.$$ See \nameref{Text:S2_Text} for the derivation of $\beta$.

\begin{figure}[!ht]
\includegraphics[width=\linewidth]{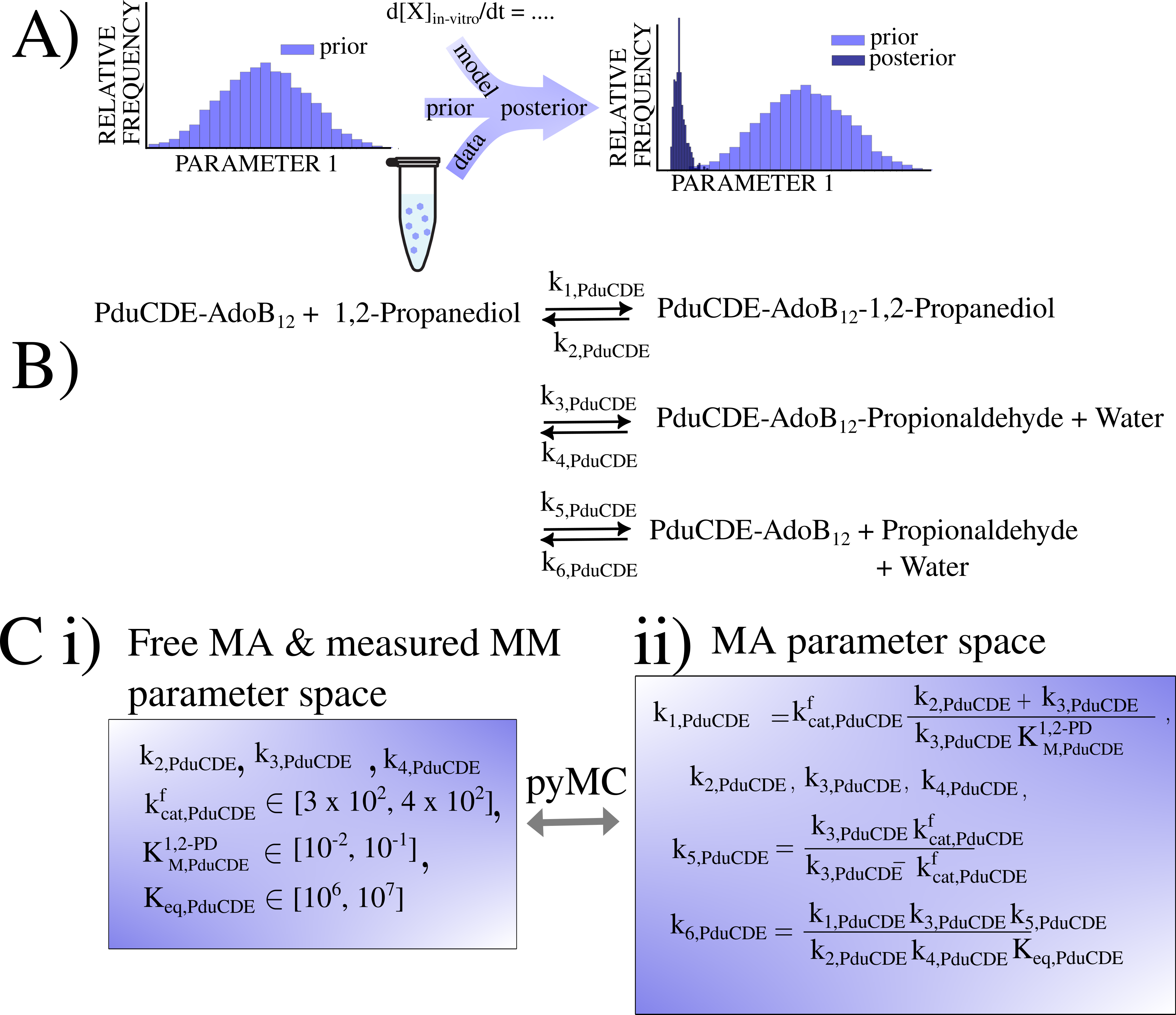}
\caption{ (A) Bayesian calibration pipeline. The {\it in vitro} model was calibrated to prior parameter beliefs and metabolite time series using PyMC \cite{pymc2023}. (B) Mass-action kinetic parameter relationship to PduCDE binding events. Odd and even indexed kinetic parameters represent reaction rates for the forward and reverse reaction, respectively. (C) Transformation between free mass-action (MA) \& measured Michaelis-Menten (MM) space, and MA kinetic parameter space. (i) PyMC computes the leading MA variables from the measured MM space and free MA parameters. (ii) Combining free and leading MA variables, PyMC computes all MA parameters. Likelihood gradients are generated in terms of the mass-action parameter set. PyMC transforms the likelihood derivative to free MA and the measured MM parameter space via fundamental back-propagation operations. 
}
\label{Fig3}
\end{figure}

We used PyMC \cite{pymc2023}, a Python-based probabilistic programming language, and its implementation of No U-Turns Hamiltonian Monte Carlo \cite{Hoffman2014} to generate parameter samples from our posterior distribution (Figure \ref{Fig3}A). No U-turns (NUTS) is a semi-automated extension of Hamiltonian Monte Carlo (HMC). HMC belongs to a family of direct Bayesian methods called Markov Chain Monte Carlo, which approximates an intractable distribution using a sequential random process. Hamiltonian Monte Carlo utilizes gradient information of the posterior distribution to avoid the highly inefficient random walk exploration common in many Markov Chain Monte Carlo (MCMC) schemes \cite{Neal2012}.  We used sunode, a Python wrapper to CVODES  \cite{Hindmarsh2002, pymc2023}, to compute model time series and their associated gradients.

\subsection*{Parameter Constraints and Prior Distributions}
\subsubsection*{Kinetic Parameters}

Mass-action (MA) kinetic parameters are difficult to measure and often not available in the literature. Michaelis-Menten (MM) kinetics, in contrast, are commonly estimated from data \cite{Johnson2011, Noor2012, Gunawardena2014}.  We defined our posterior distribution partially in terms of reported Michaelis-Menten parameters (Figure \ref{Fig3}Ci). This ensured that associated model dynamics were faithful to available literature measurements and reduced uncertainty in the reaction kinetics. The MM kinetics of all enzymes have been partially or fully experimentally characterized (parameters summarized in \nameref{S1_Table}). The MM parameters were taken to have a uniform prior over measurement confidence intervals. 

To utilize MM parameter samples in the mass-action model, we derived the analytic transformation between the MM and MA kinetics. First, we systematically expressed MM parameters in terms of MA kinetics using the QSSA approximation. The derivation is detailed in \nameref{Text:S3_Text} using PduQ as an example. See \nameref{S2_Table} for a list of formulas for all MM parameters in terms of MA. We then solved these equations for MA in terms of the measured MM parameters and a subset of MA parameters using Wolfram Mathematica 13 \cite{Mathematica}. This partitioned MA kinetics into free parameters, which are not determined by MM parameters and are included in the posterior, and leading parameters, which are directly calculated from MM parameters and free MA parameters (Figure \ref{Fig3}Cii). As leading parameters can be calculated, this process reduces the degrees of freedom in the Bayesian calculation. Free MA parameters were partly constrained by enforcing strict positivity of the leading parameters. See \nameref{S3_Table} for the list of all leading and free kinetic parameters and Th\nameref{S4_Table} for all positivity constraints on free MA parameters. Free MA parameters with strict positivity constraints were taken to have a uniform prior, and free MA parameters without constraints were taken to have a truncated normal prior.

PyMC seamlessly facilitated the parameter and gradient transformations between the MA parameter space and free MA and MM space (Figure \ref{Fig3}C). The transformation is necessary because the model and derivatives are expressed in terms of the mass-action kinetic parameter sets, but the posterior distribution from which we sampled is defined for the Michaelis-Menten and free mass-action space (Figure \ref{Fig3}Cii). To enable translation, we input the leading MA parameter expressions into PyMC via its Deterministic method. We also had to appropriately transform the model derivatives in terms of Michaelis-Menten and free mass-action parameters for NUTS sampling. PyMC computed the Jacobian transformations and transformed model gradients to posterior gradients using the relationship defined in the Deterministic method.

\subsubsection*{Thermodynamic Feasibility}

To enforce thermodynamic consistency, we also included K\textsubscript{eq}, a thermodynamic measure of reversibility, in our posterior. K\textsubscript{eq} is the ratio of the forward MA kinetic parameter product to the reverse MA kinetic parameter product.  
 \begin{equation}
 \label{Eq:KeqMA}
K_{\text{eq}} = \frac{\prod\limits_{i=1}^{n/2} k_{2i-1}}{\prod\limits_{i=1}^{n/2} k_{2i}}
 \end{equation}
 where k\textsubscript{2i-1} and  k\textsubscript{2i} are the reaction rates of the i\textsuperscript{th} forward and reverse reaction, respectively, and $n$ is the total number of reaction steps (Figure \ref{Fig3}B). 
 
We estimated the K\textsubscript{eq} of each reaction using Equilibrator, a web-based platform that utilizes the group and component contribution method to calculate the Gibbs free energy of a reaction \cite{Beber2021}. We took the K\textsubscript{eq} prior to be uniform over the confidence intervals defined by Equilibrator. 

As with the Michaelis-Menten measurements, we integrated K\textsubscript{eq} in our model by designating an MA leading parameter and solving the K\textsubscript{eq} relationship (equation \ref{Eq:KeqMA}) for the leading MA parameter. Integration of thermodynamic relationships can improve the identifiability of kinetic parameters \cite{mason2019energetic}. Previous work enforced this relationship locally by rescaling to a reference state and integration into the priors \cite{saa2016probabilistic}. See \nameref{S5_Table} for the leading parameters and transposed K\textsubscript{eq} relations for all enzymatic reactions in our study. With exception of PduQ, Michaelis-Menten leading parameters were carefully chosen to be independent of the K\textsubscript{eq} leading parameter. This avoided simultaneously solving the K\textsubscript{eq} and Michaelis-Menten equations.  See Table \nameref{S5_Table}.

\subsubsection*{MCP Parameters}
We restricted the MCP enzyme numbers using catalytic protein enumerations per MCP from Yang et al. 2020 \cite{Yang2020} (\nameref{S1_Table}). These experiments quantified protein content in purified Pdu MCPs using high-resolution liquid chromatography mass spectrometry and thus provide an excellent estimate of protein content in our system. The number of enzymes per MCP was restricted to the listed range and taken to have a uniform prior over the support. The MCP permeabilities were assumed to be between 10\textsuperscript{-11}-10\textsuperscript{-3} m/s. We set the permeability of structurally similar molecules, such as CoA \& propionyl-CoA, and NAD+ \& NADH, to have the same MCP permeability. These molecules differed by a small atom chain and, being large molecules, likely face similar barriers across the MCP shell.    

\subsection*{Compartment expression and purification}
Pdu MCPs were purified using a differential centrifugation method as previously described  \cite{Sinha2012, Nichols2019}. We inoculated single colonies of strains into 5 mL of lysogeny broth (LB) liquid media, which were incubated at 30 \degree C, 225 rpm for 24 hours. The overnight culture was subcultured 1:1,000 into 200 mL of No Carbon Essential (NCE) media comprising 29 mM potassium phosphate monobasic, 34 mM potassium phosphate dibasic, 17 mM sodium ammonium hydrogen phosphate, supplemented with 50 $\mu$M ferric citrate, one mM magnesium sulfate, 42 mM succinate as a carbon source, and 55 mM 1,2-propanediol as the {\it pdu} operon inducer in a 1 L Erlenmeyer flask. The NCE subculture was grown at 37 \degree C at 225 rpm until the cultures reached an OD\textsubscript{600} of 1.0-1.5. 200 mL of cells were spun down at 5,000 x g for 5 minutes at 4 \degree C. The supernatant was decanted into bleach and the pellet was resuspended in lysis buffer (32 mM Tris-HCl, 200 mM potassium chloride, 5 mM magnesium chloride, 0.6\% (v/v) 1,2-propanediol, 0.6\% (v/v) octylthioglucoside (OTG), 5 mM $\beta$-mercaptoethanol, 0.8 mg/mL lysozyme (Thermo Fisher Scientific), 0.04 units/mL DNase I (New England Biolabs, Inc.) pH 7.5–8.0. The resuspended cells underwent a 30-minute incubation in this lysis buffer at room temperature, with gentle rocking at 60 rpm. Following the lysis period, the lysate was kept on ice for 5 minutes and then clarified by centrifugation (12,000 x g, 5 minutes, 4 \degree C) twice. To isolate the MCPs from the clarified lysate, centrifugation was performed at 21,000 x g for 20 minutes at 4 \degree C in a swinging bucket rotor. The supernatant was discarded, and the pellet was washed with buffer (32 mM Tris-HCl, 200 mM KCl, 5 mM MgCl\textsubscript{2}, 0.6\% (v/v) 1,2-propanediol, 0.6\% (v/v) OTG, pH 7.5–8.0). Subsequent MCP pelleting was achieved by spinning at 21,000 x g for 20 minutes at 4 \degree C in a swinging bucket rotor. The supernatant was removed, and the MCP pellet was resuspended in buffer (50 mM Tris-HCl, 50 mM KCl, 5 mM MgCl\textsubscript{2}, pH 8.0) and stored at 4 \degree C until needed. The concentration of the purified Pdu MCPs was determined using a bicinchoninic acid assay (Thermo Scientific).

\subsection*{{\it in vitro} assay}

In vitro Pdu MCP assays were performed in 30 $\mu$L reactions in 2 mL Eppendorf tubes and incubated at 30 \degree C. The reaction environment consisted of 100 mM Bis-Tris buffer, acetate salts (8 mM magnesium acetate, 10 mM ammonium acetate, 134 mM potassium acetate), 55 mM 1,2-propanediol, 20 $\mu$M AdoB\textsubscript{12}, 1 U AckA (Sigma-Aldrich), and 66.66 $\mu$g purified microcompartments. Unless otherwise stated, CoA, NAD+, and NADH were added at 1.5 mM concentrations. Reactions were measured a half-hour intervals until the 6-hour mark. A final measurement was taken at 24 hours. The reactions were quenched at each time point by adding 30 $\mu$L of 10\% (v/v) trichloroacetic acid. The reactions were subsequently centrifuged at 21,000 x g for 10 minutes at 4 \degree C, and the supernatant was stored at -80 \degree C until analysis by HPLC. 

\subsection*{Metabolite quantification}
Pdu pathway metabolites were quantified using an Agilent 1260 HPLC system. The injection volume was 5 $\mu$L, and the analytical column used was a Rezex\textsubscript{TM} ROA-Organic Acid H+ (8\%) LC Column (Phenomenex) at 35 \degree C with 5 mM sulfuric acid as the mobile phase at a flow rate of 0.4 mL/min for 45 minutes. A refractive index detector (RID) was used to detect metabolites, and we compared the resulting peaks from the RID to peak areas of known dilutions of 1,2-propanediol, propionaldehyde, 1-propanol and propionate as described previously\cite{Lee2017}.

\section*{Results}
\subsection*{Calibration to {\it in vitro} data set reveals bimodal fits}

\begin{figure}[!htp]\centering
\includegraphics[width=1\linewidth]{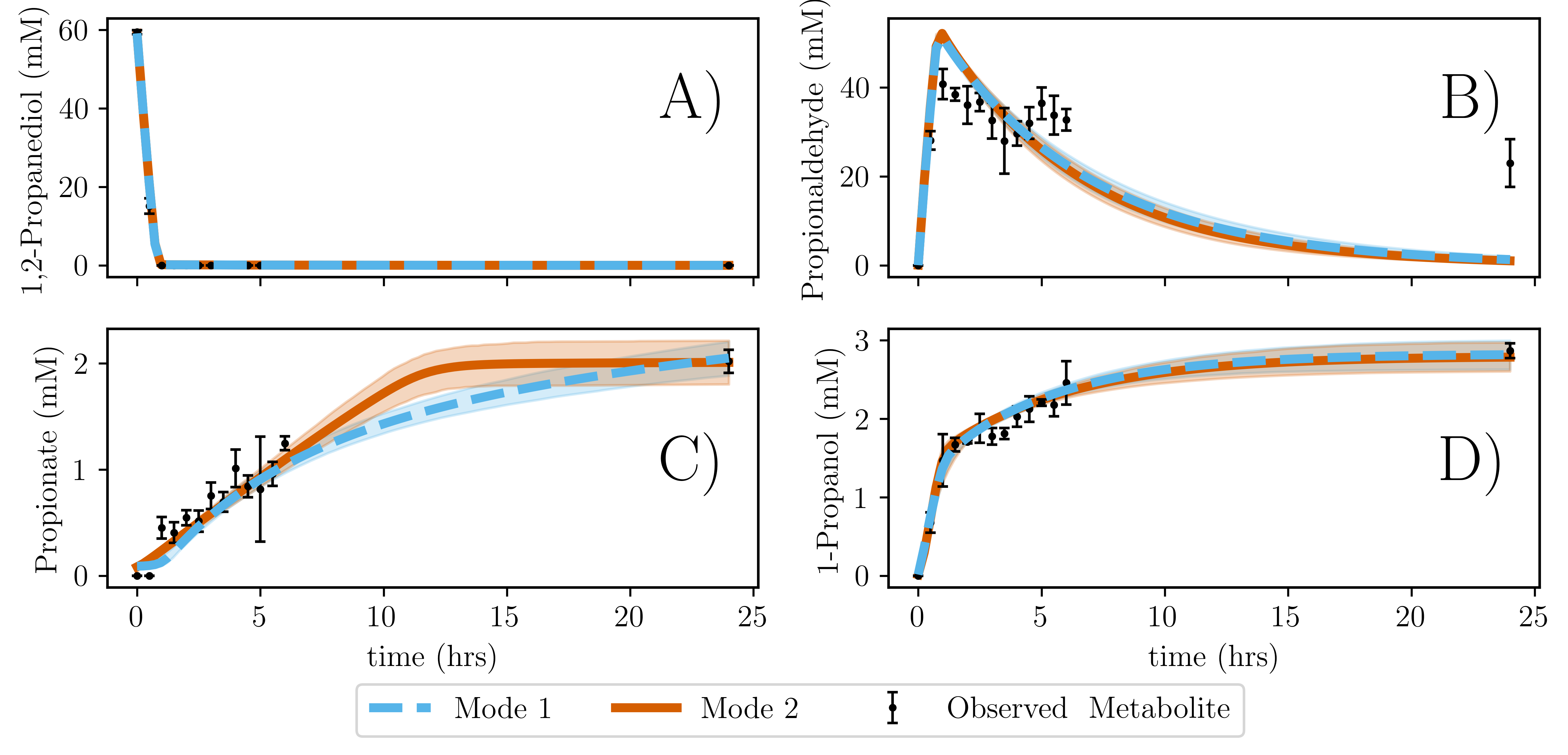}
\caption{Modes 1 and 2 fit to A) 1,2-propanediol, B) propionaldehyde, C) propionate, and D) 1-propanol. Mode 1 and mode 2 differed primarily in  propionate dynamics. There is a discrepancy between estimated propionaldehyde dynamics and experimental measurements because of unexplained mass loss at the time of study (see  Section \nameref{ssec:BayesianInference}). Later experiments showed mass loss was due to evaporation. The inclusion of mass loss did not affect downstream propionate and 1-propanol dynamics.}
    \label{Fig4}
\end{figure}

We identified two modes (mode 1 and mode 2) that fit the {\it in vitro} time series with distinct dynamics and equal log-likelihood (mean of log-likelihood $\sim$ -280) (Figure \ref{Fig4}). Posterior distribution sampling is summarized in \nameref{S6_Table}, \nameref{S7_Table} and \nameref{S1_Fig} for mode 1, and \nameref{S8_Table}, \nameref{S9_Table} and \nameref{S2_Fig} for mode 2. Mode 1 and 2 differed primarily in external propionate dynamics (Figure \ref{Fig4}C). Mode 1 propionate production stalled from 0 to 2 hrs due to a transient bottleneck in PduL turnover. Once the bottleneck was cleared, propionate production increased up to the 24-hour mark. In mode 2, propionate increased linearly until an abrupt halt around 12 hrs due to PduP-CoA accumulation (Section \nameref{ssec:CoApermeability}). 1-Propanol and propionaldehyde dynamics differed marginally across modes (Figure \ref{Fig4}D) while 1,2-propanediol dynamics were identical  (Figure \ref{Fig4}A and B). 

NUTS samplers infrequently switched between modes, due to the lack of proximity of the modes in parameter space. This resulted in poor MCMC convergence (Effective Sample Size (ESS) $<$ 100 per chain and $\hat{R} \geq 1.05$; see \nameref{S3_Fig} and \nameref{S10_Table}). To improve convergence and ensure accurate sampling of the local posterior around each mode, we restricted MCMC sampling to parameter ranges unique to each mode. The MCP permeability to CoA was one of the parameters with the clearest difference between modes.  We sampled mode 1 by constraining the CoA MCP permeability to  [10\textsuperscript{-11}, 10\textsuperscript{-8}] m/s, and mode 2 by constraining the CoA MCP permeability to  [10\textsuperscript{-7.5}, 10\textsuperscript{-3}] m/s (Figure \ref{Fig7}).  PduP inhibition parameters were set to the mean of select, well-behaving MCMC chains (See \nameref{S11_Table} and \nameref{S12_Table}).  We set k\textsubscript{1,iPduP} = 10\textsuperscript{-3.58} /mM s and k\textsubscript{2,iPduP} = 10\textsuperscript{-7.59} /s for mode 1, and k\textsubscript{1,iPduP} = 10\textsuperscript{-0.36} /mM s  and k\textsubscript{2,iPduP} = 10\textsuperscript{-3.76} /s for mode 2. Simultaneously estimating PduP inhibition kinetics and MCP CoA permeability resulted in sticky MCMC chains when sampling mode 2 (\nameref{S4_Fig} and \nameref{S13_Table}). Such poor convergence was likely due to strong correlations between PduP inhibition kinetics and MCP CoA permeability (Section \nameref{ssec:CoApermeability}). A sensitivity analysis around the inhibition parameters showed marginal differences in model fit and inferred parameter distribution for each mode, indicating the reliability of the fits reported (See \nameref{S5_Fig}, \nameref{S6_Fig}, \nameref{S7_Fig}, \nameref{S8_Fig}, \nameref{S14_Table} and \nameref{S15_Table})

 We note that there was unexplained non-conservation of mass in the {\it in vitro} experiment. The {\it in vitro} assay consumed 55 mM of 1,2-propanediol within the first hour. However, the consumed concentration was not converted into propionaldehyde, propionate, and 1-propanol (Figure \ref{Fig4}). We determined that the remaining mass could not be stored in propionyl-CoA, as there was only 1.5 mM of CoA added to the system, and unlikely to be stored in propionyl-phosphate, as {\it in vitro} propionate concentrations were observed to be insensitive to changes in ATP and phosphate concentrations \cite{Palmero2025}. 
 
 We identified evaporation as the cause of propionaldehyde mass loss (\nameref{S9_Fig}) and, subsequently, included evaporation in our {\it in vitro} model. Note that irreversible evaporation overestimated propionaldehyde between 1 and 4 hours, and underestimated it between 5 and 25 hours. Reversible mass loss is a more accurate model of evaporation in the {\it in vitro} assay, as assay vials were topped with a stopper, and, in a closed system, evaporation is a reversible process. However, a reversible evaporation model would require including propionaldehyde vapor as a latent variable and, thus, increase the computation cost of our calibration.
 
Moreover, it is unlikely that including reversible mass loss would significantly shift the results of our current calibration. 
A model with no evaporation returned similar mode fits in metabolites other than propionaldehyde (\nameref{S10_Fig} and \nameref{S11_Fig}) and MCP permeability estimates (\nameref{S12_Fig} and \nameref{S13_Fig}). The most significant change in MCP permeability estimates is a small downward shift of the MCP permeability to propionaldehyde in mode 1 when evaporation is removed (\nameref{S12_Fig}). Our methodology for estimating MCP permeability parameters is, therefore, mostly insensitive to perturbations in late-stage propionaldehyde dynamics.  

\subsection*{Modes inferred similar posterior updates for kinetic parameters}

\begin{figure}[!ht]
\centering
   \includegraphics[width=\textwidth]   {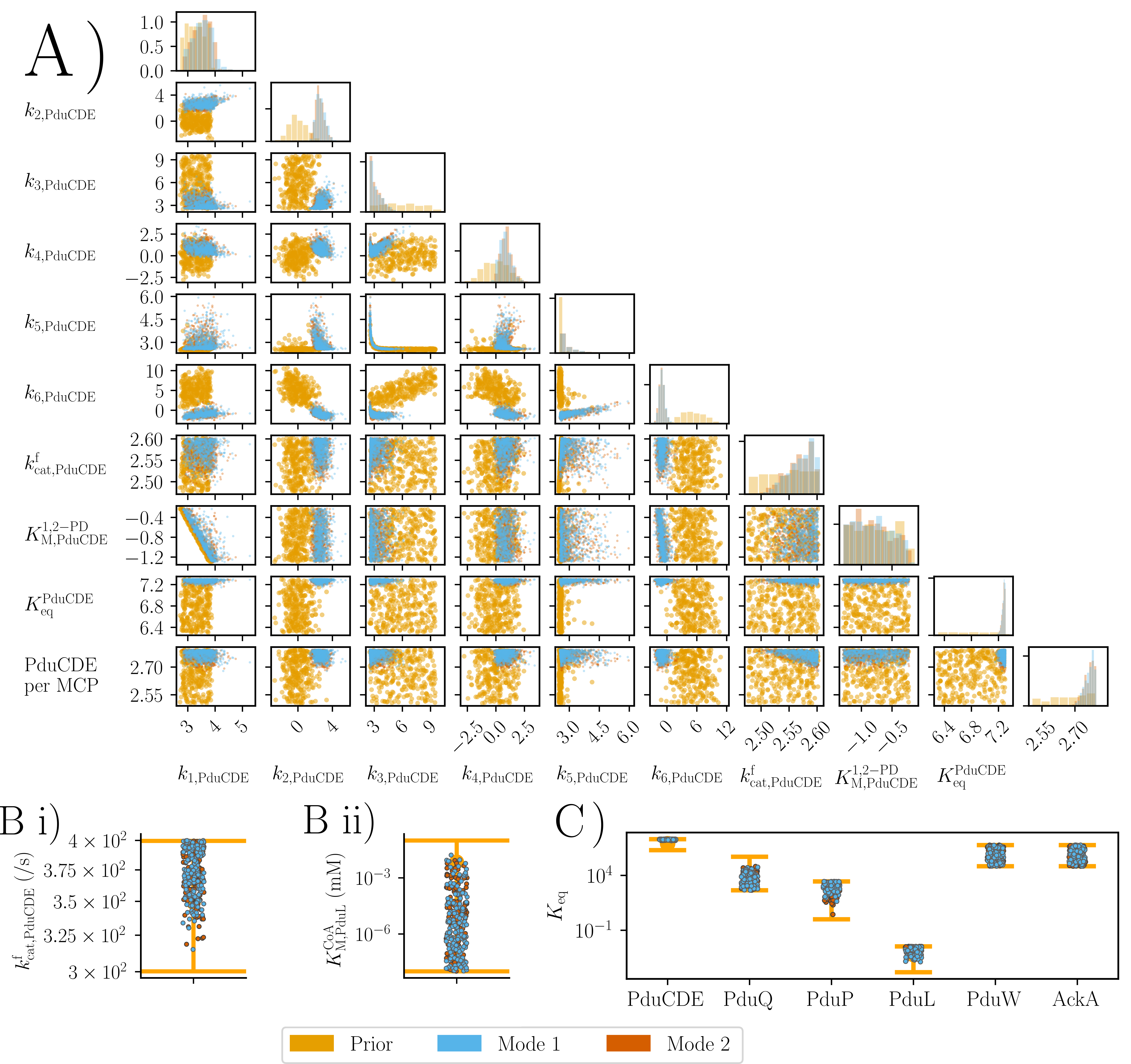}
   \caption{Prior to Posterior kinetic parameter update of PduCDE and K\textsubscript{eq} parameters. A) Corner plot of prior and posterior (mode 1 and mode 2) PduCDE parameter samples. Shown are all PduCDCE mass-action kinetic parameters, measured Michaelis-Menten kinetic parameters, and  MCP enzyme number. Strip plot comparing mode 1 and mode 2 inferred B) k\textsubscript{cat,PduCDE}\textsuperscript{f} and K\textsubscript{M,PduL}\textsuperscript{CoA}, and C) K\textsubscript{eq} of all reactions. The uniform prior range is overlaid as an orange pointplot interval. 
   }
   \label{Fig5}
   \end{figure}

The two modes experienced overlapping after posterior updates for all mass-action and Michaelis-Menten non-inhibition parameters (Figure \ref{Fig5}A, \nameref{S14_Fig}, \nameref{S15_Fig}, \nameref{S16_Fig}, \nameref{S17_Fig}, \nameref{S18_Fig}, \nameref{S19_Fig}, \nameref{S20_Fig},  and \nameref{S21_Fig}). The updated free mass-action posteriors spanned ranges and correlations unseen in the prior. This is unsurprising as the prior admits a poor fit to the {\it in vitro} data (\nameref{S22_Fig}). While similar in many enzymatic parameters, the initially discovered modes differed notably in their inferred PduP inhibition rates (See \nameref{S11_Table} and \nameref{S12_Table}). Such differences contributed to the differences in propionate dynamics across modes (Section \nameref{ssec:CoApermeability} and \nameref{ssec:NAD(H)permeability}).


The Michaelis-Menten (MM) marginal posterior distributions were wholly contained within the prior support (Figure \ref{Fig5}A and \nameref{S23_Fig}). This indicates that enzyme activity implied from literature constraints is consistent with the observed {\it in vitro} dynamics. 
We note that NUTS inferred a low 
K\textsuperscript{PduL}\textsubscript{M,CoA} (mean $\sim$ 10\textsuperscript{-3} mM) (Figure \nameref{S23_Fig}). The K\textsuperscript{PduL}\textsubscript{M,CoA} prior support spanned an extensive range due to high reported uncertainty \cite{Liu2007}. A low 
K\textsuperscript{PduL}\textsubscript{M,CoA} allowed CoA, at high enough concentrations, in our model fits to readily bind to PduL and stall propionate production (see Section \nameref{ssec:CoApermeability}). 

Posterior reaction reversibilities, K\textsubscript{eq}, spanned their prior support except PduCDE, PduL and PduP (Figure \ref{Fig5}B). These enzymes concentrated near their respective upper bound, notably shifting from their uniform priors.  K\textsubscript{eq} upper bound accumulation prevented product saturation of the enzymes and, thus, ensured that PduCDE and PduP were primarily propionaldehyde and propionyl-CoA-producing enzymes, respectively. The K\textsubscript{eq,PduL} accumlation near the upper bound ensured that PduL feasibly produced propionyl-phosphate despite being a thermodynamically unfavorable reaction (K\textsubscript{eq,PduL} $\ll$ 1).

\subsection*{{\it in vitro} Calibration reduced uncertainty in 1,2-propanediol and propionaldehyde MCP permeability}

\begin{figure}[!ht]
\includegraphics[width=\linewidth]{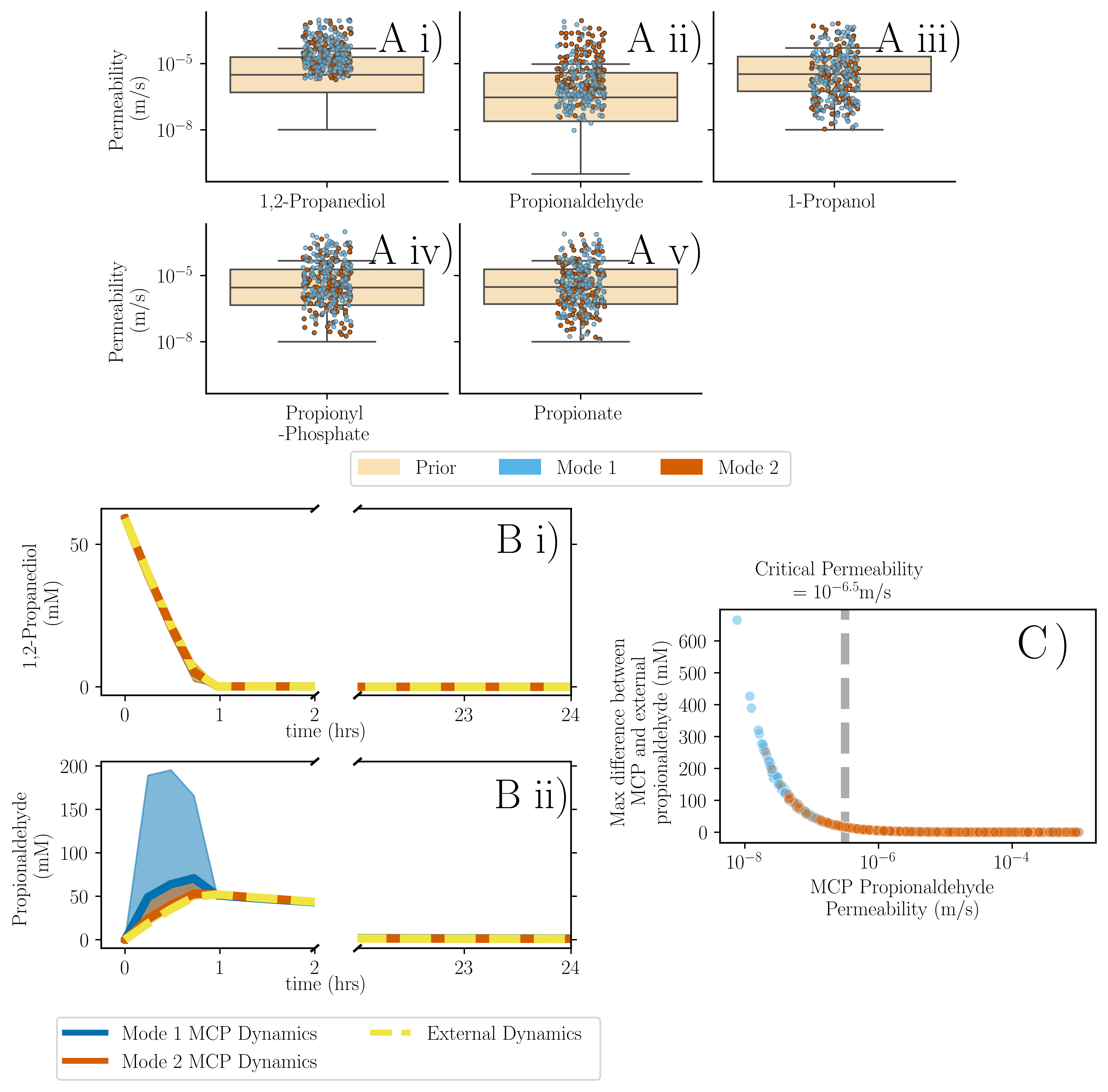}
    \caption{Prior to Posterior MCP permeability update and its impact on internal MCP dynamics. A) Prior to posterior updates of MCP permeability to i) 1,2-propanediol, ii) propionaldehyde, iii) 1-propanol, iv) propionyl-phosphate and v) propionate. Samples from mode 1 and 2 are shown and are overlayed atop a boxplot summary of the prior distribution. Note only that only 1,2-propanediol and propionaldehyde experienced a posterior update. Propionaldehyde permeability experienced different posterior updates across modes. Bi) Internal and external MCP 1,2-propanediol dynamics of mode 1 and 2. 95\% confidence intervals of the posterior mean are shown. Estimated external dynamics are nearly identical between modes, so only a mean curve is shown. Ci) Internal and external MCP 1,2-propanediol dynamics of mode 1 and 2. 95\% confidence interval of the posterior mean are shown. Note that mode 1 accumulates more propionaldehyde than mode 2. Cii) Maximum difference between MCP and external propionaldehyde concentration as a function of propionaldehyde permeability. The MCP only accumulates propionaldehyde only if the MCP permeability is less than 10\textsuperscript{-6.5}m/s.}\label{Fig6}
\end{figure}

Mode 1 and 2 updated the posterior shell permeability to 1,2-propanediol and propionaldehyde while leaving the permeability to 1-propanol, propionyl-phosphate and propionate unchanged (Figure \ref{Fig6}Ai-v). Both modes inferred 1,2-propanediol and propionaldehyde permeability to be greater than 10\textsuperscript{-6.} and 10\textsuperscript{-8.} m/s, respectively. Unchanged 1-propanol, propionyl-phosphate and propionate permeabilities indicated that, given the uncertainty in enzyme kinetics, the MCP could generate sufficient metabolite concentrations to overcome a wide range of permeability values and reproduce the propionate and 1-propanol time series.  Metabolite permeability posterior distribution were the same across modes except for propionaldehyde (Figure \ref{Fig6}Aii). Mode 1 admitted a lower mean propionaldehyde permeability (10\textsuperscript{-6.3} m/s) than mode 2 (10\textsuperscript{-5.2} m/s). The reasons for differences in the MCP propionaldehyde permeability distributions were not immediately apparent from computational experiments.

All samples from the inferred 1,2-propanediol posterior permeability range, strongly entrained 1,2-propanediol MCP concentration to the observed external dynamics (Figure \ref{Fig6}Bi). The posterior for the 1,2-propandiol permeability of both modes suggest that the permeability must be greater than $10\textsuperscript{-6} m/s$, which will allow uninhibited entry of 1,2-propandiol into the MCP and rapid 1,2-propanediol consumption. To test the impact of a 1,2-propanediol permeability much less than 10\textsuperscript{-6} m/s we evaluated our model on posterior samples with 1,2-propanediol permeability set to 10\textsuperscript{-8} m/s. Decreasing the 1,2-propanediol permeability to 10\textsuperscript{-8} m/s disrupted entry of 1,2-propanediol into the MCP and resulted in a lower quality of fit for both modes (\nameref{S24_Fig} and \nameref{S25_Fig}). This suggests that the rapid uptake of 1,2-propandiol in experiments is only consistent with a relatively high 1,2-propandiol greater than $10\textsuperscript{-6} m/s$, which is why the posterior is constrained to that range.

Across modes, the propionaldehyde permeability allowed the MCP to either match or exceed the external propionaldehyde concentration (Figure \ref{Fig6}Bii). We identified 10\textsuperscript{-6.5} m/s  as the critical value for propionaldehyde accumulation in the MCP (Figure \ref{Fig6}C). At permeabilities less than 10\textsuperscript{-6.5} m/s, the MCP maintains a concentration gradient with the external volume. At values greater than 10\textsuperscript{-6.5} m/s, the dynamics inside the MCP match external dynamics. From our simulations, when the permeability fell below 10\textsuperscript{-6.5} m/s, the MCP accumulated up to 2 orders of magnitude more propionaldehyde than the external volume and, above 10\textsuperscript{-6.5} m/s, the MCP did not accumulate propionaldehyde.

 Similiar to 1,2-propanediol permeability, we also investigated whether either mode is consistent with a propionaldehyde permeability less than 10\textsuperscript{-8} m/s by evaluating our model on posterior samples with propionaldehyde permeability set to 10\textsuperscript{-10} m/s. The model evaluations showed decreased 1,2-propanediol consumption and propionaldehyde production relative to the original fit (\nameref{S26_Fig} and  \nameref{S27_Fig}). The low permeability prevented the leakage of propionaldehyde out of MCP and increased the propionaldehyde concentration of MCP so much that it overcame the high thermodynamic barrier of PduCDE and outcompeted 1,2-propanediol to bind to PduCDE (\nameref{S28_Fig}, \nameref{S29_Fig}, \nameref{S30_Fig} and \nameref{S31_Fig}). This explained why such low permeabilities were not inferred in our calibration and, importantly, indicates that sufficiently low propionaldehyde permeabilities are inconsistent with the thermodynamic and {\it in vitro} measurements.

\subsection*{CoA permeability and PduP-CoA inhibition kinetics contribute to differences in propionate dynamics.}
\label{ssec:CoApermeability}

\begin{figure}[!ht]
\centering
\includegraphics[width=0.9\linewidth]{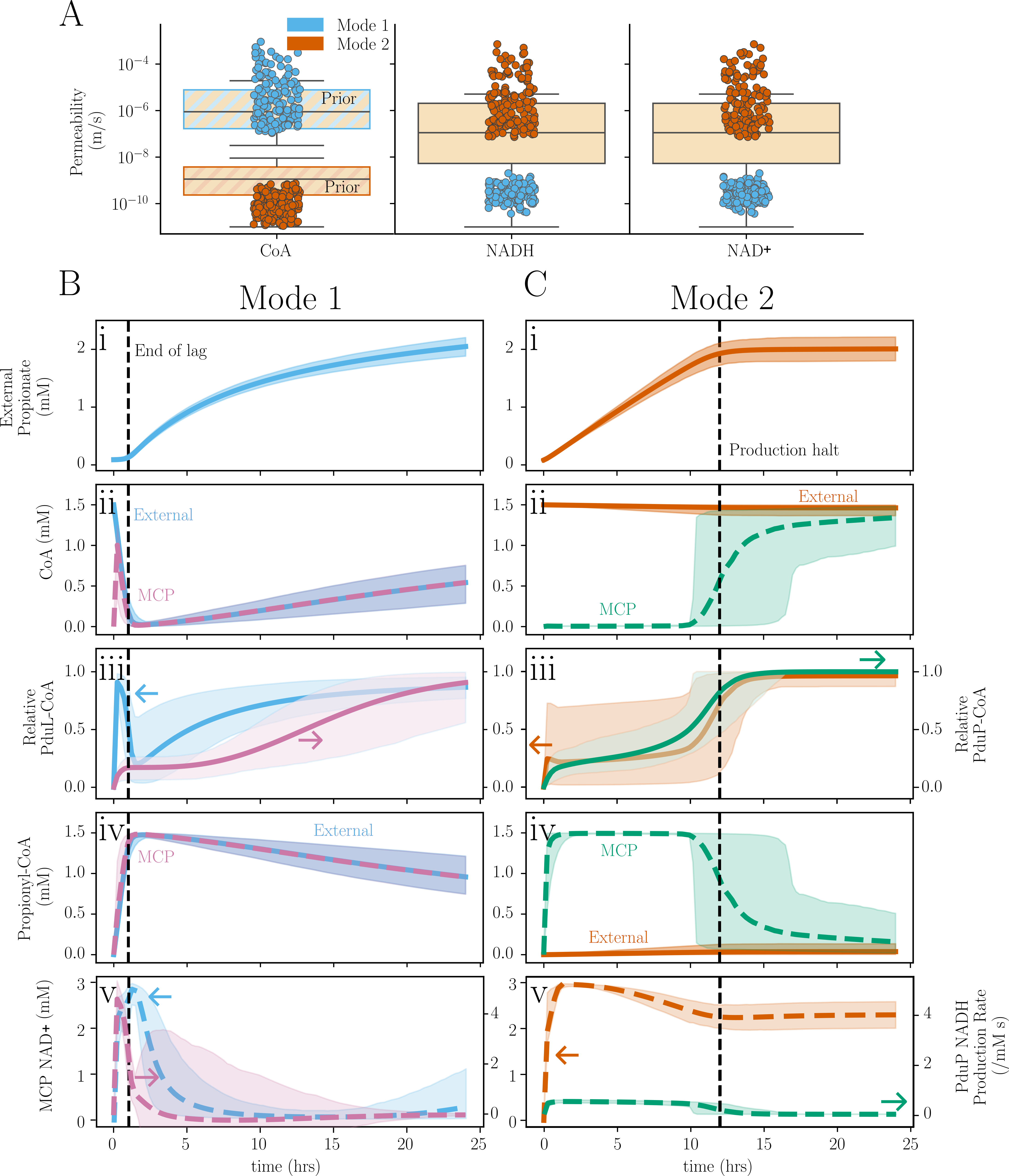}
    \caption{External Mode 1 (blue) and Mode 2 (orange) dynamics stem, in part, from differences in co-factor permeability. A) Inferred MCP CoA, NADH, and NAD+ permeability across modes, including updates from prior (box and whiskers) to posterior (circles). 
    B) Mode 1 exhibits i) a propionate lag until 1 hour (vertical dashed line). ii) MCP (dashed pink) and external (solid blue) CoA quickly reach a quasi-steady state. iii) The lag corresponds to a period of high PduL-CoA (left axis, blue) while accumulation of PduP-CoA (right axis, pink) occurs later. iv) MCP (dashed pink) and external (solid, blue) Propionyl-CoA follow the same dynamics. v) Depletion of MCP NAD+ (left axis, blue) and low PduP NADH production (right axis, pink).
    C) Mode 2 exhibits i) a termination of propionate production at 12 hours (vertical dashed line). ii) MCP (dashed green) and external (solid orange) CoA are decoupled. iii) Increases in PduL-CoA  (left axis, orange) and PduP-CoA (right axis, green) cause the halt in propionate production. iv) MCP (dashed green) and external (solid orange) propionyl-CoA are decoupled. v) MCP NAD+ (left axis, orange) is high, and PduP NADH production (right axis, green) is low; computed from PduP contribution to MCP NADH.
    PduL-CoA and PduP-CoA relative to the total respective enzyme concentration. 
    }\label{Fig7}
\end{figure}

As previously discussed, the two modes that we found differed in propionate dynamics. Mode 1 produced propionate with an initial lag and diminishing production rate while mode 2 produced propionate linearly with an abrupt halt at the 12 hour mark (Figure \ref{Fig4}C, Figure \ref{Fig7}B,C). The one of the main differences in parameters between modes were: mode 1 inferred a CoA permeability with 95\% high density intervals (HDI), [10\textsuperscript{-6.97}, 10\textsuperscript{-3.58}] m/s, while to mode 2 inferred a 95\% HDI, [10\textsuperscript{-10.94}, 10\textsuperscript{-9.34}] m/s (Figure \ref{Fig7}A). These differences in inferred CoA permeability were coupled with variation in PduP and PduL inhibition rates and mechanisms, which together produced different propionate dynamics across modes (Figure \ref{Fig7}A). Here we explain a detailed analysis of these coupled effects for Mode 1 and 2.

Mode 1 had an initial propionate lag (Figure \ref{Fig7}Bi) resulting from PduL-CoA accumulation in a high MCP CoA permeability environment. High MCP permeability to CoA (Figure \ref{Fig7}A) enabled fast equilibration of external and internal CoA pools (Figure \ref{Fig7}Bii) such that PduL-CoA complexes rapidly formed and prevented proprionate production (Figure \ref{Fig7}Biii). The PduL-CoA bottleneck cleared by the 30-minute mark once PduP converted much of the CoA pool to propionyl-CoA (Figure \ref{Fig7}iv). The subsequent decrease in propionate production rate during hours 1-24 was due to the slow accumulation of CoA and, thus, the conversion of available PduL back to PduL-CoA (Figure \ref{Fig7}Bii). PduQ oxidation and low MCP NAD+ permeability (Figure \ref{Fig7}A) reduced NAD+ availability (Figure \ref{Fig7}Bv) after the one-hour mark, consequently reducing PduP ability to convert CoA to propionyl-CoA. Lower NAD+ also reduced competitive binding between NAD+ and CoA to PduP, such that PduP-CoA began to accumulate around 10 hours (Figure \ref{Fig7}Biii), further slowing PduP output. This resulted in an additional net increase in CoA as PduL converted more propionyl-CoA to CoA than PduP could utilize (Figure \ref{Fig7}Bii). CoA accumulation further increased PduL-CoA formation (Figure \ref{Fig7}Bii) and, thus, slowed the propionate production rate over time. 

Mode 2 produced propionate linearly with an abrupt halt at around 10 hours (Figure \ref{Fig7}Ci) due to the slow formation of PduL-CoA and PduP-CoA complexes in a low MCP CoA permeability environment. The low MCP CoA permeability (Figure \ref{Fig7}A) maintained the external and MCP CoA pools out of equilibrium from each other, resulting in low CoA concentrations in the MCP before hour 10 (\ref{Fig7}Cii). Low CoA limited the formation of CoA inhibition complexes (Figure \ref{Fig7}Cii) without limiting propionyl-CoA production via PduP (Figure \ref{Fig7}Civ). This allowed PduQ \& PduP to recycle the internal MCP CoA pool and produce propionate at a constant rate without lag. PduP-CoA formation rates were higher in mode 2 than in mode 1, such that both PduL-CoA and PduP-CoA complexes formed at around the same rate (Figure \ref{Fig7}Ciii).

Oxidation of PduQ leads to a reduction in NAD+ availability during the first 10 hours (Figure \ref{Fig7}Cv). This small decrease in NAD+ was sufficient to reduce the relative rates of propionyl-CoA production compared to PduP-CoA and PduL-CoA complex formation causing sequestration of the majority of the enzyme and slowed production of propionyl-CoA (Figure \ref{Fig7}Civ)  at around 12 hours.
  CoA mass then accumulated in the MCP (Figure \ref{Fig7}Cii) and instigated rapid PduL-CoA \& PduP-CoA formation (Figure \ref{Fig7}Ciii). Cascading PduP and PduL inhibition abruptly halted propionate production.

\subsection*{NAD(H) permeability adjusted for differing PduP activity across modes}
\label{ssec:NAD(H)permeability}

As 1-propanol dynamics were consistent across modes, the differences in NAD(H) recycling between modes did not impact 1-propanol production, an NADH-dependent process. To compensate for the differing levels of NAD(H) turnover in the MCP, the modes adjusted the NADH/NAD$+$ permeabilities. Mode 1 inferred lower NAD(H) permeabilities with 95\% high density interval, [10\textsuperscript{-10.2}, 10\textsuperscript{-8.96}] m/s, relative to mode 2 with 95\% HDI, [10\textsuperscript{-7.14}, 10\textsuperscript{-3.70}] m/s (Figure \ref{Fig7}Ai-iii).

Low NAD(H) permeability in mode 1 decoupled the cytosolic and MCP NAD(H) pools. A low NADH permeability in mode 1 prevented an oversupply of NADH to PduQ, which would have resulted in higher production of 1-propanol than observed (\nameref{S32_Fig}). With high CoA MCP permeability, PduP was not CoA limited. PduP-CoA complex formation was relatively low during the first hour. With ample substrate and no complex inhibition PduP output high rates of propionyl-CoA production (Figure \ref{Fig7}B iv). High PduP activity provided PduQ with sufficient NADH to produce 1-propanol until around hour 12 when oxidation inhibition stopped production (Figure \ref{Fig7}B v).
The lack of recycling due to inhibition and low NAD+ permeability caused near-zero MCP NAD+ concentrations after 1 hour (Figure \ref{Fig7}B v). This ensured a subsequent decrease in PduP activity and propionate production, consistent with the data.  

In mode 2, the cofactor permeabilities were reversed with high permeability to NAD(H) and low permeability to CoA.  A high NADH permeability supplied PduQ with sufficient NADH to produce 1-propanol. PduP turnover was instead rate-limited by low MCP CoA availability (Figure \ref{Fig7}C ii). This prevented PduP from supplying PduQ with sufficient NADH to maintain 1-propanol production (Figure \ref{Fig7}C v) unlike mode 1 where NADH was sufficient at least initially (Figure \ref{Fig7}B v). External NADH was then necessary to compensate for low PduP activity. Conversion of this NADH through PduP activity resulted in high NAD+ production in the MCP.  
A high NAD+ permeability allowed leakage of NAD+ out of the MCP and prevented the up to $\sim$100 fold increase in MCP NAD+ concentration occuring at low MCP permeabilities to NAD+ (\nameref{S33_Fig}), which would disrupt PduQ dynamics and 1-propanol production (\nameref{S34_Fig} and \nameref{S35_Fig}). 

Thus, we found two regions of parameter space consistent with the rapid conversion of 1,2-propanediol to propionaldehyde followed by low conversion to either propionate or 1-propanol. In both modes, PduQ and PduP activity were restricted by a combination of limited co-factor permeability, PduQ oxidation, and complex inhibition. However, mode 1 slowed enzymatic activity through low NAD(H) permeability and mode 2 through low CoA permeability. Further data is needed to distinguish between either mode and elucidate the role of cofactors in MCP.

\section*{Discussion}

\subsection*{ Mass-Action kinetic modeling constrained by Michaelis-Menten measurements enabled predictions of enzyme complex dynamics}

In this work, we use mass-action mechanistic models to derive model dynamics consistent with experimentally measured {\it in vitro} MCP metabolite dynamics. Unlike published mass-action models and fitting procedures, we constrained our mass-action kinetic parameters to Michaelis-Menten measurements using quasi-steady-state derived equations. The constraints decreased the size of admissible parameter space and allowed for high-fidelity model calibration. Our use of the probabilistic programming language, PyMC, facilitated the integration of complex Michaelis-Menten expressions, and automated reparameterization and Jacobian computations \cite{pymc2023}. Our work shows that measured Michaelis-Menten kinetics are consistent with the low {\it in vitro} 1-propanol and propionate production when compared to {\it in vivo} dynamics \cite{Kennedy2022, Mills2022}, suggesting that {\it in vitro} reaction experienced inhibition which was not observed {\it in vivo}.

Mass-action modeling identified enzyme complex dynamics that would not have otherwise been allowed using Michaelis-Menten rate functions such as those of PduL and PduP, displayed in Figure \ref{Fig7}. We found two distinct modes that fit data with similar goodness of fit (Figure \ref{Fig4}). While both modes experienced similar updates of enzyme kinetic parameter from the prior to posterior, the modes differed significantly in inferred cofactor permeability and PduP inhibition rates (Figure \ref{Fig7}). Differences in inferred MCP functional and enzyme inhibition properties resulted in different propionate fits (Figure \ref{Fig4}C). Mode 1 and mode 2 dynamics have the largest differences in dynamic trajectories in propionate and 1-propanol between 6 and 24 hours, indicating additional measurements in this time-period would aid in discerning between modes. Furthermore, quantification of PduP inhibition rate, and K\textsubscript{M} of CoA for PduP and PduL could also eliminate modes and reduce uncertainty within modes.

\subsection*{Propionaldehyde permeability inference is consistent with existing literature}

 Prior observations that Pdu MCPs limit cytosolic accumulation of propionaldehyde \cite{Sampson2008} are consistent with our findings. Across modes, we inferred MCP propionaldehyde permeability greater than 10\textsuperscript{-8} m/s. Any posterior sample with propionaldehyde permeability less than 10\textsuperscript{-6.5} m/s accumulated propionaldehyde within the MCP (Figure \ref{Fig6}Cii). To increase the robustness of our results, further data, such as PduCDE kinetics, must be collected to reduce uncertainty in the propionaldehyde permeability.

Our results are also consistent with the Park et al. 2017 finding that the MCP 1,2-propanediol permeability is 3-10 times greater than the propionaldehyde permeability \cite{ParkYeates2017}. We inferred MCP 1,2-propanediol permeability to be greater than 10\textsuperscript{-6} m/s. As our 1,2-propanediol and propionaldehyde permeabilities are independent (mutual information $\approx$ 0), there exist posterior samples that satisfy the 3-10 permeability ratio (\nameref{S36_Fig}). We can then further restrict the MCP propionaldehyde permeability to be greater than 10\textsuperscript{-7} m/s, using Park et al. 2017, and less than 10\textsuperscript{-6.5} m/s, to maintain propionaldehyde accumulation in the MCP.

\subsection*{Mode 1 cofactor permeability is most consistent with previously published results}

Mode 1 and 2 inferred non-overlapping cofactor MCP permeabilities which are within range of previously published estimates for cofactors \cite{Young2025}. Mode 1 inferred NAD(H) permeability to have a 95\% HDI interval of [10\textsuperscript{-10.2}, 10\textsuperscript{-8.96}] m/s and CoA permeability to have a 95\% HDI interval of [10\textsuperscript{-6.97}, 10\textsuperscript{-3.58}] m/s. Mode 2 inferred NAD(H) permeability to have a 95\% HDI interval of [10\textsuperscript{-7.14}, 10\textsuperscript{-3.70}] m/s and CoA permeability to have a 95\% HDI interval of [10\textsuperscript{-10.94}, 10\textsuperscript{-9.34}] m/s. Such differences were due to correlation with inhibition kinetics. Relative to the prior, the combined ranges did not appreciably reduce admissible parameter space.  

Mode 1 is most consistent with the hypothesis that MCPs provide a private NAD+/NADH pool to Pdu enzymes. Cheng et al. 2012 found that purified WT MCPs produced propionyl-CoA faster than $\Delta$PduQ MCPs. The authors then concluded that MCPs have low permeability to NAD+ because externally supplied NAD+ could not compensate for internally supplied NAD+ from PduQ \cite{Cheng2012}. This conclusion is further supported by the fact that cofactors ($\sim$1 nm in diameter) would have a low probability of crossing the central pore ($\sim$0.6 nm in diameter) of PduA, the most abundant MCP shell protein. While mode 1 found permeabilities for CoA in a much higher range compared to NAD+/NADH, a subset of these are under the 10\textsuperscript{-6.5} m/s necessary to separate the MCP and cytosolic pool. It is also possible that the more linear nature of CoA compared to NAD(H) molecules allows easier transit through the MCP shell, and therefore higher permeabilities, but further validation is warranted. 

\subsection*{Limitations and Next Steps}
\subsubsection*{Bayesian Inference Framework}

Despite robust fits, the computational cost of NUTS and semiautomated grouping of mass-action kinetic parameters limits the application of the framework to medium and large-scale reaction networks. Applying NUTS with a max tree depth of 7 to a large-scale genome model of Greene et al. 2017 could take $>$1000 days to generate a similar number of samples. We also derive and simultaneously solve the known Michaelis-Menten equations for all leading and free parameter combinations. After comparing each combination, we manually import the simplest expression into Python. Such manual entry would be infeasible for larger networks and further is necessary to expand the application of NUTS to medium and large scale models. Alternative approaches, such as identifying and removing structural relationships between parameters through differential algebraic approaches are appealing \cite{linden2024increasing} but do not scale well with the complexity of the network in our experience. Overall, further work to automate the identification and incorporation of constraints between parameters would improve the scalability of Bayesian Approaches.

\subsubsection*{Mass-Action Limitations}

As binding orders for only a subset of enzymes have been studied, we constructed plausible orders for those with unknown order. An incorrect assumption can bias parameter inference and extrapolations. We assume that PduQ, PduL and PduW sequentially bind to cofactors first and their organic substrate second. These enzymes could reasonably exhibit sequential, random or Theorel-Chance binding mechanisms \cite{Ulusu2015}.  Changing the mechanism would affect Michaelis-Menten parameter integration, inferred enzyme dynamics and inhibition mechanisms. 

More consequentially, we assume that only free PduQ is oxidized. This simplifies model dynamics and reduces differential equation stiffness. Our assumption implies that oxygen and nicotinamide cofactors compete for the same active site. If not true, both modes would be affected as they both rely on competition between oxidation and NAD(H) to balance 1-propanol production. However, both modes could be admissible under PduQ allosteric inhibition at lower the oxidiation rates. Further studies of PduQ oxidation are necessary to narrow the range of oxidation rates and validate our findings.

\subsubsection*{Static ATP, ADP and phosphate concentration contribute to model error}

We assume a constant ADP:ATP ratio (1:30) in our model. This is not strictly true as PduCDE via PduGH, and, more importantly, AckA all interact with the adenosine pool (ADP and ATP). Neglecting the dynamics of the ATP pool could potentially affect propionate dynamics and lead to inaccurate inferences in enzyme dynamics. However, varying the ATP concentration experimentally did not affect propionate measurements. Thus, assuming an excess adenosine pool is a reasonable assumption. We also omit AckA phosphorylation of acetate, a component of the buffer solution. As acetate is the preferred substrate of AckA, this reaction actively competes with propionate phosphorylation and interacts adenosine pool. However, preliminary experimental results (not shown) indicate that competition between acetate and propionate is likely minimal as propionate dynamics did not improve in an acetate-deficient buffer.

We also assume a constant excess phosphate concentration (20 mM) as PduL reaction seemingly runs forward without exogenous phosphate. The source of phosphate is unclear as Pdu reactions only produce phosphate derivatives (acyl-phosphates and PPPi). A likely source is purified AckA from Sigma-Aldrich, which contains trace phosphate \cite{sigmaA7437}. As PduL and phosphate interact and both affect each other's dynamics, identification of the phosphate source is necessary to accurately model PduL dynamics.

\section*{Conclusion}

We constructed and calibrated a high-fidelity, mechanistic model of purified MCPs. Using a mass-action reaction function, we identified robust fits to metabolite time series. We used No U-Turns Hamiltonian Monte Carlo in PyMC \cite{pymc2023} for Bayesian Calibration and implemented Michaelis-Menten measurements as a hard constraint on mass-action kinetic parameters. {\it in vitro} calibration identified two modes that could equivalently fit the data dynamics. Across modes, we found that the MCP propionaldehyde permeability must be greater than $10^{-8}$ m/s. The modes identified cofactor permeabilities with little overlap and, when combined, did not appreciably reduce cofactor permeability uncertainty. We showed that collecting additional propionate time points or measuring inhibition rates could aid in deciding between modes and, thus, reduce uncertainty in cofactor permeability. 


\section*{Supporting Information}

Code to reproduce the results in this paper can be found at \url{https://github.com/aarcher07/In-vitro-calibration}

%
\paragraph*{S1 Fig.}
\label{S1_Fig}
{\bf Bar plot of divergent samples in each mode 1 chain. $99$\% of the 12000 samples were non-divergent, indicating well-behaved sampling.}

\paragraph*{S2 Fig.}
\label{S2_Fig}
{\bf Bar plot of divergent samples in each mode 1 chain. $99.8$\% of the 12000 samples were non-divergent, indicating well-behaved sampling.}

\paragraph*{S3 Fig.}
\label{S3_Fig}
{\bf Trace plot of log NAD+ and CoA permeability with unconstrained mode sampling. For these chains the ESS < 100 per chain and $\hat{R} \geq 1.05$ (S10 Table)}

\paragraph*{S4 Fig.}
\label{S4_Fig}
{\bf Trace plot of mode 2 log-likelihood. Mode 2 is sampled by restricting CoA prior support to [10\textsuperscript{-7.5}, 10\textsuperscript{-3}] m/s. Both iPduP inhibition kinetics and CoA permeability were simultaneously estimated. (S12 Table)}

\paragraph*{S5 Fig.}
\label{S5_Fig}
{\bf Mode 1 (A) 1,2-propanediol, (B) Propionaldehyde, (C) Propionate and (D) 1-Propanol posterior dynamics with k\textsubscript{1,iPduP} = 10\textsuperscript{-4} /mM s,  k\textsubscript{2,iPduP}=10\textsuperscript{-9} /s, and with k\textsubscript{1,iPduP} = 10\textsuperscript{-3.58} /mM s,  k\textsubscript{2,iPduP}=10\textsuperscript{-7.59} /s. The model fit with propionaldehyde evaporation was initialized from the prior mean and burned in for 3000 samples. 3000 samples were taken post burn-in.}

\paragraph*{S6 Fig.}
\label{S6_Fig}
{\bf Mode 2 (A) 1,2-propanediol, (B) Propionaldehyde, (C) Propionate and (D) 1-Propanol posterior dynamics with k\textsubscript{1,iPduP} = 10\textsuperscript{-1} /mM s,  k\textsubscript{2,iPduP}=10\textsuperscript{-6} /s, and with k\textsubscript{1,iPduP} = 10\textsuperscript{-0.36} /mM s,  k\textsubscript{2,iPduP}=10\textsuperscript{-3.76} /s. The model fit with propionaldehyde evaporation was initialized from the prior mean and burned in for 3000 samples. 3000 samples were taken post burn-in.}

\paragraph*{S7 Fig.}
\label{S7_Fig}
{\bf Comparison of estimated mode 1 permeabilities from fitting models with k\textsubscript{1,iPduP} = 10\textsuperscript{-4} /mM s,  k\textsubscript{2,iPduP}=10\textsuperscript{-9} /s, and with k\textsubscript{1,iPduP} = 10\textsuperscript{-3.58} /mM s,  k\textsubscript{2,iPduP}=10\textsuperscript{-7.59} /s.}

\paragraph*{S8 Fig.}
\label{S8_Fig}
{\bf Comparison of estimated mode 2 permeabilities from fitting models with k\textsubscript{1,iPduP} = 10\textsuperscript{-1} /mM s,  k\textsubscript{2,iPduP}=10\textsuperscript{-6} /s, and with k\textsubscript{1,iPduP} = 10\textsuperscript{-0.36} /mM s,  k\textsubscript{2,iPduP}=10\textsuperscript{-3.76} /s}

\paragraph*{S9 Fig.}
\label{S9_Fig}
{\bf Propionaldehyde evaporation assay was fit to the curve, y=be\textsuperscript{-a t}, using NUTS. The half-life was determined to be [2.97, 4.56] hr.}

\paragraph*{S10 Fig.}
\label{S10_Fig}
{\bf NUTS mode 1 (A) 1,2-propanediol, (B) Propionaldehyde, (C) Propionate and (D) 1-Propanol posterior dynamics with and without evaporation. The model fit with propionaldehyde evaporation was initialized from the prior mean and burned in for 3000 samples. 3000 samples were taken post burn-in.}

\paragraph*{S11 Fig.}
\label{S11_Fig}
{\bf NUTS mode 2 (A) 1,2-propanediol, (B) Propionaldehyde, (C) Propionate and (D) 1-Propanol posterior dynamics with and without evaporation. The model fit with propionaldehyde evaporation was initialized from the prior mean and burned in for 3000 samples. 3000 samples were taken post burn-in.}

\paragraph*{S12 Fig.}
\label{S12_Fig}
{\bf Comparison of estimated mode 1 permeabilities from fitting models with and without evaporation.}

\paragraph*{S13 Fig.}
\label{S13_Fig}
{\bf Comparison of estimated mode 2 permeabilities from fitting models with and without evaporation.}

\paragraph*{S14 Fig.}
\label{S14_Fig}
{\bf Corner plot of prior, mode 1 posterior and mode 2 posterior PduQ samples. All mass-action kinetic parameters are shown.}

\paragraph*{S15 Fig.}
\label{S15_Fig}
{\bf Corner plot of prior, mode 1 posterior and mode 2 posterior PduQ samples. All measured Michaelis-Menten parameters and estimated thermodynamic parameters are shown.}

\paragraph*{S16 Fig.}
\label{S16_Fig}
{\bf Corner plot of prior, mode 1 posterior and mode 2 posterior PduP samples. All mass-action kinetic parameters are shown.}

\paragraph*{S17 Fig.}
\label{S17_Fig}
{\bf Corner plot of prior, mode 1 posterior and mode 2 posterior PduP samples. All measured Michaelis-Menten parameters and estimated thermodynamic parameters are shown.}

\paragraph*{S18 Fig.}
\label{S18_Fig}
{\bf Corner plot of prior, mode 1 posterior and mode 2 posterior PduL samples. All mass-action kinetic parameters are shown.}

\paragraph*{S19 Fig.}
\label{S19_Fig}
{\bf Corner plot of prior, mode 1 posterior and mode 2 posterior PduL samples. All measured Michaelis-Menten parameters and estimated thermodynamic parameters are shown.}

\paragraph*{S20 Fig.}
\label{S20_Fig}
{\bf Corner plot of prior, mode 1 posterior and mode 2 posterior AckA samples. All mass-action kinetic parametersare shown.}

\paragraph*{S21 Fig.}
\label{S21_Fig}
{\bf Corner plot of prior, mode 1 posterior and mode 2 posterior AckA samples. All measured Michaelis-Menten parameters and estimated thermodynamic parameters are shown.}

\paragraph*{S22 Fig.}
\label{S22_Fig}
{\bf External (A) 1,2-propanediol, (B) Propionaldehyde, (C) Propionate and (D) 1-Propanol model dynamics of prior, mode 1 posterior and mode 2 posterior distribution.}


\paragraph*{S23 Fig.}
\label{S23_Fig}
{\bf  Strip plot comparing mode 1 and mode 2 inferred all measured reactions. The K\textsubscript{eq} uniform prior range is overlayed as an orange pointplot interval. }


\paragraph*{S24 Fig.}
\label{S24_Fig}
{\bf Mode 1 (A) 1,2-propanediol, (B) Propionaldehyde, (C) Propionate and (D) 1-Propanol inferred posterior dynamics and prediction with MCP 1,2-propanediol permeability set to 10\textsuperscript{-8} m/s.} 

\paragraph*{S25 Fig.}
\label{S25_Fig}
{\bf Mode 2 (A) 1,2-propanediol, (B) Propionaldehyde, (C) Propionate and (D) 1-Propanol inferred posterior dynamics and prediction with MCP 1,2-propanediol permeability set to 10\textsuperscript{-8} m/s.} 

\paragraph*{S26 Fig.}
\label{S26_Fig}
{\bf Mode 1 (A) 1,2-propanediol, (B) Propionaldehyde, (C) Propionate and (D) 1-Propanol inferred posterior dynamics and prediction with MCP Propionaldehyde permeability set to 10\textsuperscript{-10} m/s.} 

\paragraph*{S27 Fig.}
\label{S27_Fig}
{\bf Mode 2 (A) 1,2-propanediol, (B) Propionaldehyde, (C) Propionate and (D) 1-Propanol inferred posterior dynamics and prediction with MCP Propionaldehyde permeability set to 10\textsuperscript{-10} m/s.} 

\paragraph*{S28 Fig.}
\label{S28_Fig}
{\bf Mode 1 MCP propionaldehyde posterior dynamics and prediction with MCP propionaldehyde permeability set to 10\textsuperscript{-10} m/s. }

\paragraph*{S29 Fig.}
\label{S29_Fig}
{\bf Mode 2 MCP propionaldehyde posterior dynamics and prediction with MCP propionaldehyde permeability set to 10\textsuperscript{-10} m/s.} 

\paragraph*{S30 Fig.}
\label{S30_Fig}
{\bf Mode 1 PduCDE-1,2-propanediol-Propionaldehyde complex posterior dynamics and prediction with MCP propionaldehyde permeability set to 10\textsuperscript{-10} m/s. Enzyme dynamics were taken relative to total PduCDE enzyme concentration.} 

\paragraph*{S31 Fig.}
\label{S31_Fig}
{\bf Mode 2 PduCDE-1,2-propanediol-Propionaldehyde complex posterior dynamics and prediction with MCP propionaldehyde permeability set to 10\textsuperscript{-10} m/s. Enzyme dynamics were taken relative to total PduCDE enzyme concentration. }

\paragraph*{S32 Fig.}
\label{S32_Fig}
{\bf Effect of NADH permeability on mode 1 MCP cofactor dynamics. MCP NADH mode 1 dynamics with perturbed NAD+ permeability set to 0.5 m/s are shown alongside inferred mode dynamics. The perturbation estimates have likelihood greater than 10 times the mean likelihood of mode 1.}

\paragraph*{S33 Fig.}
\label{S33_Fig}
{\bf Effect of NAD+ permeability on mode 2 MCP cofactor dynamics. MCP NAD+ mode 2 dynamics with perturbed NAD+ permeability set to 10\textsuperscript{-10} m/s are shown alongside inferred mode dynamics. The perturbation estimates have likelihood greater than 10 times the mean likelihood of mode 1.}

\paragraph*{S34 Fig.}
\label{S34_Fig}
{\bf Effect of NAD+ permeability on mode 2 PduQ inhibition dynamics. (A) PduQ-NAD+ and (B) PduQ-Oxidized mode 2 dynamics with perturbed NAD+ permeability set to 10\textsuperscript{-10} m/s are shown alongside inferred mode dynamics. The perturbation estimates have likelihood greater than 10 times the mean likelihood of mode 1.}

\paragraph*{S35 Fig.}
\label{S35_Fig}
{\bf Effect of NAD+ permeability on mode 2 external concentration dynamics. (A) 1-Propanol and (B) Propionate mode 2 dynamics with perturbed NAD+ permeability set to 10\textsuperscript{-10} m/s are shown alongside inferred mode dynamics. The perturbation estimates have likelihood greater than 10 times the mean likelihood of mode 1.}


\paragraph*{S36 Fig.}
\label{S36_Fig}
{\bf Corner plot of 1,2-propanediol permeability and propionaldehyde permeability. 1000 samples from mode 1 and 2 posterior, and the 1,2-propanediol to propionaldehyde permeability ratio of 3-10 are shown. } 

\paragraph*{S37 Fig.}
\label{S37_Fig}
{\bf k\textsubscript{1,iPduP} and k\textsubscript{2,iPduP} joint distribution of select, well-sampled mode 1 chains. MCP CoA permeability was constrained to [10\textsuperscript{-11}, 10\textsuperscript{-8}] m/s. }

\paragraph*{S38 Fig.}
\label{S38_Fig}
{\bf k\textsubscript{1,iPduP} and k\textsubscript{2,iPduP} joint distribution of select, well-sampled mode 2 chains. MCP CoA permeability was constrained to [10\textsuperscript{-7.5}, 10\textsuperscript{-3}] m/s. }

\paragraph*{S1 Table.}
\label{S1_Table}
{\bf Table of available model parameters. {}\textsuperscript{a} The assay was performed with acetyl-phosphate. We assume that the kinetics of acetyl-phosphate and propionyl-phosphate are within an order of magnitude. }

\paragraph*{S2 Table.}
\label{S2_Table}
{\bf Derived Michaelis-Menten Formulae for all enzymes in the purified MCP assay.}

\paragraph*{S3 Table.}
\label{S3_Table}
{\bf Decomposition of all enzyme kinetic parameters into free and leading parameters.}

\paragraph*{S4 Table.}
\label{S4_Table}
{\bf Inequality expressions for free enzyme kinetic parameters.}

\paragraph*{S5 Table.}
\label{S5_Table}
{\bf Combined leading equations and free inequalities for all enzymes.}

\paragraph*{S6 Table.}
\label{S6_Table}
{\bf Posterior MCMC results with sampling restricted to mode 1.  We set k\textsubscript{1,iPduP} = 10\textsuperscript{-3.58} /mM s and k\textsubscript{2,iPduP} = 10\textsuperscript{-7.59} /s. CoA permeability was restricted to [10\textsuperscript{-7.5}, 10\textsuperscript{-3}] m/s. 4 chains were parameterized with 3000 samples, 3000 tuning iterations, four chains, 0.85 acceptance rate, and max tree depth of 7. ESS $>$ 400 and $\hat{R}<$1.05, indicating well-behaved sampling. The {\it in vitro} differential equation was integrated with an absolute and relative tolerance of 10\textsuperscript{-9}. The {\it in vitro} adjoint equation was integrated with an absolute and relative tolerance of 10\textsuperscript{-2}.}

\paragraph*{S7 Table.}
\label{S7_Table}
{\bf Table of BFMI for each mode 1 chain. BFMIs are all $>0.2$, indicating well behaved sampling.}

\paragraph{S8 Table.}
\label{S8_Table}
{\bf Posterior MCMC results with sampling restricted to mode 2.  We set k\textsubscript{1,iPduP} = 10\textsuperscript{-0.36} /mM s  and k\textsubscript{2,iPduP} = 10\textsuperscript{-3.76} /s. CoA permeability was restricted to [10\textsuperscript{-11}, 10\textsuperscript{-8}] m/s. 4 chains were parameterized with 3000 samples, 3000 tuning iterations, 0.85 acceptance rate, and max tree depth of 7. ESS $>$ 400 and $\hat{R}<$1.05, indicating well-behaved sampling. The {\it in vitro} differential equation was integrated with an absolute and relative tolerance of 10\textsuperscript{-8}. The {\it in vitro} adjoint equation was integrated with an absolute and relative tolerance of 10\textsuperscript{-2}.}

\paragraph*{S9 Table.}
\label{S9_Table}
{\bf Table of BFMI for each mode 2 chain. BFMIs are all $>0.2$, indicating well-behaved sampling.}

\paragraph*{S10 Table.}
\label{S10_Table}
{\bf Posterior MCMC results with unrestricted sampling. 4 chains were parameterized with 8000 samples, 3000 tuning iterations, four chains, 0.85 acceptance rate, and max tree depth of 7. The {\it in vitro} differential equation was integrated with an absolute and relative tolerance of 10\textsuperscript{-9}. The {\it in vitro} adjoint equation was integrated with an absolute and relative tolerance of 10\textsuperscript{-2}.}

\paragraph*{S11 Table.}
\label{S11_Table}
{\bf Preliminary posterior sampling of MCMC mode 1 using selected chains. 4 chains were parameterized with 4000 samples, 5000 tuning iterations, four chains, 0.85 acceptance rate, and max tree depth of 7. The {\it in vitro} differential equation was integrated with an absolute and relative tolerance of 10\textsuperscript{-8}. The {\it in vitro} adjoint equation was integrated with an absolute and relative tolerance of 10\textsuperscript{-2}. The model was fit with a PduW component. However, PduW contributed little mass flux to propionate production.}

\paragraph*{S12 Table.}
\label{S12_Table}
{\bf Preliminary posterior sampling of MCMC mode 2 using selected chains. 4 chains were parameterized with 4000 samples, 5000 tuning iterations, four chains, 0.85 acceptance rate, and max tree depth of 7. The {\it in vitro} differential equation was integrated with an absolute and relative tolerance of 10\textsuperscript{-8}. The {\it in vitro} adjoint equation was integrated with an absolute and relative tolerance of 10\textsuperscript{-2}. The model was fit with a PduW component. However, PduW contributed little mass flux to propionate production.}

\paragraph*{S13 Table.}
\label{S13_Table}
{\bf Posterior MCMC results with sampling restricted to mode 2. CoA permeability was restricted to [10\textsuperscript{-11}, 10\textsuperscript{-8}] m/s. 4 chains were parameterized with 8000 samples, 3000 tuning iterations, four chains, 0.85 acceptance rate, and max tree depth of 7. The {\it in vitro} differential equation was integrated with an absolute and relative tolerance of 10\textsuperscript{-8}. The {\it in vitro} adjoint equation was integrated with an absolute and relative tolerance of 10\textsuperscript{-2}.}

\paragraph{S14 Table.}
\label{S14_Table}
{\bf Posterior MCMC results with sampling restricted to mode 1. The associated {\it in vitro} model does not include evaporation. We set k\textsubscript{1,iPduP} = 10\textsuperscript{-4} /mM s and k\textsubscript{2,iPduP} = 10\textsuperscript{-9} /s. Inhibition parameters were taken from \nameref{S37_Fig}. CoA permeability was restricted to [10\textsuperscript{-7.5}, 10\textsuperscript{-3}] m/s. Only statistics of permeability parameters are shown. 4 chains were parameterized with 3000 samples, 3000 tuning iterations, 0.85 acceptance rate, and max tree depth of 7. ESS $>$ 400 and $\hat{R}<$1.05, indicating well-behaved sampling. The {\it in vitro} differential equation was integrated with an absolute and relative tolerance of 10\textsuperscript{-9}. The {\it in vitro} adjoint equation was integrated with an absolute and relative tolerance of 10\textsuperscript{-2}. The model was fit with a PduW component. However, PduW contributed little mass flux to propionate production.}

\paragraph{S15 Table.}
\label{S15_Table}
{\bf Posterior MCMC results with sampling restricted to mode 2. The associated {\it in vitro} model does not include evaporation.  We set k\textsubscript{1,iPduP} = 10\textsuperscript{-1} /mM s  and k\textsubscript{2,iPduP} = 10\textsuperscript{-6} /s. Inhibition parameters were taken from \nameref{S38_Fig}. CoA permeability was restricted to [10\textsuperscript{-11}, 10\textsuperscript{-8}] m/s. 4 chains were parameterized with 3000 samples, 3000 tuning iterations, 0.85 acceptance rate, and max tree depth of 7. ESS $>$ 400 and $\hat{R}<$1.05, indicating well-behaved sampling. The {\it in vitro} differential equation was integrated with an absolute and relative tolerance of 10\textsuperscript{-9}. The {\it in vitro} adjoint equation was integrated with an absolute and relative tolerance of 10\textsuperscript{-2}. The model was fit with a PduW component. However, PduW contributed little mass flux to propionate production.}

\paragraph*{S1 Text.}
\label{Text:S1_Text}
{\bf Mass action decomposition of all enzymatic reactions.}

\paragraph*{S2 Text.}
\label{Text:S2_Text}
{\bf Derivation of MCP contribution to aliquot concentration.}

\paragraph*{S3 Text.}
\label{Text:S3_Text}
{\bf Derivation of PduQ Michaelis-Menten parameters from a quasi-steady state approximation of the PduQ mass-action equation.}

%

\section*{Acknowledgments}
This work was funded in part by the US Army Contracting Command - Rock Island (grant  W52P1J-21-9-3023 to D.T.E. and N.M.M.), the Army Research Office (grant W911NF-19-1-0298 to D.T.E.), the Department of Energy (grant DE-SC0019337 to D.T.E. and N.M.M. ). BJP was partially funded by a National Science Foundation Graduate training grant (grant DGE-2021900) via the Northwestern University Synthetic Biology Across the Scales Training Program.
\nolinenumbers

%
%
%

\newgeometry{right=1in}

\noindent

\section*{Supplementary Information}

\begin{figure}[ht]
    \centering
    \includegraphics[width=\textwidth]{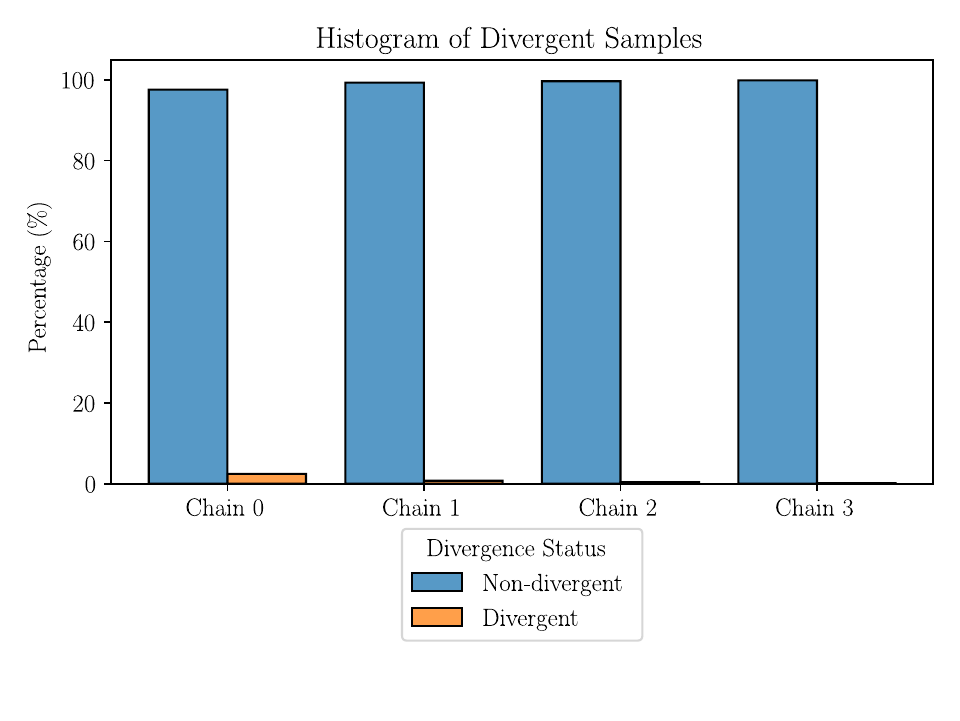}
    \caption*{SI Figure 1:  Bar plot of divergent samples in each mode 1 chain. $99$\% of the 12000 samples were non-divergent, indicating well-behaved sampling.}
\end{figure}
\newpage
\begin{figure}[ht]
    \centering
    \includegraphics[width=\textwidth]{S2Fig\_svg-tex.pdf}
    \caption*{SI Figure 2:  Bar plot of divergent samples in each mode 1 chain. $99.8$\% of the 12000 samples were non-divergent, indicating well-behaved sampling.}
\end{figure}
\newpage

\begin{figure}[ht]
    \centering
    \includegraphics[width=\textwidth]{S3Fig\_svg-tex.pdf}
    \caption*{SI Figure 3. Trace plot of log NAD+ and CoA permeability with unconstrained mode sampling. For these chains the ESS $\leq$ 100 per chain and $\hat{R} \geq 1.05$ (S10 Table).}
\end{figure}
\newpage

\begin{figure}[ht]
    \centering
    \includegraphics[width=\textwidth]{S4Fig\_svg-tex.pdf}
    \caption*{SI Figure 4 Trace plot of mode 2 log-likelihood. Mode 2 is sampled by restricting CoA prior support to [10\textsuperscript{-7.5}, 10\textsuperscript{-3}] m/s. Both iPduP inhibition kinetics and CoA permeability were simultaneously estimated. (S12 Table)}
\end{figure}
\newpage

\begin{figure}[ht]
    \centering
    \includegraphics[width=\textwidth]{S5Fig\_svg-tex.pdf}
    \caption*{SI Figure 5: {\bf Mode 1 (A) 1,2-propanediol, (B) Propionaldehyde, (C) Propionate and (D) 1-Propanol posterior dynamics with k\textsubscript{1,iPduP} = 10\textsuperscript{-4} /mM s,  k\textsubscript{2,iPduP}=10\textsuperscript{-9} /s, and with k\textsubscript{1,iPduP} = 10\textsuperscript{-3.58} /mM s,  k\textsubscript{2,iPduP}=10\textsuperscript{-7.59} /s. The model fit with propionaldehyde evaporation was initialized from the prior mean and burned in for 3000 samples. 3000 samples were taken post burn-in.}}
\end{figure}
\newpage

\begin{figure}[ht]
    \centering
    \includegraphics[width=\textwidth]{S6Fig\_svg-tex.pdf}
    \caption*{SI Figure 6: {\bf Mode 2 (A) 1,2-propanediol, (B) Propionaldehyde, (C) Propionate and (D) 1-Propanol posterior dynamics with k\textsubscript{1,iPduP} = 10\textsuperscript{-1} /mM s,  k\textsubscript{2,iPduP}=10\textsuperscript{-6} /s, and with k\textsubscript{1,iPduP} = 10\textsuperscript{-0.36} /mM s,  k\textsubscript{2,iPduP}=10\textsuperscript{-3.76} /s. The model fit with propionaldehyde evaporation was initialized from the prior mean and burned in for 3000 samples. 3000 samples were taken post burn-in.}}
\end{figure}
\newpage

\begin{figure}[ht]
    \centering
    \includegraphics[width=0.7\textwidth]{S7Fig\_svg-tex.pdf}
    \caption*{SI Figure 7: {\bf Comparison of estimated mode 1 permeabilities from fitting models with k\textsubscript{1,iPduP} = 10\textsuperscript{-4} /mM s,  k\textsubscript{2,iPduP}=10\textsuperscript{-9} /s, and with k\textsubscript{1,iPduP} = 10\textsuperscript{-3.58} /mM s,  k\textsubscript{2,iPduP}=10\textsuperscript{-7.59} /s.}}
\end{figure}
\newpage

\begin{figure}[ht]
    \centering
    \includegraphics[width=0.7\textwidth]{S8Fig\_svg-tex.pdf}
    \caption*{SI Figure 8 {\bf Comparison of estimated mode 2 permeabilities from fitting models with k\textsubscript{1,iPduP} = 10\textsuperscript{-1} /mM s,  k\textsubscript{2,iPduP}=10\textsuperscript{-6} /s, and with k\textsubscript{1,iPduP} = 10\textsuperscript{-0.36} /mM s,  k\textsubscript{2,iPduP}=10\textsuperscript{-3.76} /s}}
\end{figure}
\newpage

\begin{figure}[ht]
    \centering
    \includegraphics[width=\textwidth]{S9Fig\_svg-tex.pdf}
    \caption*{SI Figure 9: {\bf Propionaldehyde evaporation assay was fit to the curve, y=be\textsuperscript{-a t}, using NUTS. The half-life was determined to be [2.97, 4.56] hr.}}
\end{figure}
\newpage

\begin{figure}[ht]
    \centering
    \includegraphics[width=\textwidth]{S10Fig\_svg-tex.pdf}
    \caption*{SI Figure 10: {\bf NUTS mode 1 (A) 1,2-propanediol, (B) Propionaldehyde, (C) Propionate and (D) 1-Propanol posterior dynamics with and without evaporation. The model fit with propionaldehyde evaporation was initialized from the prior mean and burned in for 3000 samples. 3000 samples were taken post burn-in.}}
\end{figure}
\newpage

\begin{figure}[ht]
    \centering
    \includegraphics[width=\textwidth]{S11Fig\_svg-tex.pdf}
    \caption*{SI Figure 11 {\bf NUTS mode 2 (A) 1,2-propanediol, (B) Propionaldehyde, (C) Propionate and (D) 1-Propanol posterior dynamics with and without evaporation. The model fit with propionaldehyde evaporation was initialized from the prior mean and burned in for 3000 samples. 3000 samples were taken post burn-in.}}
\end{figure}
\newpage

\begin{figure}[ht]
    \centering
    \includegraphics[width=0.7\textwidth]{S12Fig\_svg-tex.pdf}
    \caption*{SI Figure 12 {\bf Comparison of estimated mode 1 permeabilities from fitting models with and without evaporation.}}
\end{figure}
\newpage

\begin{figure}[ht]
    \centering
    \includegraphics[width=0.7\textwidth]{S13Fig\_svg-tex.pdf}
    \caption*{SI Figure 13 {\bf Comparison of estimated mode 2 permeabilities from fitting models with and without evaporation.}}
\end{figure}
\newpage

\begin{figure}[ht]
    \centering
    \includegraphics[width=\textwidth]{S14Fig\_svg-tex.pdf}
    \caption*{SI Figure 14:
{\bf Corner plot of prior, mode 1 posterior and mode 2 posterior PduQ samples. All mass-action kinetic parameters are shown.}}
\end{figure}
\newpage

\begin{figure}[ht]
    \centering
    \includegraphics[width=\textwidth]{S15Fig\_svg-tex.pdf}
    \caption*{SI Figure 15: {\bf Corner plot of prior, mode 1 posterior and mode 2 posterior PduQ samples. All measured Michaelis-Menten parameters and estimated thermodynamic parameters are shown.}}
\end{figure}
\newpage

\begin{figure}[ht]
    \centering
    \includegraphics[width=\textwidth]{S16Fig\_svg-tex.pdf}
    \caption*{SI Figure 16: {\bf Corner plot of prior, mode 1 posterior and mode 2 posterior PduP samples. All mass-action kinetic parameters are shown.}}
\end{figure}


\begin{figure}[ht]
    \centering
    \includegraphics[width=\textwidth]{S17Fig\_svg-tex.pdf}
    \caption*{SI Figure 17: {\bf Corner plot of prior, mode 1 posterior and mode 2 posterior PduP samples. All measured Michaelis-Menten parameters and estimated thermodynamic parameters are shown.}}
\end{figure}
\newpage

\begin{figure}[ht]
    \centering
    \includegraphics[width=\textwidth]{S18Fig\_svg-tex.pdf}
    \caption*{SI Figure 18: {\bf Corner plot of prior, mode 1 posterior and mode 2 posterior PduL samples. All mass-action kinetic parameters are shown.}}
\end{figure}
\newpage

\begin{figure}[ht]
    \centering
    \includegraphics[width=\textwidth]{S19Fig\_svg-tex.pdf}
    \caption*{SI Figure 19: {\bf Corner plot of prior, mode 1 posterior and mode 2 posterior PduL samples. All measured Michaelis-Menten parameters and estimated thermodynamic parameters are shown.}}
\end{figure}
\newpage

\begin{figure}[ht]
    \centering
    \includegraphics[width=\textwidth]{S20Fig\_svg-tex.pdf}
    \caption*{SI Figure 20:{\bf Corner plot of prior, mode 1 posterior and mode 2 posterior AckA samples. All mass-action kinetic parameters are shown.}}
\end{figure}
\newpage

\begin{figure}[ht]
    \centering
    \includegraphics[width=\textwidth]{S21Fig\_svg-tex.pdf}
    \caption*{SI Figure 21: {\bf Corner plot of prior, mode 1 posterior and mode 2 posterior AckA samples. All measured Michaelis-Menten parameters and estimated thermodynamic parameters are shown.}}
\end{figure}
\newpage

\begin{figure}[ht]
    \centering
    \includegraphics[width=\textwidth]{S22Fig\_svg-tex.pdf}
    \caption*{SI Figure 22: {\bf External (A) 1,2-propanediol, (B) Propionaldehyde, (C) Propionate and (D) 1-Propanol model dynamics of prior, mode 1 posterior and mode 2 posterior distribution.}}
\end{figure}
\newpage

\begin{figure}[ht]
    \centering
    \includegraphics[width=\textwidth]{S23Fig\_svg-tex.pdf}
    \caption*{SI Figure 23: {\bf Strip plot comparing mode 1 and mode 2 inferred all measured reactions. The K\textsubscript{eq} uniform prior range is overlayed as an orange pointplot interval. }}
\end{figure}
\newpage

\begin{figure}[ht]
    \centering
    \includegraphics[width=\textwidth]{S24Fig\_svg-tex.pdf}
    \caption*{SI Figure 24: {\bf Mode 1 (A) 1,2-propanediol, (B) Propionaldehyde, (C) Propionate and (D) 1-Propanol inferred posterior dynamics and prediction with MCP 1,2-propanediol permeability set to 10\textsuperscript{-8} m/s.} }
\end{figure}
\newpage

\begin{figure}[ht]
    \centering
    \includegraphics[width=\textwidth]{S25Fig\_svg-tex.pdf}
    \caption*{SI Figure 25: {\bf Mode 2 (A) 1,2-propanediol, (B) Propionaldehyde, (C) Propionate and (D) 1-Propanol inferred posterior dynamics and prediction with MCP 1,2-propanediol permeability set to 10\textsuperscript{-8} m/s.} }
\end{figure}
\newpage

\begin{figure}[ht]
    \centering
    \includegraphics[width=\textwidth]{S26Fig\_svg-tex.pdf}
    \caption*{SI Figure 26: {\bf Mode 1 (A) 1,2-propanediol, (B) Propionaldehyde, (C) Propionate and (D) 1-Propanol inferred posterior dynamics and prediction with MCP Propionaldehyde permeability set to 10\textsuperscript{-10} m/s.} }
\end{figure}
\newpage

\begin{figure}[ht]
    \centering
    \includegraphics[width=\textwidth]{S27Fig\_svg-tex.pdf}
    \caption*{SI Figure 27: {\bf Mode 2 (A) 1,2-propanediol, (B) Propionaldehyde, (C) Propionate and (D) 1-Propanol inferred posterior dynamics and prediction with MCP Propionaldehyde permeability set to 10\textsuperscript{-10} m/s.} }
\end{figure}
\newpage

\begin{figure}[ht]
    \centering
    \includegraphics[width=\textwidth]{S28Fig\_svg-tex.pdf}
    \caption*{SI Figure 28: {\bf Mode 1 MCP propionaldehyde posterior dynamics and prediction with MCP propionaldehyde permeability set to 10\textsuperscript{-10} m/s. }}
\end{figure}
\newpage

\begin{figure}[ht]
    \centering
    \includegraphics[width=\textwidth]{S29Fig\_svg-tex.pdf}
    \caption*{SI Figure 29: {\bf Mode 2 MCP propionaldehyde posterior dynamics and prediction with MCP propionaldehyde permeability set to 10\textsuperscript{-10} m/s.} }
\end{figure}
\newpage

\begin{figure}[ht]
    \centering
    \includegraphics[width=\textwidth]{S30Fig\_svg-tex.pdf}
    \caption*{SI Figure 30: {\bf Mode 1 PduCDE-1,2-propanediol-Propionaldehyde complex posterior dynamics and prediction with MCP propionaldehyde permeability set to 10\textsuperscript{-10} m/s. Enzyme dynamics were taken relative to total PduCDE enzyme concentration.} }
\end{figure}
\newpage

\begin{figure}[ht]
    \centering
    \includegraphics[width=\textwidth]{S31Fig\_svg-tex.pdf}
    \caption*{SI Figure 31: {\bf Mode 2 PduCDE-1,2-propanediol-Propionaldehyde complex posterior dynamics and prediction with MCP propionaldehyde permeability set to 10\textsuperscript{-10} m/s. Enzyme dynamics were taken relative to total PduCDE enzyme concentration. } }
\end{figure}
\newpage

\begin{figure}[ht]
    \centering
    \includegraphics[width=\textwidth]{S32Fig\_svg-tex.pdf}
    \caption*{SI Figure 32: {\bf Effect of NADH permeability on mode 1 MCP cofactor dynamics. MCP NADH mode 1 dynamics with perturbed NAD+ permeability set to 0.5 m/s are shown alongside inferred mode dynamics. The perturbation estimates have likelihood greater than 10 times the mean likelihood of mode 1.}}
\end{figure}
\newpage

\begin{figure}[ht]
    \centering
    \includegraphics[width=\textwidth]{S33Fig\_svg-tex.pdf}
    \caption*{SI Figure 33: {\bf Effect of NAD+ permeability on mode 2 MCP cofactor dynamics. MCP NAD+ mode 2 dynamics with perturbed NAD+ permeability set to 10\textsuperscript{-10} m/s are shown alongside inferred mode dynamics. The perturbation estimates have likelihood greater than 10 times the mean likelihood of mode 1.}}
\end{figure}
\newpage

\begin{figure}[ht]
    \centering
    \includegraphics[width=\textwidth]{S34Fig\_svg-tex.pdf}
    \caption*{SI Figure 34 {\bf Effect of NAD+ permeability on mode 2 PduQ inhibition dynamics. (A) PduQ-NAD+ and (B) PduQ-Oxidized mode 2 dynamics with perturbed NAD+ permeability set to 10\textsuperscript{-10} m/s are shown alongside inferred mode dynamics. The perturbation estimates have likelihood greater than 10 times the mean likelihood of mode 1.}}
\end{figure}
\newpage

\begin{figure}[ht]
    \centering
    \includegraphics[width=\textwidth]{S35Fig\_svg-tex.pdf}
    \caption*{SI Figure 35: {\bf Effect of NAD+ permeability on mode 2 external concentration dynamics. (A) 1-Propanol and (B) Propionate mode 2 dynamics with perturbed NAD+ permeability set to 10\textsuperscript{-10} m/s are shown alongside inferred mode dynamics. The perturbation estimates have likelihood greater than 10 times the mean likelihood of mode 1.}}
\end{figure}
\newpage

\begin{figure}[ht]
    \centering
    \includegraphics[width=\textwidth]{S36Fig\_svg-tex.pdf}
    \caption*{SI Figure 36: {\bf Corner plot of 1,2-propanediol permeability and propionaldehyde permeability. 1000 samples from mode 1 and 2 posterior, and the 1,2-propanediol to propionaldehyde permeability ratio of 3-10 are shown. } }
\end{figure}
\newpage

\begin{figure}[ht]
    \centering
    \includegraphics[width=\textwidth]{S37Fig\_svg-tex.pdf}
    \caption*{SI Figure 37: {\bf  k\textsubscript{1,iPduP} and k\textsubscript{2,iPduP} joint distribution of select, well sampled mode 1 chains. MCP CoA permeability was constrained to [10\textsuperscript{-11}, 10\textsuperscript{-8}] m/s. }}
\end{figure}
\newpage

\begin{figure}[ht]
    \centering
    \includegraphics[width=\textwidth]{S38Fig\_svg-tex.pdf}
    \caption*{SI Figure 38: {\bf  k\textsubscript{1,iPduP} and k\textsubscript{2,iPduP} joint distribution of select, well sampled mode 2 chains. MCP CoA permeability was constrained to [10\textsuperscript{-7.5}, 10\textsuperscript{-3}] m/s.}}
\end{figure}

\singlespacing\renewcommand*{\arraystretch}{2}
\section*{Table S1. Table of available model parameters.}
\begin{center}
 \begin{longtable}{m{8em} >{\raggedright\arraybackslash}m{12.5em} m{7.5em} m{5em}} 
 \toprule
 Parameter& Meaning & Estimated Value &  Citation\\ 
 \midrule
\endfirsthead
\toprule
 Parameter& Meaning & Estimated Value &  Citation\\ 
\midrule
\endhead
\midrule
\multicolumn{4}{r}{Continued on next page} \\
\midrule
\endfoot
\endlastfoot
\bottomrule
 $M$ & number of MCPs in the external volume & $2.01 \times 10^{9}$ MCPs & \cite{Cheng2008} \\
 $r_m$ & radius of the microcompartment & $70$ nm & \cite{Kennedy2020}\\ 
  $V_{\text{MCP}}$ & MCP volume &  $6 \times 10^{-21}$ m$^3$  &  \\
  $S_{\text{MCP}}$ & MCP surface area volume & $6 \times 10^{-14}$ m$^2$  &  \\
Water concentration &  Static external water concentration & $5.56 \times 10^4$ mM  &  \\
ATP concentration &  ATP static concentration in the cell & $0.75$ mM  &  \\
ADP concentration &  ADP static concentration in the cell & $0.75$ mM  &  \\
Phosphate concentration &  Phosphate static concentration in the cell & $20$ mM  &  \\
 $N_{\text{PduCDE}}$ & number of PduCDE per MCP & $350-650$  & \cite{Yang2020}  \\ 
$k_{\text{cat,PduCDE}}^{\text{f}}$ & maximum forward rate reaction of PduCDE reaction & $3 \times 10^2- 4\times 10^2$ s$^{-1}$ & \cite{Bachovchin1977,Ogura2008,Kinoshita2008}   \\ 
$K^{\text{1,2-PD}}_{\text{M}, \text{PduCDE}}$ & half-max concentration of 1,2-propanediol for PduCDE reaction & $(10^{-1},1) \times 10^{-1}$ mM &  \cite{Bachovchin1977,Ogura2008,Kinoshita2008} \\ 
$K^{\text{Ado-B}_{12}}_{\text{M}, \text{PduCDE}}$ & half-max concentration of $\text{Ado-B}_{12}$ for PduCDE reaction & $8.4 \times 10^{-2}$ mM &  \cite{Ogura2008} \\ 
$K_{\text{eq, PduCDE}}$ & Dissociation constants for the PduCDE reaction & $(2.48, 18.6) \times 10^6 $ &  \cite{Beber2021} \\ 
$N_{\text{PduQ}}$ & number of PduQ per cell & $145 \pm 12$ & \cite{Yang2020} \\
$k_{\text{cat, PduQ}}^{\text{f}}$ &  maximum forward rate reaction of PduQ reaction &  $80 \pm 5 $ s$^{-1}$  & \cite{Cheng2012}  \\
$K^{\text{Propionaldehyde}}_{\text{M}, \text{PduQ}}$ & half-max concentration of propionaldehyde for PduP reaction & $(45.3  \pm 5.3) \times 10^{-3}$ mM  & \cite{Cheng2012}  \\
$K^{\text{NADH}}_{\text{M}, \text{PduQ}}$ & half-max concentration of NADH for PduQ reaction & $16 \pm 2$ mM   & \cite{Cheng2012} \\
$k_{\text{cat, PduQ}}^{\text{r}}$ & maximum forward rate reaction of PduQ reaction &  $8.75 \pm 1.25$ s$^{-1}$  & \cite{Cheng2012}  \\
$K^{\text{NAD+}}_{\text{M},\text{PduQ}}$ & half-max concentration of NAD+ for PduQ reaction & $(267.5  \pm 5.3 ) \times 10^{-3}$ mM   &  \cite{Cheng2012}  \\
$K^{ \text{1-Propanol}}_{\text{M},\text{PduQ}}$ & half-max concentration of 1-Propanol for PduQ reaction & $95.8 \pm 9.2$ mM  & \cite{Cheng2012}   \\
$K_{\text{eq, PduQ}}$ & Dissociation constants for the PduQ reaction & $(4.42, 111) \times 10^2$  & \cite{Beber2021}\\ 
$N_{\text{PduP}}$ & number of PduP per cell &  $241 \pm 31$ & \cite{Yang2020}  \\
$k_{\text{cat,PduP}}^{\text{f}}$ & maximum forward rate reaction of PduP reaction &  $32.8$ s$^{-1}$ &   \cite{SabetAzad2013} \\
$K^{\text{NAD+}}_{\text{M}, \text{PduP}}$ & half-max concentration of NAD+ for PduP reaction & $(1, 10^1) \times 10^{-1}$ mM  &  \cite{SabetAzad2013,Smith1980} \\
$K^{\text{CoA}}_{\text{M}, \text{PduP}}$ & half-max concentration of CoA for PduP reaction & $(1, 10^{1}) \times 10^{-2}$ mM   &  \cite{Smith1980} \\
$K_{\text{eq},\text{PduP}}$ & Dissociation constants for the PduP reaction & $1 - 2.83 \times 10^{3}$  &  \cite{Beber2021} \\ 
$N_{\text{PduL}}$ & number of PduP per cell & $33 \pm 3$  & \cite{Yang2020} \\
$k_{\text{cat,PduL}}^{\text{r}}$ & maximum reverse rate reaction of PduL reaction &  $20.7$ s$^{-1}$ & \cite{Liu2007}   \\
$K^{\substack{\text{Propionyl-}\\\text{Phosphate}}}_{\text{M}, \text{PduL}}$ & half-max concentration of NAD+ for PduL reaction & $(6.1 \pm 0.6)\times 10^{-1}$ mM &  \cite{Liu2007} \\
$K^{\text{CoA}}_{\text{M}, \text{PduL}}$ & half-max concentration of CoA for PduL reaction & $(3.2 \pm 6)\times 10^{-2}$ mM   &  \cite{Liu2007} \\
$K_{\text{eq},\text{PduL}}$ & Dissociation constants for the PduL reaction & $(1.29 , 339) \times 10^{-5}$  &  \cite{Beber2021} \\ 
$N_{\text{PduW}}$ & number of PduW per cell &  $11 \pm 2$   & \cite{Yang2020} \\
$K_{\text{eq, PduW}}$ & Dissociation constants for the PduW reaction & $(4.05, 579) \times 10^4$  &  \cite{Beber2021} \\ 
$k_{\text{cat, AckA}}^{\text{f}}$ & maximum forward rate reaction of AckA phosphorylation reaction &   $1000-1250$ s$^{-1}$ &  \cite{Chittori2012} \\
$K^{\substack{\text{ADP}}}_{\text{M, AckA}}$ & half-max concentration of propionate for AckA dephosphorylation reaction & $(0.5, 5.) \times 10^{-3}$  mM  & \cite{Chittori2012}   \\
$k_{\text{cat, AckA}}^{\text{r}}$ & maximum forward rate reaction of AckA dephosphorylation reaction &   $900 \pm 135$ s$^{-1}$ &  \cite{Chittori2012} \\
$K^{\substack{\text{Propionate}}}_{\text{M, AckA}}$ & half-max concentration of propionate for AckA dephosphorylation reaction & $11.2 \pm 0.1$ mM  & \cite{Chittori2012}   \\
$K^{\text{ATP}}_{\text{M, AckA}}$ & half-max concentration of ATP for AckA dephosphorylation reaction &  $(7.5 \times 10^{-2} \pm 4 \times 10^{-3})$ mM &  \cite{Chittori2012} \\
$K_{\text{eq, AckA}}$ & Dissociation constants for the AckA reaction & $(4.05, 579) \times 10^4$  &  \cite{Chittori2012} \\\hline 
 \caption{Table of Available Parameters \\
 \footnotesize{{}$^{\text{a}}$ The assay was performed with acetyl-phosphate. AckA reverse Michealis-Menten kinetics were similar $K_M$ and $k_\text{cat}$ regardless of the acid. We assume similar for the forward reaction--the kinetics of acetyl-phosphate and propionyl-phosphate are within an order of magnitude.   } }
 \label{Table:Chapter5AvailableParameters}
\end{longtable}
\end{center}
\doublespacing

\newpage

{
	\singlespacing
	\section*{Table S2 Derived Michaelis-Menten Formulae for all enzymes in the purified MCP assay.}
	\begin{longtable}
		{p{.1\textwidth}p{.25\textwidth} p{0.5\textwidth}} \hline
		
		\toprule
		Enzyme & Reaction Description & Michaelis-Menten Parameters  \\ 
		\midrule
		\endfirsthead
		\toprule
		Enzyme & Reaction Description & Michaelis-Menten Parameters  \\ 
		\midrule
		\endhead
		\midrule
		\multicolumn{3}{r}{Continued on next page} \\
		\midrule
		\endfoot
		\endlastfoot
		\bottomrule
		\multirow{2}{*}{PduCDE} &  Dehydration &    $k_{\text{cat,PduCDE}}^{\text{f}} = \frac{k_{3, \text{PduCDE}}k_{5, \text{PduCDE}}}{k_{3, \text{PduCDE}} + k_{5, \text{PduCDE}}}$
		
		$K^{\text{1,2-PD}}_{\text{M}, \text{PduCDE}} = \frac{(k_{2, \text{PduCDE}} + k_{3, \text{PduCDE}})k_{5, \text{PduCDE}}}{k_{1, \text{PduCDE}}(k_{3, \text{PduCDE}} + k_{5, \text{PduCDE}})}$
		\\\\
		& Hydration  & $
		k_{\text{cat, PduCDE}}^{\text{r}} = k_{2, \text{PduCDE}}$
		
		$K^{ \text{Propionaldehyde}}_{\text{M},\text{PduCDE}} = \frac{k_{2, \text{PduCDE}} + k_{3, \text{PduCDE}}}{k_{4, \text{PduCDE}}}$
		
		$K_{\text{M},\text{PduCDE}}^{\text{Water}} = \frac{k_{2,\text{PduCDE}}}{k_{6,\text{PduCDE}}}$\\\hline
		\multirow{2}{*}{PduQ}  & NADH Oxidation &   
		
		$k_{\text{cat, PduQ}}^{\text{f}} = \frac{k_{5,\text{PduQ}}k_{7,\text{PduQ}}}{k_{5,\text{PduQ}} + k_{5,\text{PduQ}}}$
		
		$K^{\text{NADH}}_{\text{M}, \text{PduQ}} = \frac{k_{5,\text{PduQ}}k_{7,\text{PduQ}}}{k_{1,\text{PduQ}}k_{5,\text{PduQ}} + k_{7,\text{PduQ}}k_{5,\text{PduQ}}}$
		
		$K^{\text{Propionaldehyde}}_{\text{M}, \text{PduQ}} = \frac{(k_{4,\text{PduQ}} + k_{5,\text{PduQ}})k_{7,\text{PduQ}}}{k_{3,\text{PduQ}}(k_{5,\text{PduQ}} + k_{7,\text{PduQ}})}$ \\\\
		& NAD+ Reduction &
		$k_{\text{cat, PduQ}}^{\text{r}} = \frac{k_{2,\text{PduQ}}k_{4,\text{PduQ}}}{k_{2,\text{PduQ}} + k_{4,\text{PduQ}}}$
		
		$K^{\text{NAD+}}_{\text{M},\text{PduQ}} = \frac{k_{2,\text{PduQ}}k_{4,\text{PduQ}}}{(k_{2,\text{PduQ}} + k_{4,\text{PduQ}})k_{8,\text{PduQ}}}$
		
		$K^{ \text{1-Propanol}}_{\text{M},\text{PduQ}} = \frac{k_{2,\text{PduQ}}(k_{4,\text{PduQ}} + k_{5,\text{PduQ}})}{(k_{2,\text{PduQ}} + k_{4,\text{PduQ}})k_{6, \text{PduQ}}}$ \\\hline
		\multirow{2}{*}{PduP}  & NADH Oxidation &   
		
		$k_{\text{cat,PduP}}^{\text{f}} = \frac{k_{5,\text{PduP}}k_{9,\text{PduP}}}{k_{5,\text{PduP}} + k_{9,\text{PduP}}}$ 
		
		$K^{\text{NAD+}}_{\text{M}, \text{PduP}} = \frac{k_{5,\text{PduP}}k_{9,\text{PduP}}}{k_{1,\text{PduP}}(k_{5,\text{PduP}} + k_{9,\text{PduP},\text{PduP}})}$ 
		
		$K^{\text{CoA}}_{\text{M}, \text{PduP}}= \frac{k_{5,\text{PduP}}(k_{8,\text{PduP}} + k_{9,\text{PduP}})}{k_{7,\text{PduP}}(k_{5,\text{PduP}} + k_{9,\text{PduP}})}$ 
		
		$K^{\text{Propionaldehyde}}_{\text{M}, \text{PduP}} = \frac{(k_{4,\text{PduP}} + k_{5,\text{PduP}})k_{9,\text{PduP}}}{k_{3,\text{PduP}}(k_{5,\text{PduP}} + k_{9,\text{PduP}})}$
		\\
		& NAD+ Reduction &
		$\begin{aligned}k_{\text{cat, PduP}}^{\text{r}} &= \frac{k_{2,\text{PduP}}k_{4,\text{PduP}}k_{8,\text{PduP}}}{\splitdfrac{\splitdfrac{k_{2,\text{PduP}}k_{4,\text{PduP}}}{ + k_{2,\text{PduP}}k_{8,\text{PduP}}}}{ + k_{4,\text{PduP}}k_{8,\text{PduP}}}}\\
			K^{\text{NADH}}_{\text{M}, \text{PduP}} &= \frac{k_{2,\text{PduP}}(k_{4,\text{PduP}}+k_{5,\text{PduP}})k_{8,\text{PduP}}}{\splitdfrac{\splitdfrac{k_{6,\text{PduP}}k_{4,\text{PduP}}k_{8,\text{PduP}}}{ + k_{6,\text{PduP}}k_{2,\text{PduP}}k_{4,\text{PduP}}}}{ + k_{6,\text{PduP}}k_{2,\text{PduP}}k_{8,\text{PduP}}}}\\
			K^{\text{Propionyl-CoA}}_{\text{M}, \text{PduP}} &= \frac{\splitdfrac{k_{2,\text{PduP}}k_{8,\text{PduP}}k_{4,\text{PduP}}}{+k_{2,\text{PduP}}k_{9,\text{PduP}}k_{4,\text{PduP}}}}{\splitdfrac{\splitdfrac{k_{10,\text{PduP}}k_{2,\text{PduP}}k_{4,\text{PduP}}}{ + k_{10,\text{PduP}}k_{2,\text{PduP}}k_{8,\text{PduP}}}}{ + k_{10,\text{PduP}}k_{4,\text{PduP}}k_{8,\text{PduP}}}}\end{aligned}$ \\\\\hline
		\multirow{2}{6em}{PduL} & Phosphorylation & 
		$k_{\text{cat, PduL}}^{\text{f}} = \frac{k_{5,\text{PduL}}k_{7,\text{PduL}}}{k_{5,\text{PduL}} + k_{5,\text{PduL}}}$
		
		$K^{\text{Phosphate}}_{\text{M}, \text{PduL}} = \frac{k_{5,\text{PduL}}k_{7,\text{PduL}}}{k_{1,\text{PduL}}k_{5,\text{PduL}} + k_{7,\text{PduL}}k_{5,\text{PduL}}}$
		
		$K^{\text{Propionyl-CoA}}_{\text{M}, \text{PduL}} = \frac{(k_{4,\text{PduL}} + k_{5,\text{PduL}})k_{7,\text{PduL}}}{k_{3,\text{PduL}}(k_{5,\text{PduL}} + k_{7,\text{PduL}})}$\\\\
		& Dephosphorylation &
		$k_{\text{cat, PduL}}^{\text{r}} = \frac{k_{2,\text{PduL}}k_{4,\text{PduL}}}{k_{2,\text{PduL}} + k_{4,\text{PduL}}}$
		
		$K^{\text{CoA}}_{\text{M}, \text{PduL}} = \frac{k_{2,\text{PduL}}k_{4,\text{PduL}}}{(k_{2,\text{PduL}} + k_{4,\text{PduL}})k_{8,\text{PduL}}}$
		
		$K^{\text{Propionyl Phosphate}}_{\text{M}, \text{PduL}} = \frac{k_{2,\text{PduL}}(k_{4,\text{PduL}} + k_{5,\text{PduL}})}{(k_{2,\text{PduL}} + k_{4,\text{PduL}})k_{6,\text{PduL}}}$     
		\\\hline
		
		\multirow{2}{*}{PduW} & ATP Formation & 
		$k_{\text{cat, PduW}}^{\text{f}} = \frac{k_{5,\text{PduW}}k_{7,\text{PduW}}}{k_{5,\text{PduW}} + k_{7,\text{PduW}}}$
		
		$K^{\text{ADP}}_{\text{M}, \text{PduW}} = \frac{k_{5,\text{PduW}}k_{7,\text{PduW}}}{k_{1,\text{PduW}}k_{5,\text{PduW}} + k_{7,\text{PduW}}k_{5,\text{PduW}}}$
		
		$K^{\text{Propionyl Phosphate}}_{\text{M}, \text{PduW}} = \frac{(k_{4,\text{PduW}} + k_{5,\text{PduW}})k_{7,\text{PduW}}}{k_{3,\text{PduW}}(k_{5,\text{PduW}} + k_{7,\text{PduW}})}$\\\\
		& ADP Formation &
		$k_{\text{cat, PduW}}^{\text{r}} = \frac{k_{2,\text{PduW}}k_{4,\text{PduW}}}{k_{2,\text{PduW}} + k_{4,\text{PduW}}}$
		
		$K^{\text{ATP}}_{\text{M}, \text{PduW}} = \frac{k_{2,\text{PduW}}k_{4,\text{PduW}}}{(k_{2,\text{PduW}} + k_{4,\text{PduW}})k_{8,\text{PduW}}}$
		
		$K^{\text{Propionate}}_{\text{M}, \text{PduW}} = \frac{k_{2,\text{PduW}}(k_{4,\text{PduW}} + k_{5,\text{PduW}})}{(k_{2,\text{PduW}} + k_{4,\text{PduW}})k_{6,\text{PduW}}}$ \\\hline
		
		\multirow{2}{*}{AckA} & ATP Formation & $
		k_{\text{cat, AckA}}^{\text{f}} = \frac{k_{5,\text{AckA}}k_{7,\text{AckA}}}{k_{5,\text{AckA}} + k_{7,\text{AckA}}}$
		
		$K^{\text{ADP}}_{\text{M}, \text{AckA}} = \frac{k_{5,\text{AckA}}k_{7,\text{AckA}}}{k_{1,\text{AckA}}k_{5,\text{AckA}} + k_{7,\text{AckA}}k_{5,\text{AckA}}}$
		
		$K^{\text{Propionyl Phosphate}}_{\text{M}, \text{AckA}} = \frac{(k_{4,\text{AckA}} + k_{5,\text{AckA}})k_{7,\text{AckA}}}{k_{3,\text{AckA}}(k_{5,\text{AckA}} + k_{7,\text{AckA}})} $\\\\
		
		& ADP Formation & $k_{\text{cat, AckA}}^{\text{r}} = \frac{k_{2,\text{AckA}}k_{4,\text{AckA}}}{k_{2,\text{AckA}} + k_{4,\text{AckA}}}$
		
		$K^{\text{ATP}}_{\text{M}, \text{AckA}} = \frac{k_{2,\text{AckA}}k_{4,\text{AckA}}}{(k_{2,\text{AckA}} + k_{4,\text{AckA}})k_{8,\text{AckA}}}$
		
		$K^{\text{Propionate}}_{\text{M}, \text{AckA}} = \frac{k_{2,\text{AckA}}(k_{4,\text{AckA}} + k_{5,\text{AckA}})}{(k_{2,\text{AckA}} + k_{4,\text{AckA}})k_{6,\text{AckA}}}$ \\\hline
		\label{Table:Chapter5MMFormulae}
\end{longtable}}

\newpage

\section*{Table S3. Decomposition of all enzyme kinetic parameters into free and leading parameters}
{ \singlespacing
	\begin{longtable}
		{p{0.1\textwidth} p{.1\textwidth} p{.165\textwidth} p{.5\textwidth}} 
		\toprule
		Enzyme & Kinetic Parameters & Free Parameters & Leading Parameters \\ 
		\midrule
		\endfirsthead
		\toprule
		Enzyme & Kinetic Parameters & Free Parameters & Leading Parameters \\ 
		\midrule
		\endhead
		\midrule
		\multicolumn{4}{r}{Continued on next page} \\
		\midrule
		\endfoot
		\endlastfoot
		\bottomrule
		
		PduCDE & $k_{1,\text{PduCDE}}$, $k_{2,\text{PduCDE}}$, $k_{3,\text{PduCDE}}$, $k_{4,\text{PduCDE}}$, $k_{5,\text{PduCDE}}$, $k_{6,\text{PduCDE}}$ &  $k_{2,\text{PduCDE}}$, $k_{3,\text{PduCDE}}$, $k_{4,\text{PduCDE}}$, $k^{\text{f}}_{\text{cat, PduCDE}}$, $
		K^{\text{PduCDE}}_{M,\text{1,2-PD}}$, $K_{\text{eq},\text{PduCDE}}$& 
		
		$k_{1, \text{PduCDE}} = \frac{k^{\text{f}}_{\text{cat, PduCDE}}(k_{2, \text{PduCDE}} + k_{3, \text{PduCDE}})}{k_{3}K^{\text{PduCDE}}_{\text{M, 1,2-PD}}}$ 
		
		$k_{5,\text{PduCDE}} = \frac{k_{3,\text{PduCDE}}k^f_{\text{cat,PduCDE}}}{k_{3,\text{PduCDE}} - k^f_{\text{cat,PduCDE}}}$
		
		$k_{6,\text{PduCDE}} = \frac{k_{1,\text{PduCDE}}k_{3,\text{PduCDE}}k_{5,\text{PduCDE}}}{k_{2,\text{PduCDE}}k_{4,\text{PduCDE}}K_{\text{eq, PduCDE}}}$ \\
		PduQ & $k_{1,\text{PduQ}}$, $k_{2,\text{PduQ}}$, $k_{3,\text{PduQ}}$, $k_{4,\text{PduQ}}$, $k_{5,\text{PduQ}}$, $k_{6,\text{PduQ}}$, $k_{7,\text{PduQ}}$, $k_{8,\text{PduQ}}$   & $k_{\text{cat, PduQ}}^{\text{f}}$, $K^{\text{NADH}}_{\text{M}, \text{PduQ}}$, $K^{\text{Propionaldehyde}}_{\text{M}, \text{PduQ}}$, $k_{\text{cat, PduQ}}^{\text{r}}$, $K^{\text{NAD+}}_{\text{M},\text{PduQ}}$, $K^{ \text{1-Propanol}}_{\text{M},\text{PduQ}}$, $K_{\text{eq},\text{PduQ}}$, $k_{5,\text{PduQ}}$ &
		
		{\scriptsize$\begin{aligned} k_{1, \text{PduQ}} &= \frac{k_{\text{cat, PduQ}}^{\text{f}}}{K^{\text{NADH}}_{\text{M}, \text{PduQ}}} \\
				k_{2, \text{PduQ}} &= \frac{k_{5,\text{PduQ}} (k_{\text{cat, PduQ}}^{\text{f}})^3 K^{\text{NAD+}}_{\text{M},\text{PduQ}} K^{\text{1-Propanol}}_{\text{M},\text{PduQ}}}{\splitdfrac{(k_{\text{cat, PduQ}}^{\text{r}})^2 K_{\text{eq},\text{PduQ}} K^{\text{NADH}}_{\text{M},\text{PduQ}}}{\times K^{\text{Propionaldehyde}}_{\text{M}, \text{PduQ}} (k_{5,\text{PduQ}}-k_{\text{cat, PduQ}}^{\text{f}})}}\\\\
				k_{3, \text{PduQ}} &=\frac{\splitdfrac{\splitdfrac{\splitdfrac{(k_{\text{cat, PduQ}}^{\text{f}})^2 (k_{\text{cat, PduQ}}^{\text{r}})^3 }{\times K_{\text{eq},\text{PduQ}} K^{\text{NADH}}_{\text{M},\text{PduQ}} K^{\text{Propionaldehyde}}_{\text{M}, \text{PduQ}}}}{\splitdfrac{-k_{5,\text{PduQ}}k_{\text{cat, PduQ}}^{\text{f}} (k_{\text{cat, PduQ}}^{\text{r}})^3 }{\times K_{\text{eq},\text{PduQ}}K^{\text{NADH}}_{\text{M},\text{PduQ}} K^{\text{Propionaldehyde}}_{\text{M}, \text{PduQ}}}}}{ \splitdfrac{+(k_{\text{cat, PduQ}}^{\text{f}})^4 K^{\text{NAD+}}_{\text{M},\text{PduQ}}}{\splitdfrac{\times K^{\text{1-Propanol}}_{\text{M},\text{PduQ}}}{\times (k_{5,\text{PduQ}} + k_{\text{cat, PduQ}}^{\text{r})}}}}}{\splitdfrac{\splitdfrac{\splitdfrac{k_{5,\text{PduQ}} (k_{\text{cat, PduQ}}^{\text{f}})^3 K^{\text{NAD+}}_{\text{M},\text{PduQ}}}{\times
								K^{\text{1-Propanol}}_{\text{M},\text{PduQ}}K^{\text{Propionaldehyde}}_{\text{M}, \text{PduQ}}}}{\splitdfrac{-k_{5,\text{PduQ}}(k_{\text{cat, PduQ}}^{\text{r}})^3 K_{\text{eq},\text{PduQ}}}{ \times K^{\text{NADH}}_{\text{M},\text{PduQ}} (K^{\text{Propionaldehyde}}_{\text{M}, \text{PduQ}})^2}}}{\splitdfrac{+k_{\text{cat, PduQ}}^{\text{f}} (k_{\text{cat, PduQ}}^{\text{r}})^3}{\times K_{\text{eq},\text{PduQ}} K^{\text{NADH}}_{\text{M},\text{PduQ}} (K^{\text{Propionaldehyde}}_{\text{M}, \text{PduQ}})^2}}}\\\\
				k_{4, \text{PduQ}} &= \frac{\splitdfrac{k_{5,\text{PduQ}} (k_{\text{cat, PduQ}}^{\text{f}})^3 k_{\text{cat, PduQ}}^{\text{r}}}{\times K^{\text{NAD+}}_{\text{M},\text{PduQ}} K^{\text{1-Propanol}}_{\text{M},\text{PduQ}}}}{\splitdfrac{\splitdfrac{\splitdfrac{k_{5,\text{PduQ}} (k_{\text{cat, PduQ}}^{\text{f}})^3 K^{\text{NAD+}}_{\text{M},\text{PduQ}}}{\times K^{\text{1-Propanol}}_{\text{M},\text{PduQ}}}}{\splitdfrac{-k_{5,\text{PduQ}} (k_{\text{cat, PduQ}}^{\text{r}})^3 K_{\text{eq},\text{PduQ}}}{\times K^{\text{NADH}}_{\text{M},\text{PduQ}}
								K^{\text{Propionaldehyde}}_{\text{M}, \text{PduQ}}}}}{\splitdfrac{+k_{\text{cat, PduQ}}^{\text{f}} (k_{\text{cat, PduQ}}^{\text{r}})^3 K_{\text{eq},\text{PduQ}}}{\times K^{\text{NADH}}_{\text{M},\text{PduQ}} K^{\text{Propionaldehyde}}_{\text{M}, \text{PduQ}}}}}
			\end{aligned}$}\\
		& & & 
		{\scriptsize$\begin{aligned}
				k_{6, \text{PduQ}} &=\frac{\splitdfrac{\splitdfrac{\splitdfrac{(k_{\text{cat, PduQ}}^{\text{f}})^2 (k_{\text{cat, PduQ}}^{\text{r}})^3 }{\times K_{\text{eq},\text{PduQ}} K^{\text{NADH}}_{\text{M},\text{PduQ}} K^{\text{Propionaldehyde}}_{\text{M}, \text{PduQ}}}}{\splitdfrac{-k_{5,\text{PduQ}}k_{\text{cat, PduQ}}^{\text{f}} (k_{\text{cat, PduQ}}^{\text{r}})^3 }{\times K_{\text{eq},\text{PduQ}}K^{\text{NADH}}_{\text{M},\text{PduQ}} K^{\text{Propionaldehyde}}_{\text{M}, \text{PduQ}}}}}{ \splitdfrac{+(k_{\text{cat, PduQ}}^{\text{f}})^4 K^{\text{NAD+}}_{\text{M},\text{PduQ}}}{\splitdfrac{\times K^{\text{1-Propanol}}_{\text{M},\text{PduQ}}}{\times (k_{5,\text{PduQ}} + k_{\text{cat, PduQ}}^{\text{r}})}}}}{(k_{\text{cat, PduQ}}^{\text{f}})^3 K^{\text{NAD+}}_{\text{M},\text{PduQ}} (K^{\text{1-Propanol}}_{\text{M},\text{PduQ}})^2}\\
				k_{7, \text{PduQ}} &= \frac{k_{5,\text{PduQ}} k_{\text{cat, PduQ}}^{\text{f}}}{k_{5,\text{PduQ}}-k_{\text{cat, PduQ}}^{\text{f}}}\\
				k_{8, \text{PduQ}} &= \frac{k_{\text{cat, PduQ}}^{\text{r}} }{K^{\text{NAD+}}_{\text{M},\text{PduQ}}}
			\end{aligned}$}\\\\
		
		PduP & $k_{1,\text{PduP}}$, $k_{2,\text{PduP}}$, $k_{3,\text{PduP}}$, $k_{4,\text{PduP}}$, $k_{6,\text{PduP}}$, $k_{7,\text{PduP}}$, $k_{8,\text{PduP}}$, $k_{9,\text{PduP}}$, $k_{10,\text{PduP}}$   &  $k_{2,\text{PduP}}$, $k_{3,\text{PduP}}$, $k_{4,\text{PduP}}$, $k_{6,\text{PduP}}$, $k_{7,\text{PduP}}$, $k_{8,\text{PduP}}$, $k_{\text{cat,PduP}}^{\text{f}}$, $K^{\text{NAD+}}_{\text{M}, \text{PduP}}$, $K^{\text{CoA}}_{\text{M}, \text{PduP}}$, $K_{\text{eq},\text{PduP}}$& 
		$k_{1,\text{PduP}} = \frac{k_{\text{cat,PduP}}^{\text{f}}}{K^{\text{NAD+}}_{\text{M}, \text{PduP}}}$
		
		$k_{5,\text{PduP}} = \frac{k_{8,\text{PduP}}k_{\text{cat,PduP}}^{\text{f}}}{k_{8,\text{PduP}}+ k_{\text{cat,PduP}}^{\text{f}} -k_{7,\text{PduP}}K^{\text{CoA}}_{\text{M}, \text{PduP}}}$
		
		$k_{9,\text{PduP}} \frac{k_{8,\text{PduP}}k_{\text{cat,PduP}}^{\text{f}}}{k_{7,\text{PduP}}K^{\text{CoA}}_{\text{M}, \text{PduP}} - k_{\text{cat,PduP}}^{\text{f}}}$
		
		$k_{10,\text{PduP}} = \frac{k_{1,\text{PduP}}k_{3,\text{PduP}}k_{5,\text{PduP}}k_{7,\text{PduP}}k_{9,\text{PduP}}}{k_{2,\text{PduP}}k_{4,\text{PduP}}k_{6,\text{PduP}}k_{8,\text{PduP}}K_{\text{eq, PduP}}}$\\\\
		
		PduL & $k_{1,\text{PduL}}$, $k_{2,\text{PduL}}$, $k_{3,\text{PduL}}$, $k_{4,\text{PduL}}$, $k_{5,\text{PduL}}$, $k_{6,\text{PduL}}$, $k_{7,\text{PduL}}$, $k_{8,\text{PduL}}$  & $k_{1,\text{PduL}}$, $k_{3,\text{PduL}}$, $k_{4,\text{PduL}}$, $k_{5,\text{PduL}}$, $k_{7,\text{PduL}}$, $K_{\text{eq},\text{PduL}}$, $k_{\text{cat, PduL}}^{\text{r}}$, $K^{\text{CoA}}_{\text{M}, \text{PduL}}$, $K^{\substack{\text{Propionyl-}\\\text{Phosphate}}}_{\text{M}, \text{PduL}}$ &
		$k_{2,\text{PduL}} = \frac{k_{4,\text{PduL}} k_{\text{cat, PduL}}^{\text{r}}}{k_{4,\text{PduL}}-k_{\text{cat, PduL}}^{\text{r}}}$
		
		$k_{6,\text{PduL}} = \frac{k_{\text{cat, PduL}}^{\text{r}} (k_{5,\text{PduL}}+k_{4,\text{PduL}})}{k_{4,\text{PduL}} K^{\substack{\text{Propionyl-Phosphate}}}_{\text{M}, \text{PduL}}}$
		
		$k_{8,\text{PduL}} = \frac{k_{\text{cat, PduL}}^{\text{r}}}{K^{\text{CoA}}_{\text{M}, \text{PduL}}}$
		
		$k_{7,\text{PduL}} = \frac{k_{2,\text{PduL}}k_{4,\text{PduL}}k_{8,\text{PduL}}k_{6,\text{PduL}}K_{\text{eq, PduL}}}{k_{1,\text{PduL}}k_{3,\text{PduL}}k_{5,\text{PduL}}} $ \\\\
		

		AckA & $k_{1,\text{AckA}}$, $k_{2,\text{AckA}}$, $k_{3,\text{AckA}}$, $k_{4,\text{AckA}}$, $k_{5,\text{AckA}}$, $k_{6,\text{AckA}}$, $k_{7,\text{AckA}}$, $k_{8,\text{AckA}}$   &  $k_{4,\text{AckA}}$, $k_{5,\text{AckA}}$,
		$k_{\text{cat, AckA}}^{\text{f}}$, $K^{\text{ADP}}_{\text{M}, \text{AckA}}$, $k_{\text{cat, AckA}}^{\text{r}}$, $K^{\text{ATP}}_{\text{M}, \text{AckA}}$, $K^{\text{Propionate}}_{\text{M}, \text{AckA}}$, $K_{\text{eq},\text{AckA}}$ & $k_{1,\text{AckA}} = \frac{k_{\text{cat, AckA}}^{\text{f}}}{K^{\text{ADP}}_{\text{M}, \text{AckA}}}$ 
		
		$k_{2,\text{AckA}} = \frac{k_{4,\text{AckA}} k_{\text{cat, AckA}}^{\text{r}}}{k_{4,\text{AckA}}-k_{\text{cat, AckA}}^{\text{r }}}$
		
		$k_{6,\text{AckA}} = \frac{k_{\text{cat, AckA}}^{\text{r}} (k_{5,\text{AckA}}+k_{4,\text{AckA}})}{k_{4,\text{AckA}} K^{\text{Propionate}}_{\text{M}, \text{AckA}}}$
		
		$k_{7,\text{AckA}} = \frac{k_{5,\text{AckA}} k_{\text{cat, AckA}}^{\text{f}}}{k_{5,\text{AckA}}-k_{\text{cat, AckA}}^{\text{f}}}$
		
		$k_{8,\text{AckA}} = \frac{k_{\text{cat, AckA}}^{\text{r}}}{K^{\text{ATP}}_{\text{M}, \text{AckA}}}$
		
		$k_{3,\text{AckA}} = \frac{k_{2,\text{AckA}}k_{4,\text{AckA}}k_{8,\text{AckA}}k_{6,\text{AckA}}K_{\text{eq, AckA}}}{k_{1,\text{AckA}}k_{5,\text{AckA}}k_{7,\text{AckA}}}$\\\hline
		\label{Table:Chapter5ConstrainedKinetics}
\end{longtable}}

\newpage

\section*{Table S4 Inequality expressions for free enzyme kinetic parameters.}
Column 2 (Inequalities) are constraints based on biological intuition, and column 3 (Free parameters) is how the constraint is incorporated into the leading variables. These constraints are then plugged into Table S3 to derive Table S5. Note that PduQ has multiple constraints, so the Keq and MM constraints cannot be solved independently. After solving simultaneously, you end up with two related inequalities to ensure positivity and the thermodynamic constraint.
{ \singlespacing
	\begin{longtable}
		{p{0.1\textwidth} p{.35\textwidth} p{.4\textwidth} } 
		\toprule
		Enzyme & Inequalities & Free Parameters 
		\\ 
		\midrule
		\endfirsthead
		\toprule
		Enzyme & Inequalities & Free Parameters 
		\\ 
		\midrule
		\endhead
		\midrule
		\multicolumn{2}{r}{Continued on next page} \\
		\midrule
		\endfoot
		\endlastfoot
		\bottomrule
		PduCDE & $\begin{aligned}k_{3,\text{PduCDE}} > k^{\text{f}}_{\text{cat,PduCDE}} \end{aligned}$ & 
		
		$\begin{aligned}\text{factor}_{\text{PduCDE}}&>1\text{\, where }\\
			k_{3,\text{PduCDE}} &= \text{factor}_{\text{PduCDE}} \\
			&\qquad \times k^{\text{f}}_{\text{cat,PduCDE}} \end{aligned}$
		\\\hline 
		PduQ  & {\footnotesize $\begin{aligned}k_{5,\text{PduQ}}&>k_{\text{cat, PduQ}}^{\text{f}}\\
				10^3 \leq K_{\text{eq}}^{\text{PduQ}}&<\frac{\text{factor}_{1,\text{PduQ}}}{\text{factor}_{1,\text{PduQ}} - 1} \\
				&\times \left(\frac{k_{\text{cat,PduQ}}^{\text{oxidation}}}{k_{\text{cat,PduQ}}^{\text{reduction}}}\right)^3 \\
				& \times \frac{K^{\text{NAD+}}_{\text{M},\text{PduQ}}K^{ \text{1-Propanol}}_{\text{M},\text{PduQ}}}{K^{\text{NADH}}_{\text{M}, \text{PduQ}}K^{\text{Propionaldehyde}}_{\text{M}, \text{PduQ}}}\\
				&\leq 10^6\\
				\text{where\, }k_{5,\text{PduQ}}&=\text{factor}_{1,\text{PduQ} }k_{\text{cat, PduQ}}^{\text{f}}
			\end{aligned}$ 
		}
		&
		$\begin{aligned}
			\text{factor}_{1,\text{PduQ}}  &> -\log_{10}\bigg(1\\
			&-10^{-6}\left(\frac{k_{\text{cat,PduQ}}^{\text{oxidation}}}{k_{\text{cat,PduQ}}^{\text{reduction}}}\right)^3\\
			&\times \frac{K^{\text{NAD+}}_{\text{M},\text{PduQ}}K^{ \text{1-Propanol}}_{\text{M},\text{PduQ}}}{K^{\text{NADH}}_{\text{M}, \text{PduQ}}K^{\text{Propionaldehyde}}_{\text{M}, \text{PduQ}}}\bigg)\end{aligned}
		$
		
		\\  \hline 
		
		PduP  &  $\begin{aligned}k_{7,\text{PduP}}  &> \frac{k_{\text{cat,PduP}}^{\text{f}}}{K^{\text{CoA}}_{\text{M}, \text{PduP}}}\\  
			k_{8,\text{PduP}} &> (\text{factor}_{\text{1,PduP}} - 1)\\
			&\times k_{\text{cat,PduP}}^{\text{f}}\\
			\text{where\,} k_{7,\text{PduP}} & = \text{factor}_{1,\text{PduP}}\frac{k_{\text{cat,PduP}}^{\text{f}}}{K^{\text{CoA}}_{\text{M}, \text{PduP}}}
		\end{aligned}$  
		& $\begin{aligned}\text{factor}_{1,\text{PduP}} &> 1\\
			\text{factor}_{2,\text{PduP}} &> 1\\
			\text{where\,} k_{8,\text{PduP}} &= \text{factor}_{2,\text{PduP}}\\
			&\times (\text{factor}_{1,\text{PduP}} - 1)\\
			&\times k_{\text{cat,PduP}}^{\text{f}}\end{aligned}$  
		\\\hline 
		PduL & $k_{4,\text{PduL}} > k_{\text{cat, PduL}}^{\text{r}}$ & $\begin{aligned}\text{factor}_{\text{PduL}} &> 1\\
			\text{where \,} k_{4,\text{PduL}} &= \text{factor}_{\text{PduL}}\\&\quad \times k_{\text{cat, PduL}}^{\text{r
			}}
		\end{aligned}$
		\\\hline 
		AckA & $\begin{aligned}k_{4,\text{AckA}} &> k_{\text{cat, AckA}}^{\text{r}}\\
			k_{5,\text{AckA}} &> k_{\text{cat, AckA}}^{\text{f}}\end{aligned}$  & 
		$\begin{aligned}\text{factor}_{1,\text{AckA}} &> 1\\
			\text{factor}_{2,\text{AckA}}&>1\\
			\text{where\,\,}k_{4,\text{AckA}} &= \text{factor}_{1,\text{AckA}} k_{\text{cat, AckA}}^{\text{r}}\\
			k_{5,\text{AckA}} &= \text{factor}_{2,\text{AckA}} k_{\text{cat, AckA}}^{\text{f}}\end{aligned}$ \\\hline
		\label{Table:Chapter5FreeVariableKinetics}
	\end{longtable}
}
\section*{Table S5. Combined leading equations and free inequalities for all enzymes.}
{\singlespacing
	\begin{longtable}
		{p{0.1\textwidth} p{.1\textwidth} p{.165\textwidth} p{.5\textwidth}}
		\toprule
		Enzyme & Kinetic Parameters & Free Variables & Leading Variables \\ 
		\midrule
		\endfirsthead
		\toprule
		Enzyme & Kinetic Parameters & Free Variables & Leading Variables \\ 
		\midrule
		\endhead
		\midrule
		\multicolumn{4}{r}{Continued on next page} \\
		\midrule
		\endfoot
		\endlastfoot
		\bottomrule
		
		PduCDE & $k_{1,\text{PduCDE}}$, $k_{2,\text{PduCDE}}$, $k_{3,\text{PduCDE}}$, $k_{4,\text{PduCDE}}$, $k_{5,\text{PduCDE}}$, $k_{6,\text{PduCDE}}$ &  $k_{2,\text{PduCDE}}$, $\text{factor}_{\text{PduCDE}}$, $k_{4,\text{PduCDE}}$, $k^{\text{f}}_{\text{PduCDE}}$, $
		K^{\text{PduCDE}}_{M,\text{1,2-PD}}$, $K_{\text{eq},\text{PduCDE}}$& 
		
		$k_{1, \text{PduCDE}} = \frac{k^{\text{f}}_{\text{cat, PduCDE}}(k_{2, \text{PduCDE}} + k_{3, \text{PduCDE}})}{k_{3}K^{\text{PduCDE}}_{\text{M, 1,2-PD}}}$ 
		
		$k_{5,\text{PduCDE}} = \frac{k_{3,\text{PduCDE}}k^f_{\text{cat,PduCDE}}}{k_{3,\text{PduCDE}} - k^f_{\text{cat,PduCDE}}}$
		
		$k_{6,\text{PduCDE}} = \frac{k_{1,\text{PduCDE}}k_{3,\text{PduCDE}}k_{5,\text{PduCDE}}}{k_{2,\text{PduCDE}}k_{4,\text{PduCDE}}K_{\text{eq, PduCDE}}}$ 
		
		$k_{3,\text{PduCDE}} = \text{factor}_{\text{PduCDE}} k^{\text{dehydration}}_{\text{cat,PduCDE}}$ 
		
		where $\text{factor}_{\text{PduCDE}}>1$
		\\\\
		PduQ & $k_{1,\text{PduQ}}$, $k_{2,\text{PduQ}}$, $k_{3,\text{PduQ}}$, $k_{4,\text{PduQ}}$, $k_{5,\text{PduQ}}$, $k_{6,\text{PduQ}}$, $k_{7,\text{PduQ}}$, $k_{8,\text{PduQ}}$   & $k_{\text{cat, PduQ}}^{\text{f}}$, $K^{\text{NADH}}_{\text{M}, \text{PduQ}}$, $K^{\text{Propionaldehyde}}_{\text{M}, \text{PduQ}}$, $k_{\text{cat, PduQ}}^{\text{r}}$, $K^{\text{NAD+}}_{\text{M},\text{PduQ}}$, $K^{ \text{1-Propanol}}_{\text{M},\text{PduQ}}$, $K_{\text{eq},\text{PduQ}}$, $\text{factor}_{1,\text{PduQ}}$ &
		
		{\scriptsize$\begin{aligned} k_{1, \text{PduQ}} &= \frac{k_{\text{cat, PduQ}}^{\text{f}}}{K^{\text{NADH}}_{\text{M}, \text{PduQ}}} \\
				k_{2, \text{PduQ}} &= \frac{k_{5,\text{PduQ}} (k_{\text{cat, PduQ}}^{\text{f}})^3 K^{\text{NAD+}}_{\text{M},\text{PduQ}} K^{\text{1-Propanol}}_{\text{M},\text{PduQ}}}{\splitdfrac{(k_{\text{cat, PduQ}}^{\text{r}})^2 K_{\text{eq},\text{PduQ}} K^{\text{NADH}}_{\text{M},\text{PduQ}}}{\times K^{\text{Propionaldehyde}}_{\text{M}, \text{PduQ}} (k_{5,\text{PduQ}}-k_{\text{cat, PduQ}}^{\text{f}})}}\\\\
				k_{3, \text{PduQ}} &=\frac{\splitdfrac{\splitdfrac{\splitdfrac{(k_{\text{cat, PduQ}}^{\text{f}})^2 (k_{\text{cat, PduQ}}^{\text{r}})^3 }{\times K_{\text{eq},\text{PduQ}} K^{\text{NADH}}_{\text{M},\text{PduQ}} K^{\text{Propionaldehyde}}_{\text{M}, \text{PduQ}}}}{\splitdfrac{-k_{5,\text{PduQ}}k_{\text{cat, PduQ}}^{\text{f}} (k_{\text{cat, PduQ}}^{\text{r}})^3 }{\times K_{\text{eq},\text{PduQ}}K^{\text{NADH}}_{\text{M},\text{PduQ}} K^{\text{Propionaldehyde}}_{\text{M}, \text{PduQ}}}}}{ \splitdfrac{+(k_{\text{cat, PduQ}}^{\text{f}})^4 K^{\text{NAD+}}_{\text{M},\text{PduQ}}}{\splitdfrac{\times K^{\text{1-Propanol}}_{\text{M},\text{PduQ}}}{\times (k_{5,\text{PduQ}} + k_{\text{cat, PduQ}}^{\text{r})}}}}}{\splitdfrac{\splitdfrac{\splitdfrac{k_{5,\text{PduQ}} (k_{\text{cat, PduQ}}^{\text{f}})^3 K^{\text{NAD+}}_{\text{M},\text{PduQ}}}{\times
								K^{\text{1-Propanol}}_{\text{M},\text{PduQ}}K^{\text{Propionaldehyde}}_{\text{M}, \text{PduQ}}}}{\splitdfrac{-k_{5,\text{PduQ}}(k_{\text{cat, PduQ}}^{\text{r}})^3 K_{\text{eq},\text{PduQ}}}{ \times K^{\text{NADH}}_{\text{M},\text{PduQ}} (K^{\text{Propionaldehyde}}_{\text{M}, \text{PduQ}})^2}}}{\splitdfrac{+k_{\text{cat, PduQ}}^{\text{f}} (k_{\text{cat, PduQ}}^{\text{r}})^3}{\times K_{\text{eq},\text{PduQ}} K^{\text{NADH}}_{\text{M},\text{PduQ}} (K^{\text{Propionaldehyde}}_{\text{M}, \text{PduQ}})^2}}}\\\\
				k_{4, \text{PduQ}} &= \frac{\splitdfrac{k_{5,\text{PduQ}} (k_{\text{cat, PduQ}}^{\text{f}})^3 k_{\text{cat, PduQ}}^{\text{r}}}{\times K^{\text{NAD+}}_{\text{M},\text{PduQ}} K^{\text{1-Propanol}}_{\text{M},\text{PduQ}}}}{\splitdfrac{\splitdfrac{\splitdfrac{k_{5,\text{PduQ}} (k_{\text{cat, PduQ}}^{\text{f}})^3 K^{\text{NAD+}}_{\text{M},\text{PduQ}}}{\times K^{\text{1-Propanol}}_{\text{M},\text{PduQ}}}}{\splitdfrac{-k_{5,\text{PduQ}} (k_{\text{cat, PduQ}}^{\text{r}})^3 K_{\text{eq},\text{PduQ}}}{\times K^{\text{NADH}}_{\text{M},\text{PduQ}}
								K^{\text{Propionaldehyde}}_{\text{M}, \text{PduQ}}}}}{\splitdfrac{+k_{\text{cat, PduQ}}^{\text{f}} (k_{\text{cat, PduQ}}^{\text{r}})^3 K_{\text{eq},\text{PduQ}}}{\times K^{\text{NADH}}_{\text{M},\text{PduQ}} K^{\text{Propionaldehyde}}_{\text{M}, \text{PduQ}}}}}
			\end{aligned}$}\\
		& & & 
		{\scriptsize$\begin{aligned}
				k_{6, \text{PduQ}} &=\frac{\splitdfrac{\splitdfrac{\splitdfrac{(k_{\text{cat, PduQ}}^{\text{f}})^2 (k_{\text{cat, PduQ}}^{\text{r}})^3K_{\text{eq},\text{PduQ}} }{\times  K^{\text{NADH}}_{\text{M},\text{PduQ}} K^{\text{Propionaldehyde}}_{\text{M}, \text{PduQ}}}}{\splitdfrac{-k_{5,\text{PduQ}}k_{\text{cat, PduQ}}^{\text{f}} (k_{\text{cat, PduQ}}^{\text{r}})^3 }{\times K_{\text{eq},\text{PduQ}}K^{\text{NADH}}_{\text{M},\text{PduQ}} K^{\text{Propionaldehyde}}_{\text{M}, \text{PduQ}}}}}{ \splitdfrac{+(k_{\text{cat, PduQ}}^{\text{f}})^4 K^{\text{NAD+}}_{\text{M},\text{PduQ}}}{\times K^{\text{1-Propanol}}_{\text{M},\text{PduQ}}(k_{5,\text{PduQ}} + k_{\text{cat, PduQ}}^{\text{r})}}}}{(k_{\text{cat, PduQ}}^{\text{f}})^3 K^{\text{NAD+}}_{\text{M},\text{PduQ}} (K^{\text{1-Propanol}}_{\text{M},\text{PduQ}})^2}\\
				k_{7, \text{PduQ}} &= \frac{k_{5,\text{PduQ}} k_{\text{cat, PduQ}}^{\text{f}}}{k_{5,\text{PduQ}}-k_{\text{cat, PduQ}}^{\text{f}}}\\
				k_{8, \text{PduQ}} &= \frac{k_{\text{cat, PduQ}}^{\text{r}} }{K^{\text{NAD+}}_{\text{M},\text{PduQ}}}\\
				k_{5,\text{PduQ}}&=\text{factor}_{1,\text{PduQ} }k_{\text{cat, PduQ}}^{\text{f}}\\
				\text{factor}_{1,\text{PduQ} }&>1\\
				10^3 \leq K_{\text{eq}}^{\text{PduQ}}&<\frac{\text{factor}_{1,\text{PduQ}}}{\text{factor}_{1,\text{PduQ}} - 1} \\
				&\times \left(\frac{k_{\text{cat,PduQ}}^{\text{oxidation}}}{k_{\text{cat,PduQ}}^{\text{reduction}}}\right)^3 \\
				& \times \frac{K^{\text{NAD+}}_{\text{M},\text{PduQ}}K^{ \text{1-Propanol}}_{\text{M},\text{PduQ}}}{K^{\text{NADH}}_{\text{M}, \text{PduQ}}K^{\text{Propionaldehyde}}_{\text{M}, \text{PduQ}}}\\
				&\leq 10^6
			\end{aligned}$}\\\\
		
		PduP & $k_{1,\text{PduP}}$, $k_{2,\text{PduP}}$, $k_{3,\text{PduP}}$, $k_{4,\text{PduP}}$, $k_{6,\text{PduP}}$, $k_{7,\text{PduP}}$, $k_{8,\text{PduP}}$, $k_{9,\text{PduP}}$, $k_{10,\text{PduP}}$   &  $k_{2,\text{PduP}}$, $k_{3,\text{PduP}}$, $k_{4,\text{PduP}}$, $k_{6,\text{PduP}}$, $\text{factor}_{1,\text{PduP}}$, $\text{factor}_{2,\text{PduP}}$, $k_{\text{cat,PduP}}^{\text{f}}$, $K^{\text{NAD+}}_{\text{M}, \text{PduP}}$, $K^{\text{CoA}}_{\text{M}, \text{PduP}}$, $K_{\text{eq},\text{PduP}}$& 
		$k_{1,\text{PduP}} = \frac{k_{\text{cat,PduP}}^{\text{f}}}{K^{\text{NAD+}}_{\text{M}, \text{PduP}}}$
		
		$k_{5,\text{PduP}} = \frac{k_{8,\text{PduP}}k_{\text{cat,PduP}}^{\text{f}}}{k_{8,\text{PduP}}+ k_{\text{cat,PduP}}^{\text{f}} -k_{7,\text{PduP}}K^{\text{CoA}}_{\text{M}, \text{PduP}}}$
		
		$k_{7,\text{PduP}} = \text{factor}_{1,\text{PduP}}\frac{k_{\text{cat,PduP}}^{\text{f}}}{K^{\text{CoA}}_{\text{M}, \text{PduP}}}$  
		
		$k_{8,\text{PduP}} = \text{factor}_{2,\text{PduP}} (\text{factor}_{\text{PduP}} - 1)k_{\text{cat,PduP}}^{\text{f}}$
		
		$k_{9,\text{PduP}} \frac{k_{8,\text{PduP}}k_{\text{cat,PduP}}^{\text{f}}}{k_{7,\text{PduP}}K^{\text{CoA}}_{\text{M}, \text{PduP}} - k_{\text{cat,PduP}}^{\text{f}}}$
		
		$k_{10,\text{PduP}} = \frac{k_{1,\text{PduP}}k_{3,\text{PduP}}k_{5,\text{PduP}}k_{7,\text{PduP}}k_{9,\text{PduP}}}{k_{2,\text{PduP}}k_{4,\text{PduP}}k_{6,\text{PduP}}k_{8,\text{PduP}}K_{\text{eq, PduP}}}$
		
		$\text{factor}_{1,\text{PduP}} > 1$ 
		
		$\text{factor}_{2,\text{PduP}} > 1$
		\\\\

		PduL & $k_{1,\text{PduL}}$, $k_{2,\text{PduL}}$, $k_{3,\text{PduL}}$, $k_{4,\text{PduL}}$, $k_{5,\text{PduL}}$, $k_{6,\text{PduL}}$, $k_{7,\text{PduL}}$, $k_{8,\text{PduL}}$  & $k_{1,\text{PduL}}$, $k_{3,\text{PduL}}$, $\text{factor}_{\text{PduL}}$, $k_{5,\text{PduL}}$, $k_{7,\text{PduL}}$, $K_{\text{eq},\text{PduL}}$, $k_{\text{cat, PduL}}^{\text{r}}$, $K^{\text{CoA}}_{\text{M}, \text{PduL}}$, $K^{\substack{\text{Propionyl-}\\\text{Phosphate}}}_{\text{M}, \text{PduL}}$ &
		$k_{2,\text{PduL}} = \frac{k_{4,\text{PduL}} k_{\text{cat, PduL}}^{\text{r}}}{k_{4,\text{PduL}}-k_{\text{cat, PduL}}^{\text{r}}}$
		
		$k_{4,\text{PduL}} = \text{factor}_{\text{PduL}} k_{\text{cat, PduL}}^{\text{r}}$ 
		
		$k_{6,\text{PduL}} = \frac{k_{\text{cat, PduL}}^{\text{r}} (k_{5,\text{PduL}}+k_{4,\text{PduL}})}{k_{4,\text{PduL}} K^{\substack{\text{Propionyl-Phosphate}}}_{\text{M}, \text{PduL}}}$
		
		$k_{8,\text{PduL}} = \frac{k_{\text{cat, PduL}}^{\text{r}}}{K^{\text{CoA}}_{\text{M}, \text{PduL}}}$
		
		$k_{7,\text{PduL}} = \frac{k_{2,\text{PduL}}k_{4,\text{PduL}}k_{8,\text{PduL}}k_{6,\text{PduL}}K_{\text{eq, PduL}}}{k_{1,\text{PduL}}k_{3,\text{PduL}}k_{5,\text{PduL}}} $
		$\text{factor}_{\text{PduL}}  > 1$
		\\\\
		
		PduW & $k_{1,\text{PduW}}$, $k_{2,\text{PduW}}$, $k_{3,\text{PduW}}$, $k_{4,\text{PduW}}$, $k_{5,\text{PduW}}$, $k_{6,\text{PduW}}$, $k_{7,\text{PduW}}$, $k_{8,\text{PduW}}$ & $k_{1,\text{PduW}}$, $k_{2,\text{PduW}}$, $k_{3,\text{PduW}}$, $k_{4,\text{PduW}}$, $k_{5,\text{PduW}}$, $k_{6,\text{PduW}}$, $k_{7,\text{PduW}}$, $K_{\text{eq, PduW}}$   &  
		$k_{8,\text{PduW}} = \frac{k_{1,\text{PduW}}k_{3,\text{PduW}}k_{5,\text{PduW}}k_{7,\text{PduW}}}{k_{2,\text{PduW}}k_{4,\text{PduW}k_{6,\text{PduW}}K_{\text{eq, PduW}}}}$
		
		\\\\

		AckA & $k_{1,\text{AckA}}$, $k_{2,\text{AckA}}$, $k_{3,\text{AckA}}$, $k_{4,\text{AckA}}$, $k_{5,\text{AckA}}$, $k_{6,\text{AckA}}$, $k_{7,\text{AckA}}$, $k_{8,\text{AckA}}$   &  $\text{factor}_{1,\text{AckA}}$, $\text{factor}_{2,\text{AckA}}$,
		$k_{\text{cat, AckA}}^{\text{f}}$, $K^{\text{ADP}}_{\text{M}, \text{AckA}}$, $k_{\text{cat, AckA}}^{\text{r}}$, $K^{\text{ATP}}_{\text{M}, \text{AckA}}$, $K^{\text{Propionate}}_{\text{M}, \text{AckA}}$, $K_{\text{eq},\text{AckA}}$ & $k_{1,\text{AckA}} = \frac{k_{\text{cat, AckA}}^{\text{f}}}{K^{\text{ADP}}_{\text{M}, \text{AckA}}}$ 
		
		$k_{2,\text{AckA}} = \frac{k_{4,\text{AckA}} k_{\text{cat, AckA}}^{\text{r}}}{k_{4,\text{AckA}}-k_{\text{cat, AckA}}^{\text{r}}}$
		
		$k_{4,\text{AckA}} = \text{factor}_{1,\text{AckA}} k_{\text{cat, AckA}}^{\text{r}}$  
		
		$k_{5,\text{AckA}} = \text{factor}_{2,\text{AckA}} k_{\text{cat, AckA}}^{\text{f}}$ 
		
		$k_{6,\text{AckA}} = \frac{k_{\text{cat, AckA}}^{\text{r}} (k_{5,\text{AckA}}+k_{4,\text{AckA}})}{k_{4,\text{AckA}} K^{\text{Propionate}}_{\text{M}, \text{AckA}}}$
		
		$k_{7,\text{AckA}} = \frac{k_{5,\text{AckA}} k_{\text{cat, AckA}}^{\text{f}}}{k_{5,\text{AckA}}-k_{\text{cat, AckA}}^{\text{f}}}$
		
		$k_{8,\text{AckA}} = \frac{k_{\text{cat, AckA}}^{\text{r}}}{K^{\text{ATP}}_{\text{M}, \text{AckA}}}$
		
		$k_{3,\text{AckA}} = \frac{k_{2,\text{AckA}}k_{4,\text{AckA}}k_{8,\text{AckA}}k_{6,\text{AckA}}K_{\text{eq, AckA}}}{k_{1,\text{AckA}}k_{5,\text{AckA}}k_{7,\text{AckA}}}$
		
		$\text{factor}_{1,\text{AckA}} > 1$
		
		$\text{factor}_{2, \text{AckA}} > 1$
		
		\\\hline
		\label{Table:Chapter5ConstrainedKineticswithFactors}
\end{longtable}}

\section*{SI Table 6. Posterior MCMC results with sampling restricted to mode 1 and fixed PduP inhibition rates.}

We set k\textsubscript{1,iPduP} = 10\textsuperscript{-3.58} /mM s and k\textsubscript{2,iPduP} = 10\textsuperscript{-7.59} /s. CoA permeability was restricted to [10\textsuperscript{-7.5}, 10\textsuperscript{-3}] m/s. 4 chains were parameterized with 3000 samples, 3000 tuning iterations, four chains, 0.85 acceptance rate, and max tree depth of 7. ESS $>$ 400 and $\hat{R}<$1.05, indicating well-behaved sampling. The {\it in vitro} differential equation was integrated with an absolute and relative tolerance of 10\textsuperscript{-9}. The {\it in vitro} adjoint equation was integrated with an absolute and relative tolerance of 10\textsuperscript{-2}.

\singlespacing\renewcommand*{\arraystretch}{2}
\begin{longtable}{>{\raggedright\arraybackslash}p{2.25cm}p{0.9cm}p{0.85cm}p{1.cm}p{1cm}p{1.cm}p{1.cm}p{1.35cm}p{1.35cm}p{0.75cm}}
	\toprule
	& Mean & SD & HDI 2.5\% & HDI 97.5\% & MCSE Mean & MCSE SD & ESS Bulk & ESS Tail & $\hat{R}$ \\
	\midrule
	\endfirsthead
	\toprule
	& Mean & SD & HDI 2.5\% & HDI 97.5\% & MCSE Mean & MCSE SD & ESS Bulk & ESS Tail & $\hat{R}$ \\
	\midrule
	\endhead
	\midrule
	\multicolumn{10}{r}{Continued on next page} \\
	\midrule
	\endfoot
	\bottomrule
	\endlastfoot
	$P_{\text{MCP, 1,2-PD}}$ & -4.57 & 0.66 & -5.71 & -3.32 & 0.01 & 0.01 & 3581.00 & 2027.00 & 1.00 \\
	$P_{\text{MCP, }\substack{\text{Propion-}\\\text{aldehyde}}}$ & -6.21 & 0.91 & -7.72 & -4.30 & 0.02 & 0.01 & 3041.00 & 1994.00 & 1.00 \\
	$P_{\text{MCP, 1-Propanol}}$ & -5.50 & 1.11 & -7.49 & -3.42 & 0.01 & 0.01 & 8085.00 & 4876.00 & 1.00 \\
	$P_{\text{MCP, }\substack{\text{Propionyl-}\\\text{Phosphate}}}$ & -5.25 & 0.97 & -6.94 & -3.38 & 0.01 & 0.01 & 6305.00 & 3377.00 & 1.00 \\
	$P_{\text{MCP, Propionate}}$ & -5.50 & 1.11 & -7.60 & -3.48 & 0.01 & 0.01 & 8941.00 & 5468.00 & 1.00 \\
	$P_{\text{MCP, NADH}}$ & -9.60 & 0.34 & -10.20 & -8.96 & 0.00 & 0.00 & 5694.00 & 4764.00 & 1.00 \\
	$P_{\text{MCP, NAD+}}$ & -9.60 & 0.34 & -10.20 & -8.96 & 0.00 & 0.00 & 5694.00 & 4764.00 & 1.00 \\
	$P_{\text{MCP, CoA}}$ & -5.48 & 0.98 & -6.97 & -3.58 & 0.01 & 0.01 & 5668.00 & 3733.00 & 1.00 \\
	$k_{\text{cat,PduCDE}}^{\text{dehydration}}$ & 373.36 & 18.45 & 341.00 & 399.98 & 0.23 & 0.16 & 5772.00 & 4611.00 & 1.00 \\
	$K^{\text{1,2-PD}}_{\text{M}, \text{PduCDE}}$ & -0.80 & 0.30 & -1.30 & -0.31 & 0.00 & 0.00 & 9285.00 & 6626.00 & 1.00 \\
	$k_{2,\text{PduCDE}}$ & 2.65 & 0.55 & 1.67 & 3.66 & 0.01 & 0.01 & 4464.00 & 4589.00 & 1.00 \\
	$\text{factor}_{\text{PduCDE}}$ & 0.69 & 0.72 & 0.00 & 2.05 & 0.01 & 0.01 & 2731.00 & 3922.00 & 1.00 \\
	$k_{4,\text{PduCDE}}$ & 0.85 & 0.55 & -0.12 & 1.86 & 0.01 & 0.01 & 4446.00 & 3423.00 & 1.00 \\
	$K_{\text{eq, PduCDE}}$ & 7.27 & 0.03 & 7.22 & 7.30 & 0.00 & 0.00 & 8155.00 & 5247.00 & 1.00 \\
	$k_{\mathrm{cat,PduQ}}^{\mathrm{f}}$ & 54.82 & 2.77 & 50.19 & 59.10 & 0.03 & 0.02 & 7348.00 & 4999.00 & 1.00 \\
	$K_{\mathrm{M,PduQ}}^{\mathrm{NADH}}$ & 0.04 & 0.00 & 0.04 & 0.05 & 0.00 & 0.00 & 7747.00 & 6498.00 & 1.00 \\
	$K_{\mathrm{M,PduQ}}^{\mathrm{Propionaldehyde}}$ & 15.89 & 1.17 & 14.00 & 17.74 & 0.01 & 0.01 & 7700.00 & 6100.00 & 1.00 \\
	$k_{\mathrm{cat,PduQ}}^{\mathrm{r}}$ & 5.83 & 0.58 & 4.85 & 6.74 & 0.01 & 0.00 & 7710.00 & 5753.00 & 1.00 \\
	$K_{\mathrm{M,PduQ}}^{\mathrm{NAD}}$ & 0.27 & 0.01 & 0.25 & 0.29 & 0.00 & 0.00 & 10431.00 & 6627.00 & 1.00 \\
	$K_{\mathrm{M,PduQ}}^{\mathrm{Propanol}}$ & 95.64 & 5.30 & 86.61 & 103.79 & 0.05 & 0.04 & 9323.00 & 6706.00 & 1.00 \\
	$\text{factor}_{\text{PduQ}}$ & 3.42 & 2.03 & 0.02 & 6.54 & 0.02 & 0.01 & 9936.00 & 6382.00 & 1.00 \\
	$K_{\text{eq, PduQ}}$ & 3.91 & 0.47 & 3.01 & 4.69 & 0.01 & 0.01 & 4002.00 & 3306.00 & 1.00 \\
	$k_{1,\text{Evap}}$ & -4.37 & 0.03 & -4.42 & -4.32 & 0.00 & 0.00 & 10160.00 & 7998.00 & 1.00 \\
	$k_{1,\text{iPduQ}}$ & 0.94 & 0.17 & 0.58 & 1.21 & 0.00 & 0.00 & 3679.00 & 2152.00 & 1.00 \\
	$k_{\mathrm{cat,PduP}}^{\mathrm{f}}$ & 31.28 & 0.58 & 30.35 & 32.20 & 0.01 & 0.00 & 8160.00 & 5767.00 & 1.00 \\
	$K_{\mathrm{M,PduP}}^{\mathrm{NAD}}$ & 0.89 & 0.06 & 0.80 & 0.98 & 0.00 & 0.00 & 7604.00 & 5932.00 & 1.00 \\
	$K_{\mathrm{M,PduP}}^{\mathrm{CoA}}$ & 0.04 & 0.01 & 0.03 & 0.04 & 0.00 & 0.00 & 7599.00 & 5596.00 & 1.00 \\
	$k_{2,\text{PduP}}$ & 1.04 & 0.77 & -0.41 & 2.49 & 0.01 & 0.01 & 5086.00 & 6583.00 & 1.00 \\
	$k_{3,\text{PduP}}$ & 1.66 & 0.66 & 0.50 & 2.89 & 0.01 & 0.01 & 4852.00 & 4430.00 & 1.00 \\
	$k_{4,\text{PduP}}$ & 1.00 & 0.77 & -0.47 & 2.43 & 0.01 & 0.01 & 6394.00 & 7024.00 & 1.00 \\
	$k_{6,\text{PduP}}$ & 1.13 & 0.73 & -0.26 & 2.42 & 0.01 & 0.01 & 6505.00 & 6995.00 & 1.00 \\
	$\text{factor}_{1,\text{PduP}}$ & 3.52 & 2.01 & 0.29 & 6.83 & 0.02 & 0.01 & 8682.00 & 5958.00 & 1.00 \\
	$\text{factor}_{2,\text{PduP}}$ & 3.90 & 1.80 & 1.05 & 6.95 & 0.02 & 0.02 & 6783.00 & 4515.00 & 1.00 \\
	$K_{\text{eq, PduP}}$ & 2.80 & 0.57 & 1.74 & 3.45 & 0.01 & 0.01 & 6704.00 & 5243.00 & 1.00 \\
	$k_{\mathrm{cat,PduL}}^{\mathrm{r}}$ & 26.94 & 3.46 & 20.65 & 31.78 & 0.04 & 0.03 & 6931.00 & 5888.00 & 1.00 \\
	$K_{\mathrm{M,PduL}}^{\substack{\text{Propionyl-}\\\text{Phosphate}}}$ & 0.60 & 0.04 & 0.55 & 0.66 & 0.00 & 0.00 & 10251.00 & 6095.00 & 1.00 \\
	$\text{factor}_{\text{2,PduL}}$ & 4.25 & 1.62 & 1.71 & 7.00 & 0.02 & 0.01 & 6548.00 & 4311.00 & 1.00 \\
	$k_{1,\text{PduL}}$ & 1.30 & 0.54 & 0.39 & 2.33 & 0.01 & 0.01 & 7965.00 & 5335.00 & 1.00 \\
	$k_{3,\text{PduL}}$ & 1.80 & 0.37 & 1.21 & 2.51 & 0.01 & 0.00 & 4295.00 & 3843.00 & 1.00 \\
	$\text{factor}_{1, \text{PduL}}$ & 0.57 & 0.56 & 0.00 & 1.62 & 0.01 & 0.01 & 4152.00 & 4747.00 & 1.00 \\
	$k_{5,\text{PduL}}$ & 2.20 & 0.49 & 1.40 & 3.11 & 0.01 & 0.01 & 5028.00 & 4358.00 & 1.00 \\
	$K_{\text{eq, PduL}}$ & -2.61 & 0.13 & -2.85 & -2.47 & 0.00 & 0.00 & 5573.00 & 3773.00 & 1.00 \\
	$k_{1,\text{PduW}}$ & 4.01 & 1.01 & 2.11 & 5.85 & 0.01 & 0.01 & 6062.00 & 3901.00 & 1.00 \\
	$k_{2,\text{PduW}}$ & 2.00 & 1.01 & 0.09 & 3.90 & 0.01 & 0.01 & 10500.00 & 6517.00 & 1.00 \\
	$k_{3,\text{PduW}}$ & 4.01 & 0.98 & 2.14 & 5.83 & 0.01 & 0.01 & 4770.00 & 2592.00 & 1.00 \\
	$k_{4,\text{PduW}}$ & 2.02 & 0.99 & 0.19 & 3.94 & 0.01 & 0.01 & 9189.00 & 6742.00 & 1.00 \\
	$k_{5,\text{PduW}}$ & 3.99 & 0.98 & 2.16 & 5.85 & 0.01 & 0.01 & 9168.00 & 4431.00 & 1.00 \\
	$k_{6,\text{PduW}}$ & 1.99 & 1.00 & 0.13 & 3.84 & 0.01 & 0.01 & 9202.00 & 6082.00 & 1.00 \\
	$k_{7,\text{PduW}}$ & 4.00 & 0.99 & 2.19 & 5.89 & 0.01 & 0.01 & 7092.00 & 3713.00 & 1.00 \\
	$K_{\text{eq, PduW}}$ & 5.80 & 0.56 & 4.90 & 6.70 & 0.01 & 0.00 & 8099.00 & 6821.00 & 1.00 \\
	$k_{\mathrm{cat,AckA}}^{\mathrm{f}}$ & 1125.42 & 71.82 & 1015.26 & 1249.55 & 0.75 & 0.53 & 7882.00 & 6297.00 & 1.00 \\
	$K_{\mathrm{M,AckA}}^{\mathrm{ADP}}$ & 0.00 & 0.00 & 0.00 & 0.01 & 0.00 & 0.00 & 9285.00 & 6539.00 & 1.00 \\
	$k_{\mathrm{cat,AckA}}^{\mathrm{r}}$ & 899.61 & 77.73 & 777.90 & 1030.41 & 0.83 & 0.59 & 7620.00 & 6438.00 & 1.00 \\
	$K_{\mathrm{M,AckA}}^{\mathrm{Propionate}}$ & 11.20 & 0.06 & 11.10 & 11.29 & 0.00 & 0.00 & 9514.00 & 6975.00 & 1.00 \\
	$K_{\mathrm{M,AckA}}^{\mathrm{ATP}}$ & 0.07 & 0.00 & 0.07 & 0.08 & 0.00 & 0.00 & 8458.00 & 6188.00 & 1.00 \\
	$\text{factor}_{1, \text{AckA}}$ & 3.51 & 2.02 & 0.40 & 6.96 & 0.02 & 0.01 & 9485.00 & 6405.00 & 1.00 \\
	$\text{factor}_{2, \text{AckA}}$ & 3.48 & 2.04 & 0.01 & 6.59 & 0.02 & 0.01 & 8257.00 & 5860.00 & 1.00 \\
	$K_{\text{eq, AckA}}$ & 5.79 & 0.56 & 4.95 & 6.76 & 0.01 & 0.00 & 8134.00 & 6245.00 & 1.00 \\
	Number of MCP PduCDE in WT & 583.14 & 28.10 & 533.21 & 621.97 & 0.35 & 0.25 & 5202.00 & 4500.00 & 1.00 \\
	Number of MCP PduQ in WT & 146.37 & 6.78 & 135.03 & 156.99 & 0.07 & 0.05 & 8582.00 & 6055.00 & 1.00 \\
	Number of MCP PduP in WT & 227.33 & 13.83 & 201.42 & 245.00 & 0.14 & 0.10 & 8773.00 & 5668.00 & 1.00 \\
	Number of MCP PduL in WT & 33.47 & 1.70 & 30.56 & 36.00 & 0.02 & 0.01 & 6300.00 & 5262.00 & 1.00 \\
	Number of MCP PduW in WT & 11.01 & 1.15 & 9.22 & 12.97 & 0.01 & 0.01 & 7872.00 & 6067.00 & 1.00 \\
	External AckA concentration in in-vitro assay & -2.41 & 1.17 & -4.63 & -0.32 & 0.01 & 0.01 & 6166.00 & 3911.00 & 1.00 \\
	Initial 1,2-PD in-vitro & 0.19 & 0.17 & 0.00 & 0.50 & 0.00 & 0.00 & 7986.00 & 4813.00 & 1.00 \\
	Initial Propionaldehyde in-vitro & 0.41 & 0.27 & 0.00 & 0.88 & 0.00 & 0.00 & 9469.00 & 6227.00 & 1.00 \\
	Initial Propanol in-vitro & 0.15 & 0.14 & 0.00 & 0.40 & 0.00 & 0.00 & 8136.00 & 5049.00 & 1.00 \\
	Initial Propionate in-vitro & 0.90 & 0.10 & 0.72 & 1.00 & 0.00 & 0.00 & 8297.00 & 5137.00 & 1.00 \\
\end{longtable}
\doublespacing
\newpage
\section*{SI Table 7. Table of BFMI for each mode 1 chain}
BFMIs are all $>0.2$, indicating well behaved sampling

\hspace{1cm}
\singlespacing
\begin{table}[H]
	\centerline{
		\begin{tabular}{p{3em}p{3em}} 
			\hline
			Chain & BFMI \\ [0.5ex] 
			\hline\hline
			0 & 0.92\\
			1 & 0.86 \\
			2 & 0.85\\
			3 & 0.84\\\hline
	\end{tabular}}
\end{table}
\doublespacing
\section*{SI Table 8. Posterior MCMC results with sampling restricted to mode 2 and fixed PduP inhibition kinetics}
We set k\textsubscript{1,iPduP} = 10\textsuperscript{-0.36} /mM s  and k\textsubscript{2,iPduP} = 10\textsuperscript{-3.76} /s. CoA permeability was restricted to [10\textsuperscript{-11}, 10\textsuperscript{-8}] m/s. 4 chains were parameterized with 3000 samples, 3000 tuning iterations, 0.85 acceptance rate, and max tree depth of 7. ESS $>$ 400 and $\hat{R}<$1.05, indicating well-behaved sampling. The {\it in vitro} differential equation was integrated with an absolute and relative tolerance of 10\textsuperscript{-9}. The {\it in vitro} adjoint equation was integrated with an absolute and relative tolerance of 10\textsuperscript{-2}.
\singlespacing\renewcommand*{\arraystretch}{2}
\begin{longtable}{>{\raggedright\arraybackslash}p{2.25cm}p{0.9cm}p{0.85cm}p{1.cm}p{1cm}p{1.cm}p{1.cm}p{1.35cm}p{1.35cm}p{0.75cm}}
	\toprule
	& Mean & SD & HDI 2.5\% & HDI 97.5\% & MCSE Mean & MCSE SD & ESS Bulk & ESS Tail & $\hat{R}$ \\
	\midrule
	\endhead
	\midrule
	\multicolumn{10}{r}{Continued on next page} \\
	\midrule
	\endfoot
	\bottomrule
	\endlastfoot
	$P_{\text{MCP, 1,2-PD}}$ & -4.57 & 0.66 & -5.71 & -3.35 & 0.01 & 0.01 & 3360.00 & 1720.00 & 1.00 \\
	$P_{\text{MCP, }\substack{\text{Propion-}\\\text{aldehyde}}}$ & -5.20 & 0.90 & -6.66 & -3.35 & 0.01 & 0.01 & 3706.00 & 2260.00 & 1.00 \\
	$P_{\text{MCP, 1-Propanol}}$ & -5.50 & 1.08 & -7.44 & -3.45 & 0.01 & 0.01 & 4793.00 & 4211.00 & 1.00 \\
	$P_{\text{MCP, }\substack{\text{Propionyl-}\\\text{Phosphate}}}$ & -5.49 & 1.06 & -7.43 & -3.52 & 0.02 & 0.01 & 2663.00 & 2697.00 & 1.00 \\
	$P_{\text{MCP, Propionate}}$ & -5.51 & 1.10 & -7.51 & -3.49 & 0.01 & 0.01 & 4744.00 & 4102.00 & 1.00 \\
	$P_{\text{MCP, NADH}}$ & -5.63 & 1.04 & -7.14 & -3.70 & 0.02 & 0.01 & 2504.00 & 2876.00 & 1.00 \\
	$P_{\text{MCP, NAD+}}$ & -5.63 & 1.04 & -7.14 & -3.70 & 0.02 & 0.01 & 2504.00 & 2876.00 & 1.00 \\
	$P_{\text{MCP, CoA}}$ & -10.09 & 0.45 & -10.94 & -9.34 & 0.01 & 0.01 & 3908.00 & 2985.00 & 1.00 \\
	$k_{\text{cat,PduCDE}}^{\text{dehydration}}$ & 369.90 & 20.10 & 335.11 & 400.00 & 0.31 & 0.22 & 3821.00 & 3685.00 & 1.00 \\
	$K^{\text{1,2-PD}}_{\text{M}, \text{PduCDE}}$ & -0.80 & 0.30 & -1.30 & -0.31 & 0.00 & 0.00 & 4898.00 & 4496.00 & 1.00 \\
	$k_{2,\text{PduCDE}}$ & 2.54 & 0.49 & 1.66 & 3.47 & 0.01 & 0.01 & 2746.00 & 3288.00 & 1.00 \\
	$\text{factor}_{\text{PduCDE}}$ & 0.66 & 0.64 & 0.00 & 1.84 & 0.01 & 0.01 & 2013.00 & 3249.00 & 1.00 \\
	$k_{4,\text{PduCDE}}$ & 0.93 & 0.51 & 0.01 & 1.92 & 0.01 & 0.01 & 3247.00 & 3589.00 & 1.00 \\
	$K_{\text{eq, PduCDE}}$ & 7.27 & 0.03 & 7.23 & 7.30 & 0.00 & 0.00 & 4545.00 & 3588.00 & 1.00 \\
	$k_{\mathrm{cat,PduQ}}^{\mathrm{f}}$ & 53.63 & 2.84 & 49.20 & 58.31 & 0.04 & 0.03 & 4340.00 & 4363.00 & 1.00 \\
	$K_{\mathrm{M,PduQ}}^{\mathrm{NADH}}$ & 0.04 & 0.00 & 0.04 & 0.05 & 0.00 & 0.00 & 4208.00 & 4058.00 & 1.00 \\
	$K_{\mathrm{M,PduQ}}^{\mathrm{Propionaldehyde}}$ & 16.01 & 1.15 & 14.14 & 17.88 & 0.02 & 0.01 & 4756.00 & 4635.00 & 1.00 \\
	$k_{\mathrm{cat,PduQ}}^{\mathrm{r}}$ & 5.91 & 0.59 & 4.97 & 6.88 & 0.01 & 0.01 & 6212.00 & 5886.00 & 1.00 \\
	$K_{\mathrm{M,PduQ}}^{\mathrm{NAD}}$ & 0.27 & 0.01 & 0.24 & 0.29 & 0.00 & 0.00 & 3879.00 & 3687.00 & 1.00 \\
	$K_{\mathrm{M,PduQ}}^{\mathrm{Propanol}}$ & 95.88 & 5.35 & 86.70 & 104.03 & 0.07 & 0.05 & 5532.00 & 5281.00 & 1.00 \\
	$\text{factor}_{\text{PduQ}}$ & 3.64 & 1.96 & 0.64 & 7.00 & 0.03 & 0.02 & 3973.00 & 4655.00 & 1.00 \\
	$K_{\text{eq, PduQ}}$ & 3.73 & 0.50 & 2.80 & 4.54 & 0.01 & 0.01 & 2830.00 & 3998.00 & 1.00 \\
	$k_{1,\text{Evap}}$ & -4.34 & 0.03 & -4.39 & -4.29 & 0.00 & 0.00 & 4787.00 & 6501.00 & 1.00 \\
	$k_{1,\text{iPduQ}}$ & 0.20 & 0.11 & 0.00 & 0.37 & 0.00 & 0.00 & 2324.00 & 2046.00 & 1.00 \\
	$k_{\mathrm{cat,PduP}}^{\mathrm{f}}$ & 31.22 & 0.58 & 30.31 & 32.18 & 0.01 & 0.01 & 4346.00 & 4842.00 & 1.00 \\
	$K_{\mathrm{M,PduP}}^{\mathrm{NAD}}$ & 0.89 & 0.06 & 0.80 & 0.99 & 0.00 & 0.00 & 4463.00 & 4941.00 & 1.00 \\
	$K_{\mathrm{M,PduP}}^{\mathrm{CoA}}$ & 0.03 & 0.01 & 0.03 & 0.04 & 0.00 & 0.00 & 4984.00 & 5004.00 & 1.00 \\
	$k_{2,\text{PduP}}$ & 1.03 & 0.76 & -0.40 & 2.47 & 0.01 & 0.01 & 3220.00 & 4387.00 & 1.00 \\
	$k_{3,\text{PduP}}$ & 1.25 & 0.70 & -0.00 & 2.54 & 0.01 & 0.01 & 3126.00 & 3881.00 & 1.00 \\
	$k_{4,\text{PduP}}$ & 1.18 & 0.75 & -0.23 & 2.57 & 0.01 & 0.01 & 4070.00 & 5627.00 & 1.00 \\
	$k_{6,\text{PduP}}$ & 0.96 & 0.65 & -0.27 & 2.10 & 0.01 & 0.01 & 3202.00 & 4866.00 & 1.00 \\
	$\text{factor}_{1,\text{PduP}}$ & 3.55 & 2.02 & 0.38 & 6.93 & 0.03 & 0.02 & 5473.00 & 5559.00 & 1.00 \\
	$\text{factor}_{2,\text{PduP}}$ & 3.59 & 1.96 & 0.53 & 6.90 & 0.03 & 0.02 & 4617.00 & 5033.00 & 1.00 \\
	$K_{\text{eq, PduP}}$ & 2.92 & 0.42 & 2.14 & 3.45 & 0.01 & 0.00 & 3863.00 & 3465.00 & 1.00 \\
	$k_{\mathrm{cat,PduL}}^{\mathrm{r}}$ & 25.22 & 3.74 & 19.57 & 31.78 & 0.05 & 0.04 & 4614.00 & 4225.00 & 1.00 \\
	$K_{\mathrm{M,PduL}}^{\substack{\text{Propionyl-}\\\text{Phosphate}}}$ & 0.61 & 0.04 & 0.55 & 0.66 & 0.00 & 0.00 & 5215.00 & 4807.00 & 1.00 \\
	$\text{factor}_{\text{2,PduL}}$ & 4.25 & 1.63 & 1.70 & 7.00 & 0.03 & 0.02 & 3335.00 & 3526.00 & 1.00 \\
	$k_{1,\text{PduL}}$ & 1.03 & 0.58 & 0.06 & 2.06 & 0.01 & 0.01 & 3315.00 & 2882.00 & 1.00 \\
	$k_{3,\text{PduL}}$ & 1.95 & 0.49 & 1.10 & 2.83 & 0.01 & 0.01 & 2998.00 & 3610.00 & 1.00 \\
	$\text{factor}_{1, \text{PduL}}$ & 0.71 & 0.56 & 0.00 & 1.73 & 0.01 & 0.01 & 2549.00 & 2889.00 & 1.00 \\
	$k_{5,\text{PduL}}$ & 2.05 & 0.52 & 1.24 & 2.99 & 0.01 & 0.01 & 2816.00 & 2886.00 & 1.00 \\
	$K_{\text{eq, PduL}}$ & -3.05 & 0.35 & -3.60 & -2.47 & 0.01 & 0.00 & 2653.00 & 2466.00 & 1.00 \\
	$k_{1,\text{PduW}}$ & 3.99 & 0.99 & 2.15 & 5.85 & 0.01 & 0.01 & 4324.00 & 3163.00 & 1.00 \\
	$k_{2,\text{PduW}}$ & 2.01 & 0.98 & 0.14 & 3.85 & 0.01 & 0.01 & 4728.00 & 5206.00 & 1.00 \\
	$k_{3,\text{PduW}}$ & 3.99 & 0.99 & 2.14 & 5.87 & 0.01 & 0.01 & 5530.00 & 3342.00 & 1.00 \\
	$k_{4,\text{PduW}}$ & 1.99 & 0.98 & 0.19 & 3.90 & 0.01 & 0.01 & 6422.00 & 5527.00 & 1.00 \\
	$k_{5,\text{PduW}}$ & 3.98 & 0.99 & 2.14 & 5.88 & 0.01 & 0.01 & 4545.00 & 3205.00 & 1.00 \\
	$k_{6,\text{PduW}}$ & 2.00 & 1.00 & 0.09 & 3.86 & 0.01 & 0.01 & 5350.00 & 4944.00 & 1.00 \\
	$k_{7,\text{PduW}}$ & 4.04 & 1.01 & 2.04 & 5.88 & 0.02 & 0.01 & 3197.00 & 1953.00 & 1.00 \\
	$K_{\text{eq, PduW}}$ & 5.79 & 0.56 & 4.95 & 6.75 & 0.01 & 0.01 & 5335.00 & 5741.00 & 1.00 \\
	$k_{\mathrm{cat,AckA}}^{\mathrm{f}}$ & 1124.03 & 72.17 & 1000.09 & 1234.49 & 0.88 & 0.62 & 5829.00 & 4987.00 & 1.00 \\
	$K_{\mathrm{M,AckA}}^{\mathrm{ADP}}$ & 0.00 & 0.00 & 0.00 & 0.01 & 0.00 & 0.00 & 5862.00 & 5414.00 & 1.00 \\
	$k_{\mathrm{cat,AckA}}^{\mathrm{r}}$ & 901.15 & 78.00 & 780.01 & 1032.20 & 1.07 & 0.76 & 4507.00 & 4479.00 & 1.00 \\
	$K_{\mathrm{M,AckA}}^{\mathrm{Propionate}}$ & 11.20 & 0.06 & 11.10 & 11.29 & 0.00 & 0.00 & 4097.00 & 3677.00 & 1.00 \\
	$K_{\mathrm{M,AckA}}^{\mathrm{ATP}}$ & 0.07 & 0.00 & 0.07 & 0.08 & 0.00 & 0.00 & 5627.00 & 5026.00 & 1.00 \\
	$\text{factor}_{1, \text{AckA}}$ & 3.52 & 2.01 & 0.29 & 6.86 & 0.03 & 0.02 & 4464.00 & 4567.00 & 1.00 \\
	$\text{factor}_{2, \text{AckA}}$ & 3.50 & 2.03 & 0.40 & 6.99 & 0.03 & 0.02 & 5561.00 & 5366.00 & 1.00 \\
	$K_{\text{eq, AckA}}$ & 5.80 & 0.55 & 4.96 & 6.76 & 0.01 & 0.01 & 4874.00 & 4883.00 & 1.00 \\
	Number of MCP PduCDE in WT & 579.54 & 30.27 & 526.17 & 621.91 & 0.51 & 0.36 & 2922.00 & 2773.00 & 1.00 \\
	Number of MCP PduQ in WT & 144.35 & 6.89 & 133.02 & 155.35 & 0.11 & 0.08 & 3622.00 & 3754.00 & 1.00 \\
	Number of MCP PduP in WT & 218.83 & 16.90 & 190.25 & 245.00 & 0.26 & 0.18 & 3784.00 & 4332.00 & 1.00 \\
	Number of MCP PduL in WT & 33.17 & 1.73 & 30.45 & 36.00 & 0.02 & 0.02 & 5460.00 & 5421.00 & 1.00 \\
	Number of MCP PduW in WT & 11.02 & 1.15 & 9.26 & 13.00 & 0.02 & 0.01 & 4067.00 & 3672.00 & 1.00 \\
	External AckA concentration in in-vitro assay & -2.43 & 1.16 & -4.64 & -0.39 & 0.02 & 0.01 & 4784.00 & 3365.00 & 1.00 \\
	Initial 1,2-PD in-vitro & 0.18 & 0.16 & 0.00 & 0.48 & 0.00 & 0.00 & 5441.00 & 4453.00 & 1.00 \\
	Initial Propionaldehyde in-vitro & 0.42 & 0.28 & 0.00 & 0.89 & 0.00 & 0.00 & 3789.00 & 4255.00 & 1.00 \\
	Initial Propanol in-vitro & 0.14 & 0.14 & 0.00 & 0.39 & 0.00 & 0.00 & 4167.00 & 3274.00 & 1.00 \\
\end{longtable}

\restoregeometry

\end{document}